\documentclass[5p]{elsarticle}

\usepackage{lineno,hyperref}
\usepackage[titlenumbered,ruled]{algorithm2e}
\usepackage{graphicx, subfigure}
\usepackage{color}
\usepackage{xcolor}
\usepackage{booktabs}
\usepackage{longtable}
\usepackage{lipsum}
\usepackage{supertabular}
\usepackage{mathtools}
\usepackage{amssymb}
\usepackage[flushleft]{threeparttable}
\usepackage{multirow,array}

\usepackage{url}
\urldef{\mailsa}\path|{alfred.hofmann, ursula.barth, ingrid.haas, frank.holzwarth,|
\urldef{\mailsb}\path|anna.kramer, leonie.kunz, christine.reiss, nicole.sator,|
\urldef{\mailsc}\path|erika.siebert-cole, peter.strasser, lncs}@springer.com|

\begin{document}
\begin{frontmatter}

\title{Android Malware Detection using Deep Learning on API Method Sequences}

\author{ElMouatez Billah Karbab, Mourad Debbabi, Abdelouahid Derhab, Djedjiga Mouheb}
\address{Concordia University, Concordia University, King Saud University, University of Sharjah}
\ead{e\_karbab@encs.concordia.ca}
%\ead{{e_karbab},{debbabi},{d_mouheb}@encs.concordia.ca}

\begin{abstract}
Android OS experiences a blazing popularity since the last few years. This predominant platform has established itself not only in the mobile world but also in the Internet of Things (IoT) devices. This popularity, however, comes at the expense of security, as it has become a tempting target of malicious apps. Hence, there is an increasing need for sophisticated, automatic, and portable malware detection solutions. In this paper, we propose \textsf{MalDozer}, an automatic Android malware detection and family attribution framework that relies on sequences classification using deep learning techniques. Starting from the raw sequence of the app's API method calls, \textsf{MalDozer} automatically extracts and learns the malicious and the benign patterns from the actual samples to detect Android malware. \textsf{MalDozer} can serve as a ubiquitous malware detection system that is not only deployed on servers, but also on mobile and even IoT devices. We evaluate \textsf{MalDozer} on multiple Android malware datasets ranging from $1$K to $33$K malware apps, and $38$K benign apps. The results show that \textsf{MalDozer} can correctly detect malware and attribute them to their actual families with an \textit{F1-Score} of $96\%-99\%$ and a \textit{false positive} rate of $0.06\%- 2\%$, under all tested datasets and settings.
\end{abstract}


\end{frontmatter}

%%%%%%%%%%%%%%%%%%%%%%%%%%%%%%%%%%%%%%%%%%%%%%%%%%%%
\section{Introduction}
Mobile apps have become an inherent part of our everyday life since many of the services are provided to us through mobile apps.  The latter change the way we communicate, as they are installed in most cases on smart devices.  In contrast to personal computers, smart devices are equipped with sophisticated sensors, from cameras and microphones to gyroscopes and GPS \cite{delmastro2016people}. These various sensors open a whole new world of applications for end-users \cite{delmastro2016people}, and generate huge amounts of data, which contain highly sensitive information. Consequently, this raises the need for security solutions to protect users from malicious apps, which exploit the sophistication of the smart devices and their sensitive data. On the other hand, the Internet of Things (IoT) smart systems have become equally, if not more, important than the mobile ones: (i) IoT systems are not only installed on conventional devices such as phones but are also considered in critical systems such as industrial IoT devices \cite{Gilchrist:2016:III:2994178}\cite{Yan:2008:ITR:1796470}. (ii) According to Ericsson \cite{Ericsson2016}, the number of IoT devices is expected to surpass the number of mobile devices by 2018 and could reach 16 billion by 2021. In this setting, security solutions should defend against malicious apps targeting both mobile and IoT devices. Android OS is phenomenally growing by powering a vast spectrum of smart devices. It has the biggest share in the mobile computing industry with $85\%$ in 2017-Q1 \cite{Smartphone2016} due to its open-source distribution and sophistication. Besides, it has become not only the dominant platform for mobile phones and tablets but is also gaining increasing attention and penetration in the IoT realm \cite{Android2016},\cite{IoT2016},\cite{rasp_3_iot}. In this context, Google has launched Android Things \cite{brillokey}, an Android OS for IoT devices, where developers benefit from the mature Android stack to develop IoT apps targeting thin devices \cite{android_auto}, \cite{brillokey}, \cite{android_wear}, \cite{rasp_2_iot}. Therefore, protecting Android devices from malicious apps is of parmount importance. 

\paragraph{Problem Statement}
To address the above challenges, there is a clear need for a solution that defends against malicious apps in mobile and IoT devices with specific requirements to overcome the limitations of existing Android malware detection systems. First, the Android malware detection system should ensure a high accuracy with minimum false alarms. Second, it should be able to operate at different deployment scales: (i) Server machines, (ii) Personal machines, (iii) Smartphones and tablets, and (iv) IoT devices. Third, detecting that a given app is malicious may not be enough, as more information about the threat is needed to prioritize the mitigation actions. The type of attack could be crucial to prevent the intended damage. Therefore, it is essential to have a solution that goes a step further and attributes the malware to a specific family, which defines the potential threat that our system is exposed to. Finally, it is necessary to minimize manual human intervention to the largest extent and make the detection dependent mainly on the app sample for automatic feature extraction and pattern recognition. As malicious apps are quickly getting stealthier, the security analyst should be able to catch up with this pace. This is due to the fact that for every new malware family, a manual analysis of the samples is required to identify its pattern and features that distinguish it from benign apps. 

\paragraph{Solution} In this paper, we propose \textsf{MalDozer}, a simple, yet effective and efficient framework for Android malware detection based on sequences mining using neural networks. \textsf{MalDozer} framework is based on an artificial neural network that takes, as input, the raw sequences of API method calls, as they appears in the DEX file, to enable malware detection and family attribution. During the training, \textsf{MalDozer} can automatically recognize malicious patterns using only the sequences of raw method calls in the assembly code. \textsf{MalDozer} achieves a high accuracy in malware detection under multiple datasets, including Malgenome \cite{malgenome_dataset}  ($1$K samples), Drebin \cite{Drebin_Dataset} ($5.5$K samples),  our \textsf{MalDozer} dataset ($20$K samples), and a merged dataset of $33$K malware samples. Additionally, $38$K benign apps downloaded from Google Play \cite{google_play} are also used in the evaluation. \textsf{MalDozer} achieves an F1-score between $96\%$ and $99\%$ in the detection task. Furthermore, using the same datasets, \textsf{MalDozer} can correctly attribute the Android malware to the actual family with an F1-score between $96\%$ and $98\%$ in the family attribution task. \textsf{MalDozer} is both effective and also efficient. We evaluate the efficiency of \textsf{MalDozer} under multiple deployment architectures, ranging from high-end servers to very small IoT devices \cite{rasp_2_iot}. The results of our evaluation confirm that \textsf{MalDozer} can efficiently run on all these devices. The key idea of \textsf{MalDozer} relies on using neural networks on the API assembly method invocations to identify Android malware. More precisely, the input of \textsf{MalDozer} is the sequences of the API method calls as they appear in the \textit{DEX} file, where a sequence represents the Android app. First, we map each method in the sequence invocation to a fixed length high-dimensional vector that semantically represents the method invocation \cite{Mikolov2013Distributed} and replace the sequence of the Android app methods by a sequence of vectors. Afterward, we feed the sequence of vectors to a neural network with multiple layers. In this paper, we make the following contributions:

\begin{itemize}
\item \textsf{MalDozer}, a novel, effective, and efficient Android malware detection framework using the raw sequences of API method calls based on neural networks. We take a step beyond malware detection by attributing the detected Android malware to its family with a high accuracy.
\item We propose an automatic feature extraction technique during the training using \textit{method embedding}, where the input is the raw sequence of API method calls, extracted from DEX assembly.
\item We conduct an extensive evaluation on different data- sets real Android malware and benign apps. The results demonstrate that \textsf{MalDozer} is very efficient and effective. It is also resilient against API evolution over time and against changing the order of API method calls. Additionally, \textsf{MalDozer} could be deployed and run properly, at various scales.
\end{itemize}

\section{Background}
In this section, we provide the necessary background that is relevant to our framework. We start by defining the cornerstone of \textsf{MalDozer}, namely neural network, and why it is interesting in the context of Android malware detection (Section \ref{background:1}). Afterward, we present the threat model as well as the assumptions considered in \textsf{MalDozer} design (Section \ref{background:2}). Next, we enumerate the main use cases of \textsf{MalDozer} framework (Section \ref{background:3}). 

\subsection{Deep Learning and Neural Network}\label{background:1}
A neural network is a machine learning computation model, which relies on a large number of neural units. The latter are approximate abstractions of the brain neurons, which could solve a very complex problem using highly dense neurons connected to each other by axons. Typically, Artificial Neuron Network (ANN) is composed of multiple layers, where each layer has many artificial neurons. The first layer is the input layer, and the last layer is the output one. The rest of the layers are called hidden layers. Notice that the neurons in each layer $i$ are connected to layer $i+1$, but the connection method could differ from a model to another. To this end, in the deep learning terminology, a neural network consists of multiple hidden layers, i.e., the more layers there are, the deeper the neural network is.  The conventional machine learning methods are limited by the manually-crafted features from the raw data. Here, the security expert analyzes the malicious apps and extracts the relevant features. The latter will be fed to a classifier to produce a learning model. The main advantage of a neural network is that it could automatically learn the representation (features) from the raw data to perform the detection task. In this paper, we aim at taking a step further towards Android malware detection with automatic representation learning. To achieve this aim, we leverage deep learning techniques and only consider the raw API method calls from Android DEX files for the purpose of malware detection and attribution with automatic feature extraction.
 
\subsection{Threat Model and Assumptions}\label{background:2}
We position \textsf{MalDozer} as an anti-malware system that detects Android malware and attributes it to a known family with a high accuracy and minimal false positive and negative rates. We assume that the analyzed Android apps, whether malicious or benign, are developed mainly in Java or any other language that is translated to DEX bytecode. Therefore, Android apps developed by other means, e.g., web-based, are out of the scope of the current design of \textsf{MalDozer}. Also, we assume that apps' core functionalities are in the DEX bytecode and not in C/C++ native code \cite{android_ndk}, i.e., the attacker is mainly using the DEX bytecode for the malicious payload. Furthermore, we assume that \textsf{MalDozer} detection results could not be affected by malicious activities. In the case of a server, Android malicious apps have no effect on the server system. However, in the case of deployment on infected mobiles or IoT devices, \textsf{MalDozer} should be protected from malicious activities to avoid tampering its results. 
 
\subsection{Usage Scenarios} \label{background:3}
The effectiveness of \textsf{MalDozer}, i.e., its high accuracy, makes it a suitable choice for malware detection in large-scale app store systems, especially that its update only requires very minimal manual intervention. We only need to train \textsf{MalDozer} model on new samples  without a \textit{feature engineering}, since \textsf{MalDozer} can automatically extract and learn the malicious and benign features during the training. Notice that \textsf{MalDozer} could detect unknown malware based on our evaluation as presented in Section \ref{sec:evaluation}. Furthermore, due to the efficiency of \textsf{MalDozer}, it could be deployed on mobile devices such as phones and tablets. As for mobile devices, \textsf{MalDozer} acts as a detection component in the anti-malware system, where the goal is to scan new apps. The family attribution is very handy when detecting new malware apps. Indeed, \textsf{MalDozer} helps the anti-malware system to take the necessary precautions and actions based on the malware family, which could have some specific malicious threats such as ransomware. It is also important to mention that we were able to run \textsf{MalDozer} on resource-limited IoT devices considered by Android Things such as Raspberry PI \cite{rasp_2_iot}.

\subsection{Android Architecture} \label{background:4}
Android has been settled by the Android Open Source Project (AOSP) team, maintained by Google and supported by the Open Handset Alliance (OHA)~\cite{oha_handset}. It encompasses the Original Equipment Manufacturers (OEMs), chip-makers, carriers and application developers. Android apps are written in Java. However, the native code and shared libraries are generally developed in C/C++~\cite{android_ndk}. The current Android architecture \cite{android_arch} consists of a Linux kernel, which is designed for an embedded environment consisting of limited resources. On top of the Linux kernel, there is a Hardware Abstraction Layer (HAL), which provides standard interfaces that expose device hardware capabilities to the higher-level Java API framework, by allowing programmers to create software hooks between the Android platform stack and the hardware. There is also Android Runtime (ART), which  is an application runtime environment used by the Android OS. It replaced Dalvik starting from Android 5.0. ART translates the app's bytecode into native instructions that are later executed by the device's runtime environment. ART introduces the ahead-of-time (AOT) compilation feature, which allows compiling entire applications into native machine code upon their installation. The native libraries developed in C/C++ support high-performance third-party reusable shared libraries. The Java API Framework provides APIs form the building blocks the user need to create Android apps. The System Apps, are the apps hat are included within the system, and are required to make the device run.

\subsubsection{Android APK Format} \label{background:5}

Android Application Package (\textit{APK}) is the file format adopted by Android for apps distribution and installation. It comes as a \textit{ZIP} archive file, which contains all the components needed to run the app.  By analogy, \textit{APK} files are similar to Windows \textit{EXE} installation files or Linux \textit{RPM}/\textit{DEB} files. The \textit{APK} package is organized into different directories (namely \textbf{lib}, \textbf{res}, and \textbf{assets}) and files (namely \textbf{AndroidManifest.xml} and \textsf{classes.dex}). More precisely, i) The \textbf{AndroidManifest.xml} file contains the app meta-data, e.g., name, version, required permissions, and used libraries. ii) The \textsf{classes.dex} file contains the compiled Java classes. iii) The \textbf{lib} directory stores C/C++ native libraries \cite{android_ndk}.  iv) The resources directory (\textbf{res}) contains the non-source code files, such as video, image, and audio files, which are packaged during compilation. 

\section{Methodology}
\begin{figure*}[!htb]
  \centering
      \includegraphics[width=0.95\textwidth]{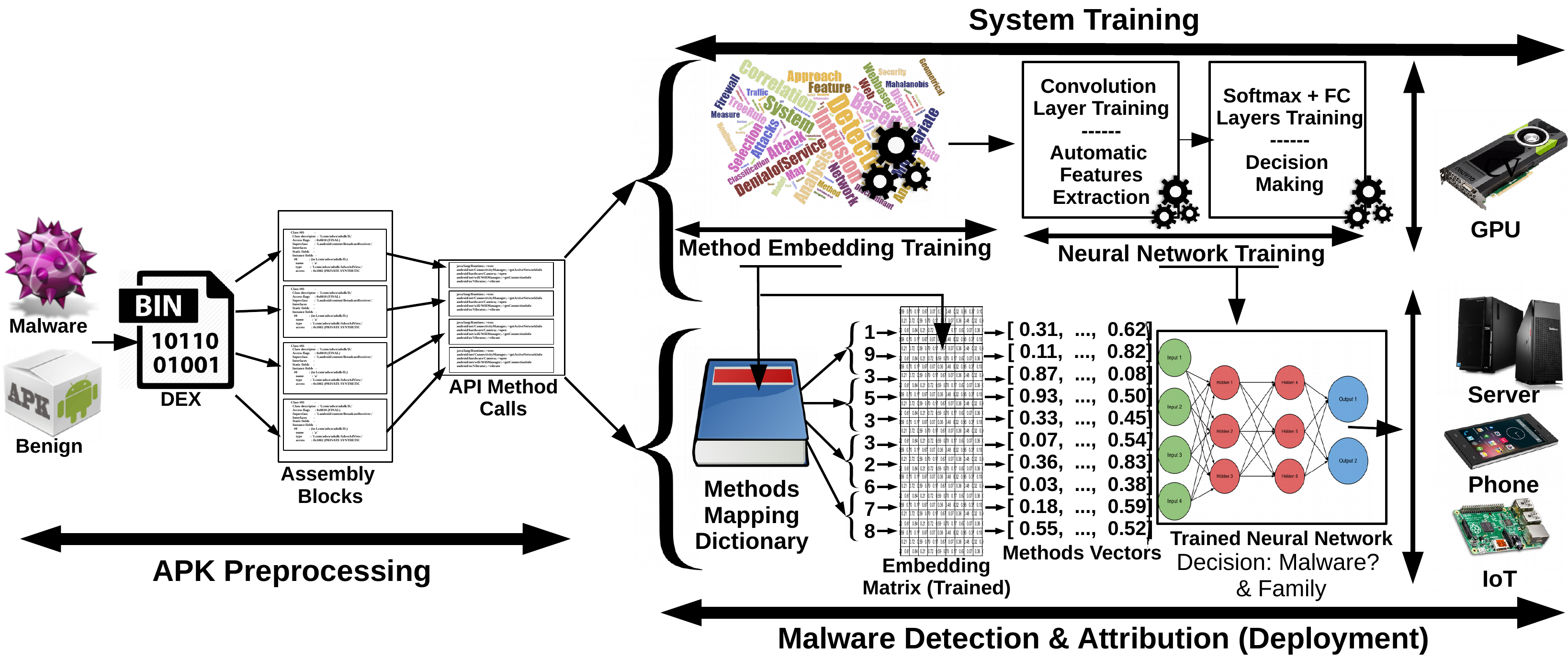}
  \caption{Approach Overview}
  \label{fig:approach_overview}
\end{figure*}

In this section, we present \textsf{MalDozer} framework and its components (Figure \ref{fig:approach_overview}). \textsf{MalDozer} has a simple design, where a minimalistic preprocessing is employed to get the assembly methods. As for the feature extraction (representation learning) and detection/attribution, they are based on the actual neural network. This permits \textsf{MalDozer} to be very efficient with fast preprocessing and neural network execution. Since \textsf{MalDozer} is based on a supervised machine learning, we first need to train our model. Afterward, we deploy this model along with a preprocessing procedure on the targeted devices. Notice that the preprocessing procedure is common between the training and the deployment phases to ensure the correctness of the detection results (Figure \ref{fig:approach_overview}).

\paragraph{1- Extraction of API Method Calls} \textsf{MalDozer} workflow starts by extracting the sequences of API calls from Android app packages, in which we consider only the \textit{DEX} file. We disassemble the \textsf{classes.dex} to produce the Dalvik VM assembly. Our goal is to formalize the assembly to keep the maximum raw information with minimum noise. Notice here that we could use Android APIs (such as \texttt{android/net/ConnectivityManager} in Figure \ref{fig:android_api}) instead of permission to have a granular view that helps distinguishing a malware app.

\begin{figure}[!htb]
\begin{scriptsize}
\begin{center}
\fbox{\begin{minipage}{20em}
\texttt{android/net/ConnectivityManager}\\
\texttt{android/net/ConnectivityManager}\\
\texttt{android/telephony/SmsManager}\\
\texttt{android/telephony/SmsManager}\\
\texttt{android/location/LocationManager}\\
\texttt{android/location/LocationManager}
\end{minipage}}
\end{center}

\end{scriptsize}
  \caption{Android API from a Malware Sample }
  \label{fig:android_api}
\end{figure}

However, quantifying Android API  could be noisy because there are plenty of common API calls shared between apps. Some solutions tend to filter only dangerous APIs and use them for detection. In this case, we require a manual  categorization of dangerous APIs. Moreover, Android API gives an abstract view of the actual malicious activity that could deceive the malware detection. For this reason, we leverage Android API method calls as \texttt{android/net/ConnectivityManager;->} \texttt{getNetworkInfo} in Figure \ref{fig:android_api_calls}. By doing so, the malware detector will have a more granular view of the app activity. In our case, we address this problem from another angle; we treat Android apps as a sequence of API method calls. We consider all the API calls with no filtering, where the order is part of the information we use to identify malware. It represents the temporal relationship between two API method calls (in a given basic block), and defines the intended sub-tasks of the app. The sequence of API method calls  preserves the temporal relationship over individual basic blocks of the linear disassembly and ignores the order between these blocks. The obtained result  is a merged sequence (Figure \ref{fig:approach_overview}).

\begin{figure}[!htb]
\begin{scriptsize}
\begin{center}
\fbox{\begin{minipage}{29em}
\texttt{android/net/ConnectivityManager;->getNetworkInfo}
\texttt{android/net/ConnectivityManager;->getAllNetworkInfo}
\texttt{android/telephony/SmsManager;->sendTextMessage}
\texttt{android/telephony/SmsManager;->sendMultipartTextMessage}
\texttt{android/location/LocationManager;->getLastKnownLocation}
\texttt{android/location/LocationManager;->getBestProvider}
\end{minipage}}
\end{center}
\end{scriptsize}
  \caption{Granular View using API Method Calls}
  \label{fig:android_api_calls}
\end{figure}

In other words, a \textit{DEX} file, denoted by $cd$, is composed of a set of $n$ compiled Java classes, $cd=\{cl_1, \cdots, cl_n \}$. Each Java class $cl_i$ is, in turn, composed of a set of $m$ methods, which are basic blocks, $cl_i=\{mt_1^{i} \cdots, mt_m^{i}\}$. By going down to the API method level, $mt^{i}_{j}$ is a sequence of $k$ API method calls,  Formally $mt^{i}_{j}=(P^{i,j}_1 , \cdots, P^{i,j}_k)$ , where $P^{i,j}_l$ is the l$^{th}$ API method call in method $mt^i_j$.

\begin{center}
\begin{algorithm}
  \footnotesize {
    \SetKwFunction{EL}{EmptyList}
    
    \SetKwInOut{Input}{Input}\SetKwInOut{Output}{Output}

    \Input{$cd$: Java Assembly}
    \Output{$MSeq$: Methods Sequence}
    \BlankLine
    \BlankLine
    \Begin{
      $MSeq$ = \EL{}\;
      \BlankLine
      \BlankLine
      \ForEach {$cl \in cd$}{
      \BlankLine
      	\ForEach {$mt \in cl$}{
      	\BlankLine
      		\ForEach {$P \in mt$}{
      		%\BlankLine
 			$MSeq.Add(p)$\;
      		%\BlankLine
      		}
      		\BlankLine
      		%\BlankLine
        }
        \BlankLine
        %\BlankLine
      }
      \BlankLine
      %\BlankLine
      \Return ${MSeq}$\;
      \BlankLine
      %\BlankLine
    }
  }
\caption{Extraction}
\label{alg:methods_extraction}
\end{algorithm}
\end{center}

\paragraph{2- Discretization of API Method Calls} In this step, we discretize the sequences of API method calls that are in an Android app (Algorithm \ref{alg:methods_discretization}). More precisely, we replace each API method with an identifier, resulting in a sequence of numbers. We also build a dictionary that maps each API call to its identifier.  Notice that in the current implementation, the mapping dictionary is deployed with the learning model to map the API calls of the analyzed apps. In the deployment, we could find unknown API calls related to third party libraries. To overcome this problem: (i) We consider a big dataset that covers most of the API calls. (ii) In the deployment phase, we replace unknown API calls with fixed identifiers. Afterward, we unify the length of the sequences $L$ (hyperparameter) and  pad a given sequence with zeros if its length $l < L$.

\begin{center}
\begin{algorithm}
  \footnotesize {
    \SetKwFunction{EL}{EmptyList}
    
    \SetKwInOut{Input}{Input}\SetKwInOut{Output}{Output}

    \Input{$MSeq$: Methods Sequence \newline
    	   $MapDict$: Mapping Dict}
    \Output{$DSeq$: Discrete Sequence }
     %\BlankLine

    \Begin{
      $DSeq$ = \EL{}\;
      
      \ForEach {$P \in MSeq$}{
        \eIf {$m \in MapDict.Keys()$}{
          $Dvalue \gets MapDict[P]$\;
          $DSeq.Add(Dvalue)$\;
        }{
          $Dvalue \gets 0$\;
          $DSeq.Add(Dvalue)$\;        
        }
      }
      \Return ${DSeq}$\;
    }
  }
\caption{Discretization}
\label{alg:methods_discretization}
\end{algorithm}
\end{center}

\paragraph{3- Unification of the Sequences' Size} The length of the sequences varies from one app to another. Hence, it is important to unify the length of the sequences. There are two cases depending on the length of the sequence and the hyper-parameter. We choose a uniform sequence size as follows: i) If the length of a given sequence is greater than the uniform sequence size $L$, we take only the first $L$ items to represent the apps. ii) In case the length of the sequence is less than $L$, we pad the sequence with zeros. It is important to mention that the uniform sequence size hyper-parameter has an influence on the accuracy of \textsf{MalDozer}. A simple rule is that the larger is the size, the better is, but this will require a lot of computation power and a long time to train the neural network. 

\paragraph{3- Generation of the Semantic Vectors.} The identifier in the sequences needs to be shaped  to fit as input to our neural network. This could be solved by representing each identifier by a vector. The question that arises is \textit{how are such vectors produced?} A straightforward solution is to use one-hot vectors, where a vector has one in the interface value row, and zero in the rest. Such a vector is very sparse because its size is equal to the number of API calls, which makes it impractically and computationally prohibitive for the training and the deployment. To address this issue, we resort to a dense vector that uses a continuous space. These vectors are semantically related, and we could express their relation by computing a distance. The smaller the distance is, the more related the vectors are (i.e., the API calls). We describe word embedding in Section \ref{sec:word_embedding}. The output of this step is sequences of vectors for each app that keeps the order of the original API calls; each vector has a fixed size $K$ (hyper-parameter).

\paragraph{4- Prediction using a Neural Network} The final component in \textsf{MalDozer} framework is the neural network, which is composed of several layers. The number of layers and the complexity of the model are hyper-parameters. However, we aim to keep the neural network model as simple as possible to gain in the execution time during its deployment, especially on IoT devices. In our design, we rely on the convolution layers \cite{Kim2014Convolutional} to automatically discover the pattern in the raw method calls. The input to the neural network is a sequence of vectors, i.e., a matrix of $L\times K$ shape. In the training phase, we train the neural network parameters (layers weight) based on the app vector sequence and its labels: (i) malware or benign for the detection task, and (ii) malware families for the attribution task. In the deployment phase, we extract the sequence of methods and use the embedding model to produce the vector sequence. Finally, the neural network takes the vector sequence to decide about the given Android app.

\section{MalDozer Method Embedding} \label{sec:word_embedding}
The neural network takes vectors as input. Therefore, we represent our Android API method calls as vectors. As a result, we formalize an Android app as a sequence of vectors with fixed size ($L$). We could use one-hot vector. However, its size is the number of unique API method calls in our dataset. This makes such a solution not scalable to large-scale training. Also, the word embedding technique outperforms the results of the one-hot vector technique  in our case \cite{Mikolov2013Distributed}, \cite{Pennington201GloVe}, \cite{Kim2014Convolutional}. Therefore, we seek a compact vector, which also has a semantic value. To fulfill these requirements, we choose the word embedding techniques, namely, word2vec \cite{Mikolov2013Distributed} and GloVe \cite{Pennington201GloVe}. Our primary goal is to have a  dense vector for each Android API method that keeps track of its contexts in a large dataset of Android apps. Thus, in contrast with one-hot vectors, each word embedding vector contains a numerical summary of the Android API call meaning representation. Moreover, we could apply geometric techniques on the API call vectors to measure the semantic relationship between their functionalities, i.e.,  developers tend to use certain API method calls in the same context. In our context, we learn these vectors from our dataset that contains benign and malicious apps by using word2vec \cite{Mikolov2013Distributed}. The latter is a computationally efficient predictive model from learning word embedding vectors, which are applied on the raw Android API method calls. The output obtained from training the embedding word model is a matrix $K\times A$, where $K$ is the size of the embedding vector, and $A$ is the number of unique Android API method calls. Both $K$ and $A$ are hyper-parameters; we use $K=64$ in all our models. In contrast, the hyper-parameter $A$ is a major factor in the accuracy of \textsf{MalDozer}. The more API calls we consider, the more accurate and robust our model is. Notice that, our word embedding is trained along with the neural network, where we tune both of them for a given task such as detection. Despite that, it can be trained separately to generate the embedding word vector independently of the detection task. In the deployment phase (Figure \ref{fig:approach_overview}), \textsf{MalDozer} uses the word embedding model and looks up for each API method call identifier to find the corresponding embedding vector .   

\begin{figure*}[!htb]
  \centering
      \includegraphics[width=0.8\textwidth]{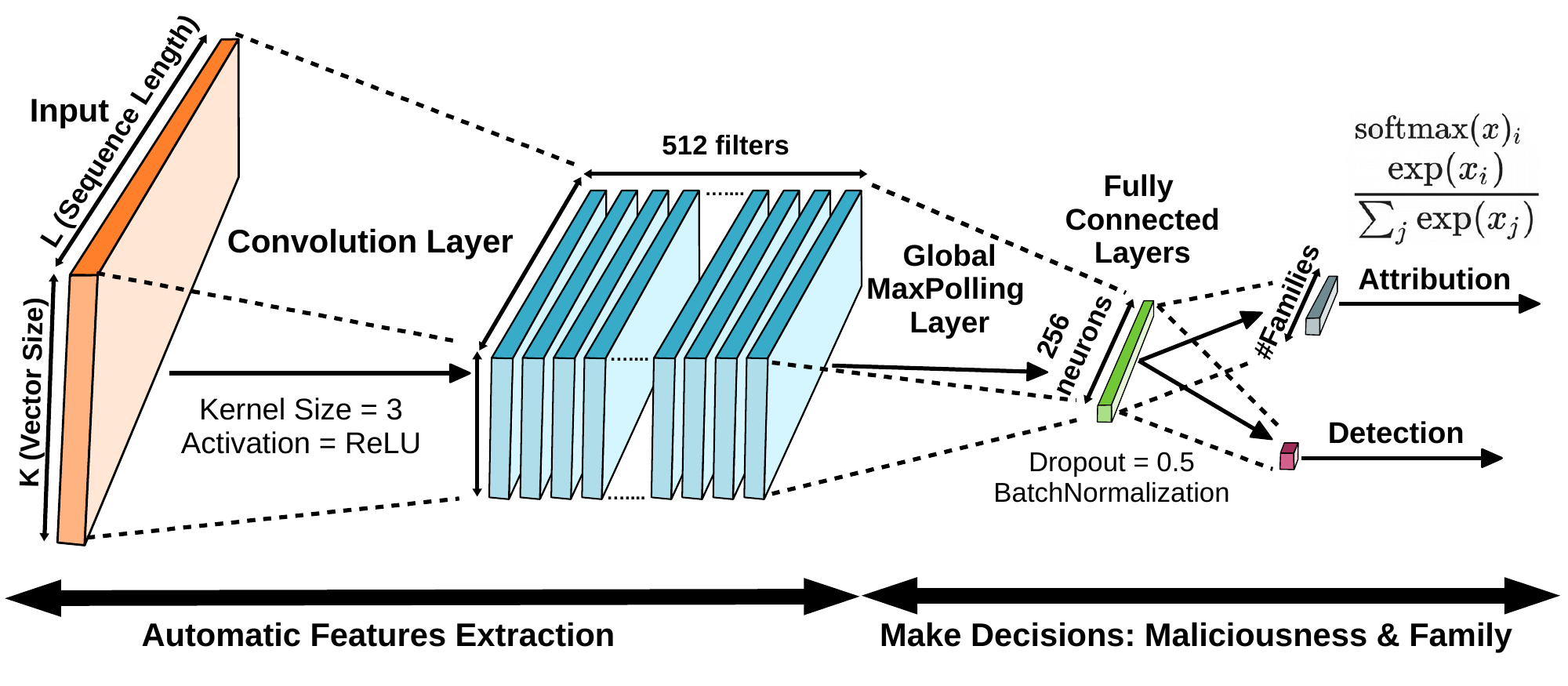}
  \caption{Neural Network Architecture}
  \label{fig:maldozer_neural_nets}
\end{figure*}

\section{MalDozer Neural Network} \label{sec:neural_network}
\textsf{MalDozer} neural network is inspired by \cite{Kim2014Convolutional}, where the authors use a neural network for sentence classification task such as sentiment analysis.  The proposed architecture shows high results and outperforms many of state-of-the-art benchmarks with a relatively simple neural network design. Here, we raise the following questions: \textit{Why could such a Natural Language Processing (NLP) model be useful in Android malware detection?} And \textit{why do we choose to build it on top of this design \cite{Kim2014Convolutional}?} We formulate our answers as follows: i) NLP is a challenging field where we deal with text. So, there is an enormous number of vocabularies; also we could express the same meaning in different ways. We also have the same semantics with many combinations of words, which we call the \textit{natural language obfuscation}. In our context, we deal with sequences of Android API method calls and want to find the combination of patterns of method calls, which produces the same (malicious) activity. We use the API method calls as they appear in the binary, i.e., there is a temporal relationship between API methods in basic blocks but we ignore the order among these blocks. By analogy to NLP, the basic blocks are the sentences and the API method calls are the words. Further, the app (paragraph) is a list of basic blocks (unordered sentences). This task looks easier compared to the NLP one because of the huge difference in the vocabulary, i.e., the number of Android API method calls is significantly less than the number of words in natural language. Also, the combination in the NLP is much complex compared to Android API calls. ii) We choose to use this model due to its efficiency and ability to run our model on resource-constrained devices. Table \ref{tab:detection_neuralnet} depicts the neural network architecture of \textsf{MalDozer}'s detection and attribution tasks. Both networks are very similar; the only notable difference is in the output layer. 
In the detection task, we need only one neuron in the output layer because the network decides whether the app is malware or not. As for the attribution task, there are multiple neurons, one for each Android malware family.  Having the same architecture for the detection and attribution makes the development and the evaluation of a given design more simple. Because the network architecture achieves good results in one task, it will have very similar results in the other one. 
As presented in Figure \ref{fig:maldozer_neural_nets}, the first layer is a convolution layer \cite{Kim2014Convolutional} with rectified linear unit (ReLU) activation function ($f(x) = max(0, x)$). Afterward, we use global max pool \cite{Kim2014Convolutional} and connect it to a fully-connected layer. Notice that in addition to Dropout \cite{Goodfellow-et-al-2016} used to prevent overfitting, we also utilize Benchnormalization   \cite{Goodfellow-et-al-2016} to improve our results. Finally, we have an output layer, where the number of neurons depends on the detection or attribution tasks. 

\begin{table}[!h]l
\centering
\begin{threeparttable}
\begin{tabular}{|c||c|c|c|}
    \hline \hline
     \#&  \textit{Layers}     & \textit{Options}        & \textit{Activ}  \\ \hline\hline
     1 & Convolution & Filter=512, FilterSize=3 &  ReLU	\\ \hline 
     2 & MaxPooling  & / & /	\\ \hline 
     3 & FC  & \#Neurons=256, Dropout=0.5 &  	ReLU	\\ \hline
     4 & FC  & \#Neurons=\{1,\#Families\tnote{1} \} &  Softmax 	\\ \hline\hline 
\end{tabular}
\begin{tablenotes}
\item[1] The number of malware families in the training dataset.
\end{tablenotes}
\end{threeparttable}
\caption{MalDozer  Malware Neural Network} 
\label{tab:detection_neuralnet}
\end{table}

\begin{table}[!h]
\centering
%\begin{scriptsize}
\begin{threeparttable}
\begin{tabular}{|c||c|c|c|}
    \hline \hline
     		 &   \textbf{Server (1/2)}  & \textbf{Laptop} & \textbf{Raspberry PI 2}  \\ \hline\hline
     \textbf{GPU}  &  TITAN X / no &  no & no	\\ \hline
     \textbf{CPU}  &  Intel E5-2630&   Intel T6400 &	 ARM Cortex A7	\\ \hline 
     \textbf{RAM} &  128GB &  3GB & 	1GB	\\ \hline 
\end{tabular}
\end{threeparttable}
\caption{Hardware Specifications} 
\label{tab:hardware}
%\end{scriptsize}
\end{table}

\section{Implementation}
In this section, we present the software \& hardware components  of  \textsf{MalDozer} evaluation. 
 
\paragraph{Software}
We implement \textsf{MalDozer} using \textit{Python} and \textit{Bash} scripting languages. First, Python zip library extracts the \textit{DEX} file from the \textit{APK} file. We use \textit{dexdump} command-line to produce the assembly from the DEX file.  \textit{Dexdump} is available through the Android SDK, but in the case of Raspberry PI, we built it from its source code. Regular expressions are employed to extract API  method calls from the assembly. To develop the neural network, we use Tensorflow \cite{tensorflow}. Notice that there is no optimization in the preprocessing; in the run-time evaluation,  we use only a single thread app. 
 
\paragraph{Hardware}
To evaluate the efficiency of \textsf{MalDozer}, we evaluate multiple types of hardware, as shown in Table \ref{tab:hardware}, starting from servers to \textit{Raspberry PI} \cite{rasp_2_iot}. For training, the Graphic Processing Unit (GPU) is a vital component because the neural network training needs immense computational power. The training takes hours under \textit{NVIDIA TitanX}. However, the deployment could be virtually on any device including IoT devices. To this end, we consider \textit{Raspberry PI} as IoT device because it is one of the hardware platforms supported by Android Things \cite{brillokey}. We also use low-end laptops in our evaluation, as shown in Table \ref{tab:hardware}.

\section{Evaluation} \label{sec:evaluation}
In this section, we conduct our` evaluation using different datasets that primarily cover the following performance aspects: 
\begin{itemize}
	\item (I) \textit{Detection Performance}: We evaluate how effectively \textsf{MalDozer} can distinguish between malicious and benign apps in terms of F1-measure, precision, recall, and false positive rate. 
	\item (II) \textit{Attribution Performance}: We evaluate how effectively \textsf{MalDozer} can correctly attribute a given malicious app to its malware family. 
	\item (III) \textit{Runtime Performance}: We measure the preprocessing and the detection runtime on different types of hardware.
\end{itemize}
\subsection{Evaluation Metrics}
 
The evaluation results are presented under the following metrics: 

\begin{itemize}
	\item \textit{True positives} (TP): This metric measures the number of malicious apps that are successfully detected.
	\item \textit{False negatives} (FN):  This metric measures the number of malicious apps that are incorrectly classified.
	\item \textit{False positives} (FP):  This metric measures the number of benign apps that are incorrectly classified.
	\item \textit{Precision} (P):  It is the percentage of positive prediction, i.e., the percentage  of the detected malware out of all sample apps. Formally,  $ P = \frac{TP}{TP + FP}$
	\item \textit{Recall} (R): It is the percentage  of correct malicious apps detected out of all malware samples. Formally, $R = \frac{TP}{TP + FN}$
	\item \textit{F1-Score} (F1): It is a measure that considers precision and recall. Formally, $F1= 2 ~\textsf{x}~ \frac{P \textsf{x} R}{P + R}$. 
\end{itemize}

We also measure \textit{False Positive Rate} (FPR), \textit{False Negative Rate} (FNR) and \textit{Accuracy} (ACC), which are given as follows:

\begin{equation}
\begin{array}[b]{r}
FPR = \frac{FP}{FP + TP}
\end{array},
\begin{array}[b]{r}
FNR = \frac{FN}{FN + TP}
\end{array},
\begin{array}[b]{r}
ACC = \frac{TP + TN}{P + N}
\end{array}
\nonumber
\end{equation}

\subsection{Datasets} \label{sec:datasets}
In our evaluation, we have two main tasks: i) Detection, which aims at checking if a given app is malware or not, ii) Attribution, which aims at determining the family of the detected malware. We conduct the evaluation experiments under two types of datasets: i) Mixed dataset, which contains malicious apps and benign apps, as presented in Table \ref{tab:detection_dataset}. ii) Malware dataset, which  contains only malware, as shown in Table \ref{tab:attribution_dataset}. As for the malware dataset, we leverage reference datasets such as \textit{Malgenome} \cite{malgenome_dataset} and \textit{Drebin} \cite{arp2014drebin}. We also collect two other datasets from different sources, e.g.,  \textit{virusshare.com}, \textit{Contagio Minidump} \cite{contagiominidump}. The total number of malware samples is $33K$, including Malgenome and Drebin datasets. As for the attribution task, we use only malware from the previous datasets, where each family has at least 40 samples, as presented in Tables \ref{tab:maldozer_dataset}, \ref{tab:malgenome_dataset}, \ref{tab:drebin_dataset}. To this end, we propose \textsf{MalDozer} dataset, as in Table \ref{tab:maldozer_dataset}, which contains 20K malware samples from 32 malware families. We envision to make \textsf{MalDozer} dataset  available upon request for the research community. The benign app samples have been collected from \textit{Playdrone} dataset \cite{playdrone}. We leverage the top $38K$ apps that are ranked by the number of downloads.

\begin{table}[!h]
\centering
\begin{threeparttable}
\begin{tabular}{|l||c|c|c|}
\hline \hline
Dataset & \textbf{\#Malware} & \textbf{\#Benign} & \textbf{Total} \\\hline\hline
\textbf{Malgenome} &  1,258   & 37,627   & 38.885\\\hline
\textbf{Drebin}  &  5,555   & 37,627  & 43,182\\\hline
\textbf{MalDozer}  &  20,089  & 37,627   & 57,716\\\hline
\textbf{All}  &  33,066  & 37,627     & 70,693\\\hline \hline
\end{tabular}
\end{threeparttable}
\caption{Datasets for Detection Task} 
\label{tab:detection_dataset}
\end{table}

\begin{table}[!h]
\centering
\begin{threeparttable}
\begin{tabular}{|l||c|c|c|}
\hline \hline
Dataset & \textbf{\#Malware} & \textbf{\#Family} \\\hline \hline
\textbf{Malgenome} &  985   & 9    \\\hline
\textbf{Drebin} &  4,661   & 20      \\\hline
\textbf{MalDozer}  &  20,089  & 32 \\\hline \hline
\end{tabular}
\end{threeparttable}
\caption{Datasets for Attribution Task} 
\label{tab:attribution_dataset}
\end{table}

\subsection{Malware Detection Performance}
We evaluate \textsf{MalDozer} on different cross-validation settings, two, three, five and ten-fold, to examine the detection performance under different training/test set percentages ($50\%, 66\%, 80\%, 90\%$) from the actual dataset ($10$ training epochs). Table \ref{tab:detecton_scores_malgenome} depicts the detection results on Malgenome dataset. \textsf{MalDozer} achieves excellent results, F1-Score=$99.84\%$, with a small \textit{False Positive Rate} (FPR), $0.04\%$, despite the unbalanced dataset, where benign app samples are the most dominant in the dataset. The detection results are  similar under all cross-validation settings. Table \ref{tab:detecton_scores_drebin} presents the detection results on Drebin dataset, which are very similar to the Malgenome ones.  \textsf{MalDozer} reaches F1-Score=$99.21\%$, with FPR=$0.45\%$. Similar detection results are shown in Table \ref{tab:detecton_scores_maldozer} on \textsf{MalDozer} dataset ( F1-Score=$98.18\%$ and FPR=$1.15\%$). Table \ref{tab:detecton_scores_all} shows the results related to all datasets, where \textsf{MalDozer} achieves a good result (F1-Score=$96.33\%$). However, it has a higher false positive rate compared to the previous results (FPR=$3.19\%$). This leads us to manually investigate the false postives. We discover, by correlating with \textit{virusTotal.com}, that several false positive apps are already detected by many vendors as malware.

\begin{table}[!h]
\centering
\begin{threeparttable}
\begin{tabular}{|c||c|c|c|c|c|}
\hline \hline
 %& \textbf{F1-Score\%} & \textbf{Precision\%} & \textbf{Recall\%} & \textbf{FPR\%} \\
 & \textbf{F1\%} & \textbf{P\%} & \textbf{R\%} & \textbf{FPR\%} \\
\hline \hline
\textbf{2-Fold}	& 99.6600 & 99.6620 & 99.6656 & 0.06 \\\hline
\textbf{3-Fold}	& 98.1926 & 98.6673 & 97.9812 & 1.97 \\\hline
\textbf{5-Fold}	& 99.8044 & 99.8042 & 99.8045 & 0.09 \\\hline
\textbf{10-Fold} & \textbf{99.8482} & \textbf{99.8474} & \textbf{99.8482} & \textbf{0.04}	\\
\hline \hline
\end{tabular}
\end{threeparttable}
\caption{Detection  on Malgenome Dataset} 
\label{tab:detecton_scores_malgenome}
\end{table}

\begin{table}[!h]
\centering
\begin{threeparttable}
\begin{tabular}{|c||c|c|c|c|c|}
\hline \hline
 %& \textbf{F1-Score\%} & \textbf{Precision\%} & \textbf{Recall\%} & \textbf{FPR\%} \\
 & \textbf{F1\%} & \textbf{P\%} & \textbf{R\%} & \textbf{FPR\%} \\
\hline \hline
\textbf{2-Fold}	& 98.8834 & 98.9015 & 98.9000 & 0.13 \\\hline
\textbf{3-Fold}	& 99.0142 & 99.0130 & 99.01579 & 0.51 \\\hline
\textbf{5-Fold}	& 99.1174 & 99.1173 & 99.1223 & 0.31 \\\hline
\textbf{10-Fold} & \textbf{99.2173} & \textbf{99.2173} & \textbf{99.2172} & \textbf{0.45} \\
\hline \hline
\end{tabular}
\end{threeparttable}
\caption{Detection on Drebin Dataset} 
\label{tab:detecton_scores_drebin}
\end{table}

\begin{table}[!h]
\centering
\begin{threeparttable}
\begin{tabular}{|c||c|c|c|c|c|}
\hline \hline
 %& \textbf{F1-Score\%} & \textbf{Precision\%} & \textbf{Recall\%} & \textbf{FPR\%} \\
 & \textbf{F1\%} & \textbf{P\%} & \textbf{R\%} & \textbf{FPR\%} \\
\hline \hline
\textbf{2-Fold}	& 96.8576 & 96.9079 & 96.8778 & 1.01 \\\hline
\textbf{3-Fold}	& 97.6229 & 97.6260 & 97.6211 & 2.00 \\\hline
\textbf{5-Fold}	& 97.7804 & 97.7964 & 97.7753 & 2.25 \\\hline
\textbf{10-Fold} & \textbf{98.1875} & \textbf{98.1876} & \textbf{98.1894} & \textbf{1.15} \\
\hline \hline
\end{tabular}
\end{threeparttable}
\caption{Detection on MalDozer Dataset} 
\label{tab:detecton_scores_maldozer}
\end{table}

\begin{table}[!h]
\centering
\begin{threeparttable}
\begin{tabular}{|c||c|c|c|c|c|}
\hline \hline
 %& \textbf{F1-Score\%} & \textbf{Precision\%} & \textbf{Recall\%} & \textbf{FPR\%} \\
 & \textbf{F1\%} & \textbf{P\%} & \textbf{R\%} & \textbf{FPR\%} \\
\hline \hline
\textbf{2-Fold}	& 96.0708 & 96.0962 & 96.0745 & 2.53 	\\\hline
\textbf{3-Fold}	& 95.0252 & 95.0252 & 95.0278 & 4.01 \\\hline
\textbf{5-Fold}	& 96.3326 & 96.3434 & 96.3348 & 2.67 \\\hline
\textbf{10-Fold} & \textbf{96.2958} & \textbf{96.2969} & \textbf{96.2966} & \textbf{3.19} 	\\
\hline \hline
\end{tabular}
\end{threeparttable}
\caption{Detection on All Dataset} 
\label{tab:detecton_scores_all}
\end{table}

\begin{scriptsize}
\begin{figure*}[ht!]
     \begin{center}
        \subfigure[\scriptsize BaseBridge]{%
            \label{fig:ben_fscore}
            \includegraphics[width=0.20\textwidth]{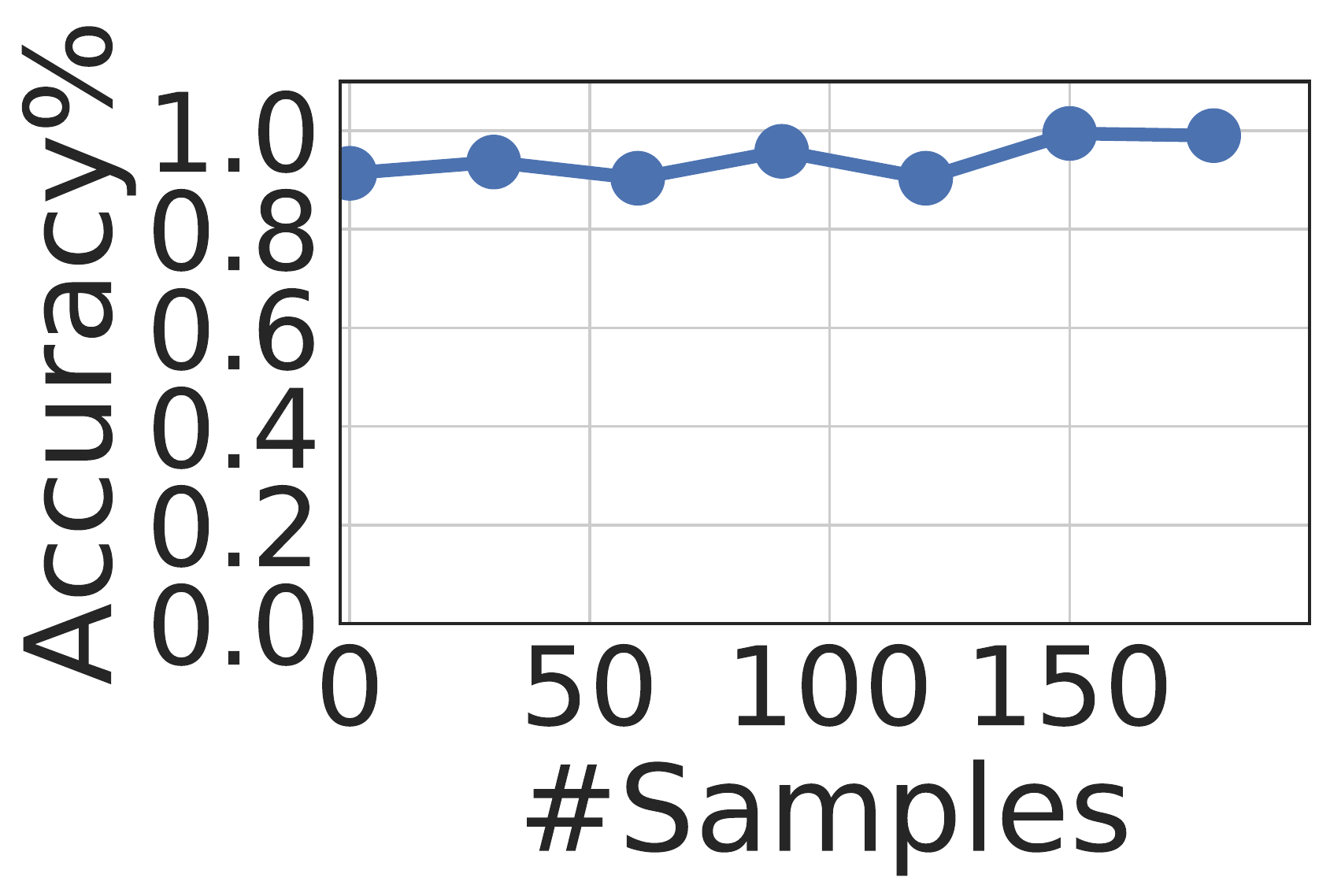}
        }
        \subfigure[\scriptsize DrKungFu] {%
           \label{fig:ben_precision}
           \includegraphics[width=0.20\textwidth]{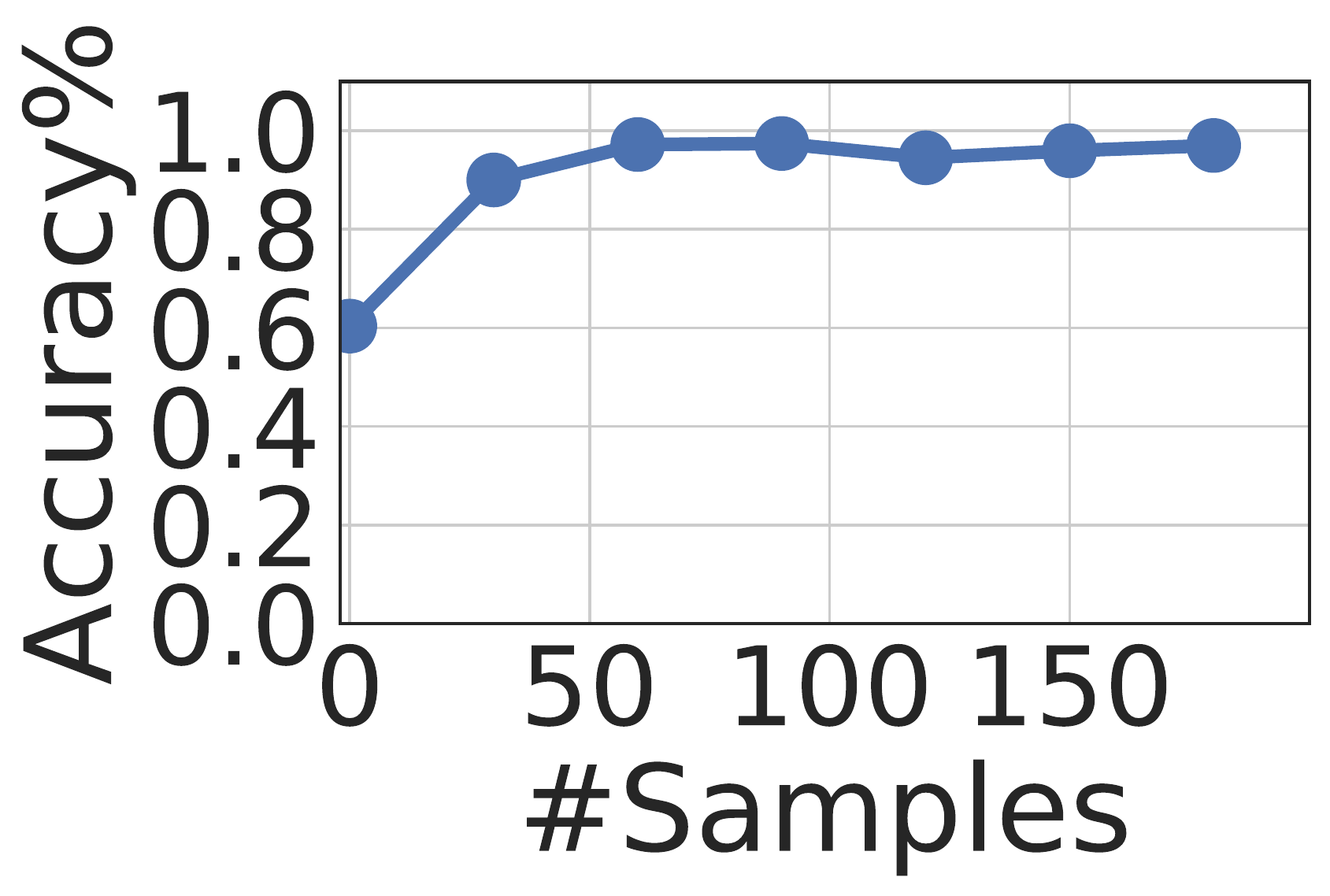}
        }
        \subfigure[\scriptsize FakeInst]{%
            \label{fig:ben_recall}
            \includegraphics[width=0.20\textwidth]{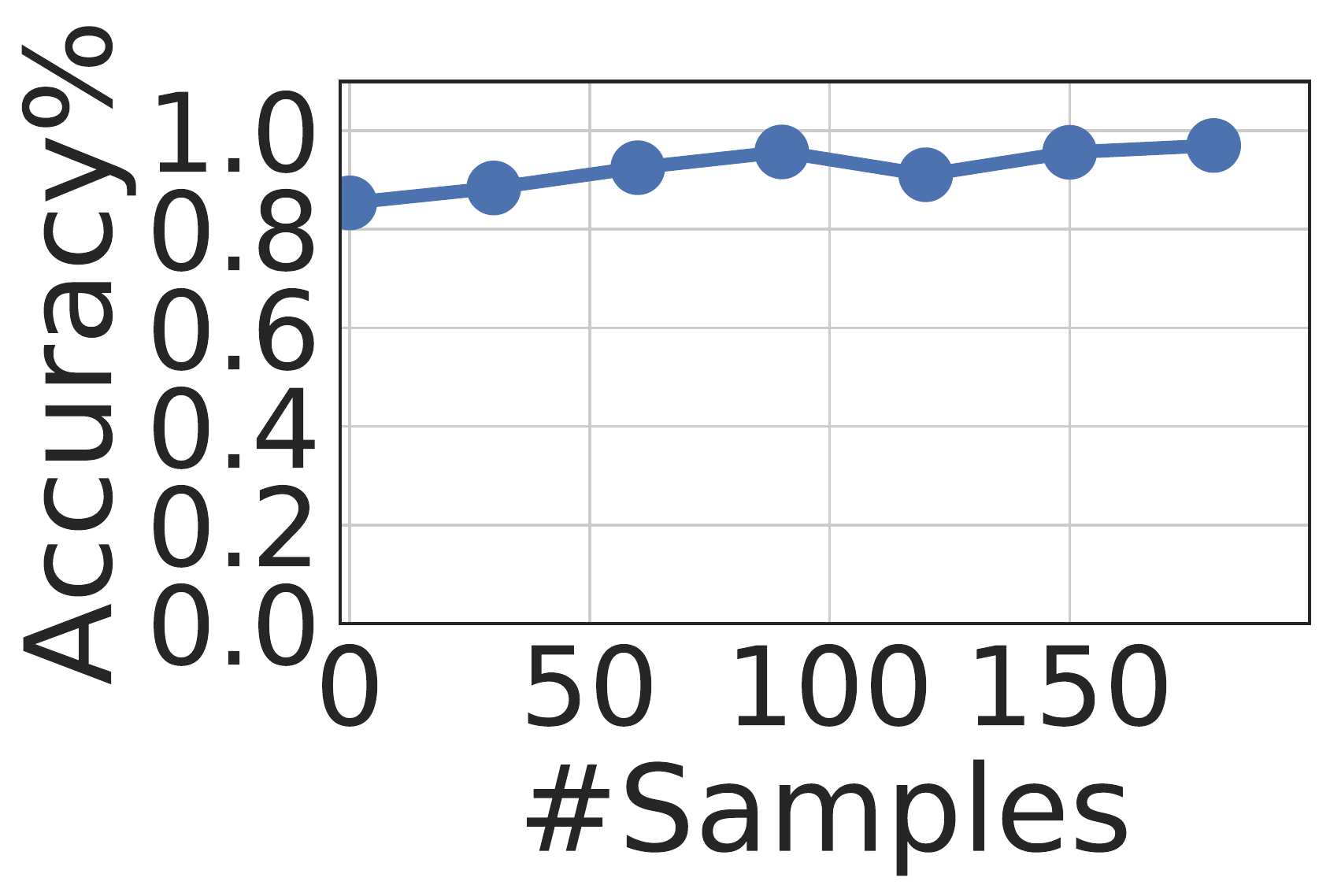}
        }
        \subfigure[\scriptsize GinMaster]{%
            \label{fig:ben_fscore}
            \includegraphics[width=0.20\textwidth]{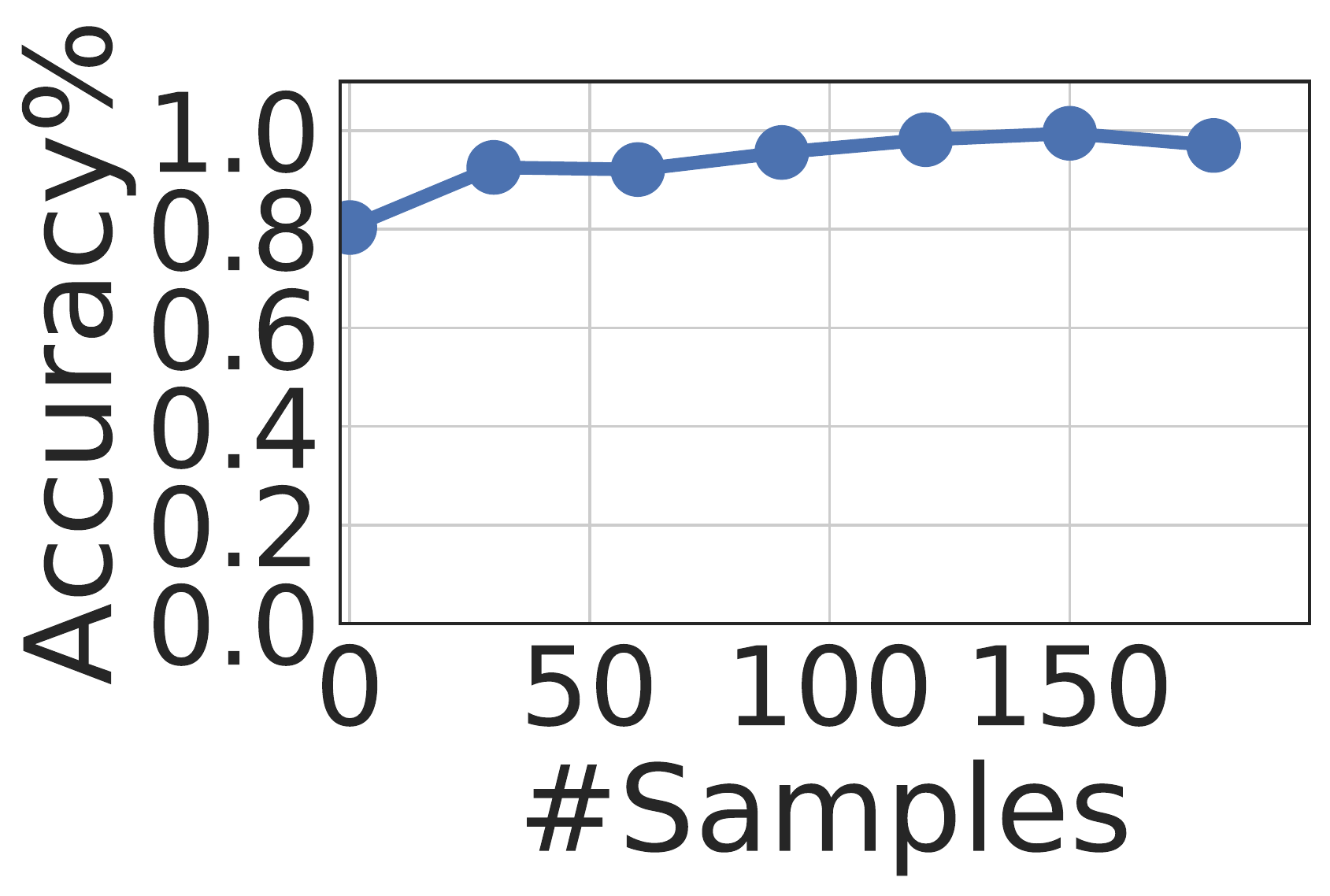}
        }
        \subfigure[\scriptsize Opfake] {%
           \label{fig:ben_precision}
           \includegraphics[width=0.20\textwidth]{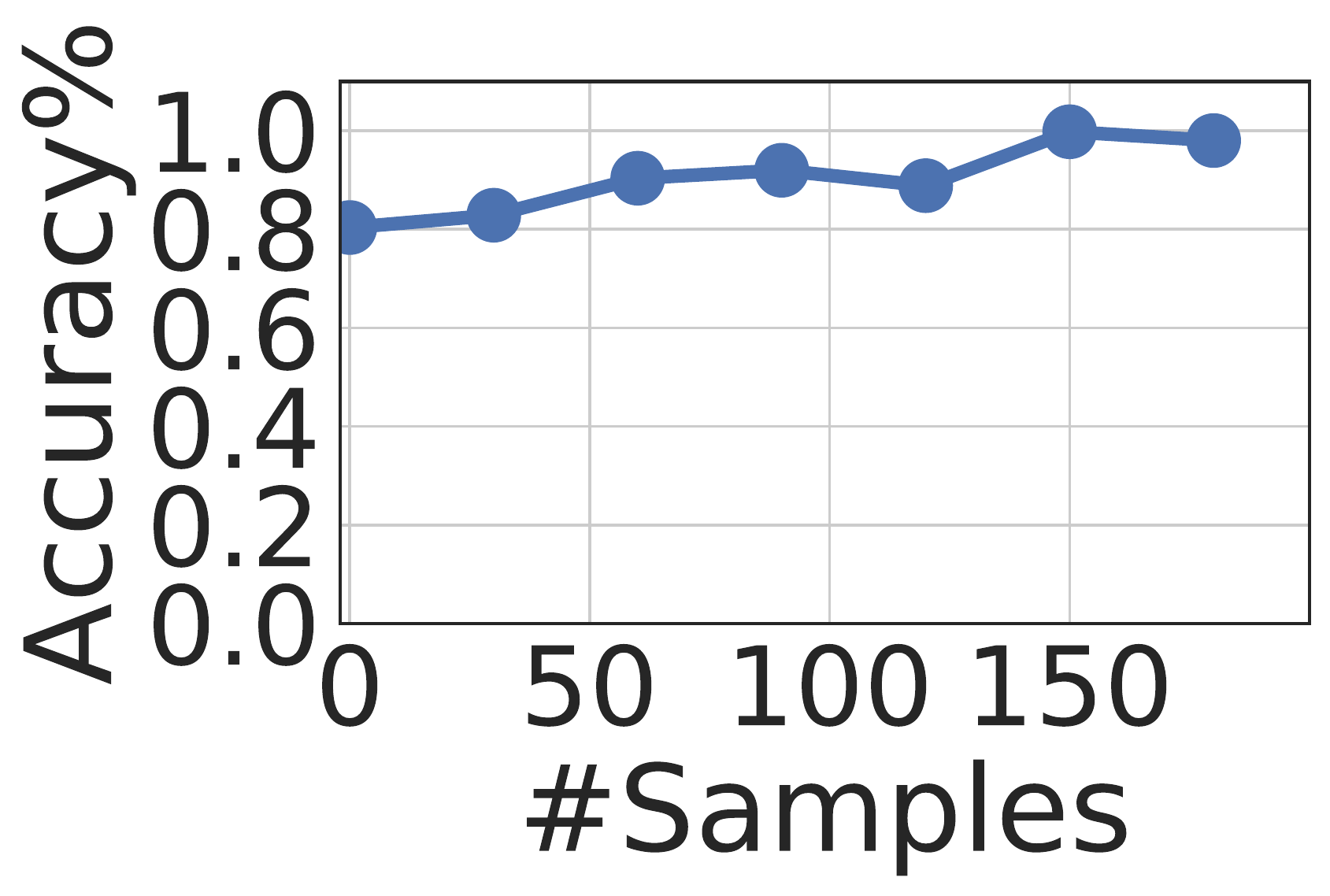}
        }
        \subfigure[\scriptsize Plankton]{%
            \label{fig:ben_recall}
            \includegraphics[width=0.20\textwidth]{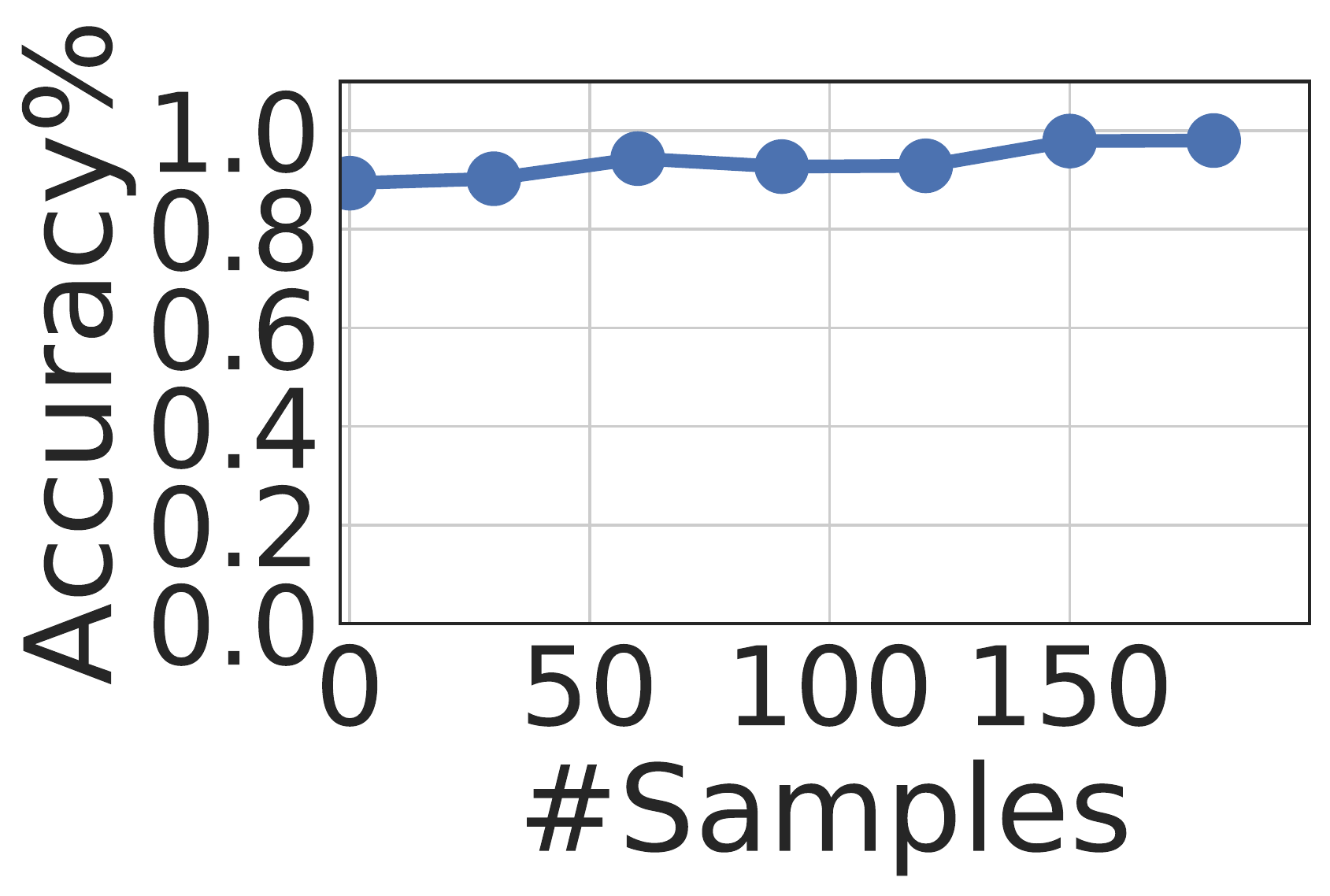}
        }
    \end{center}
    \caption{Evaluation of Unknown Malware Detection}
   \label{fig:unknown_need}
\end{figure*}
\end{scriptsize}

\subsubsection{Unknown Malware Detection}

Although \textsf{MalDozer} demonstrates very good detection results, some questions  still arise: (i) \textit{Can \textsf{MalDozer} detect samples of unknown malware families?} And (ii) \textit{How many samples are needed for a given family to achieve a good accuracy?} To answer these questions, we conduct the following experiment on Drebin mixed dataset (Malware + Benign), where we focus on top malware families (i.e., BaseBridge, DroidKungFu, FakeInstaller, GinMaster, Opfake, Plankton). For each family, we train ($5$ epochs) our model on a subset dataset, which does not include samples of that family. These samples are used as a test set. Afterward, we train with few samples from the family and evaluate the model on the rest of them. Progressively, we add more samples to the training and assess the accuracy of our model on detecting the rest of the family samples. Answering the above questions: \textit{(i) Can \textsf{MalDozer} detect unknown malware family samples?} Yes, Figure \ref{fig:unknown_need} shows the accuracy versus the number of samples in the training dataset. We see that \textsf{MalDozer} (zero sample vs. accuracy) could detect the unknown malware family sample without previous training. The accuracy varies from $60\%$ to $90\%$. \textit{(ii) How many samples for a given family to achieve a good accuracy?} \textsf{MalDozer}  needs only about $10$ to $20$ samples to reach $90\%$ (Figure \ref{fig:unknown_need}).  In the case of DroidKungFu, \textsf{MalDozer} needs $20$ samples to reach $90\%$. Considering $10$ to $20$ samples from a malware family is relatively a small number to get high results. 

\subsubsection{Resiliency Against API Evolution over Time}
\begin{figure*}[ht!]
     \begin{center}        
        \subfigure[\scriptsize 2013 Dataset]{%
            \label{fig:dataset_2013}
            \includegraphics[width=0.35\textwidth]{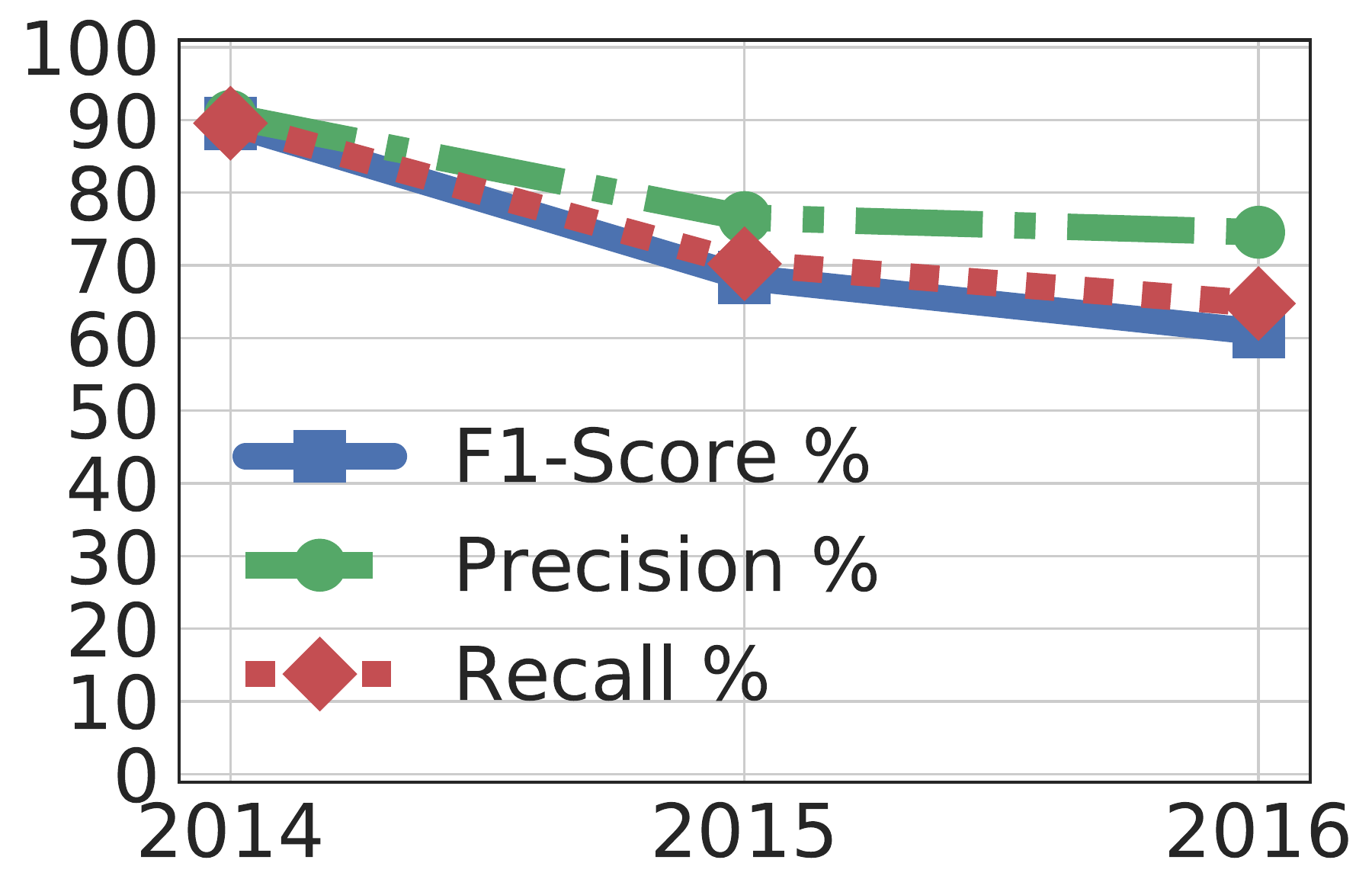}
        }
        \subfigure[\scriptsize 2014 Dataset] {%
           \label{fig:dataset_2014}
           \includegraphics[width=0.35\textwidth]{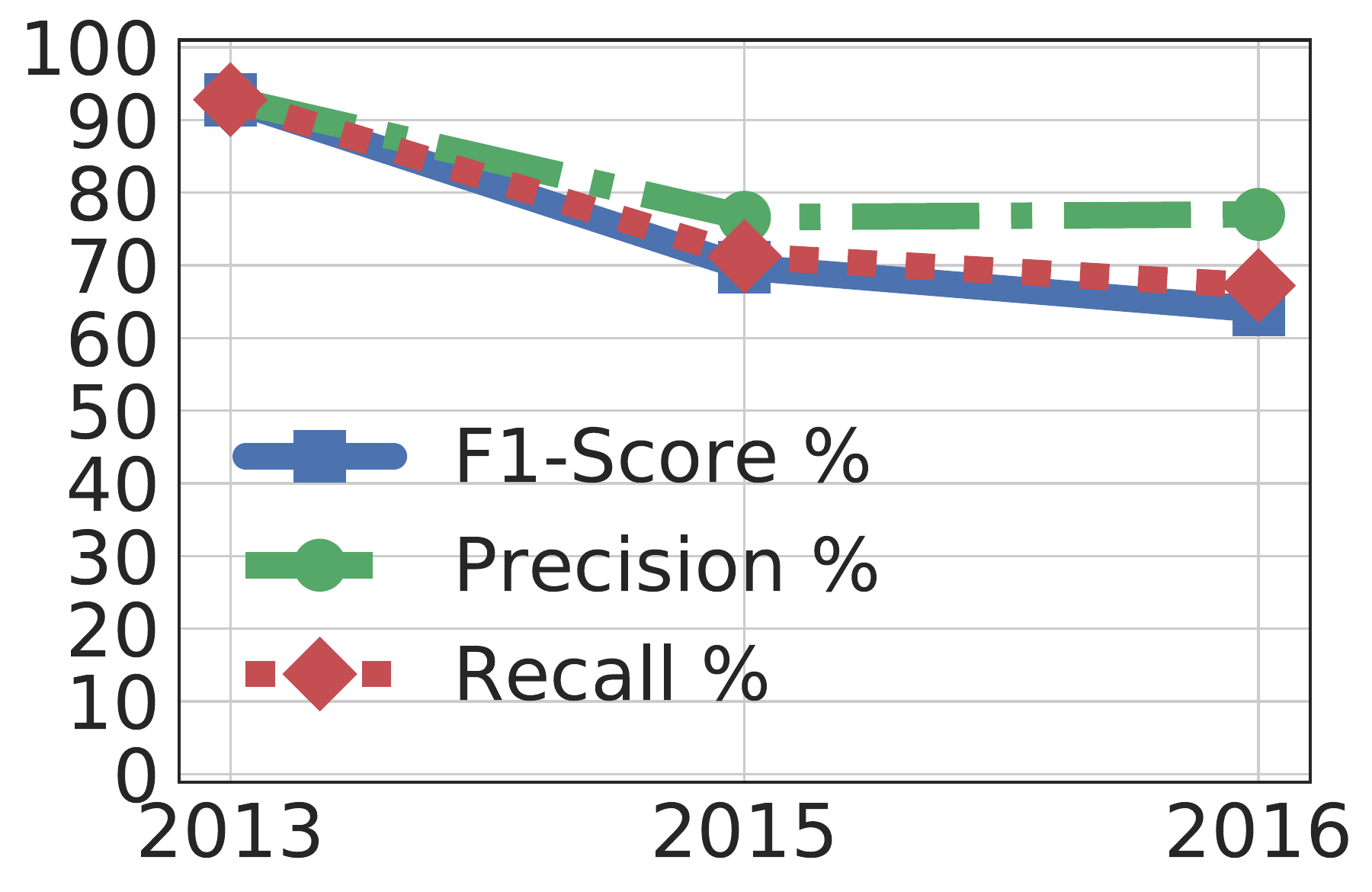}
        }
        \subfigure[\scriptsize 2015 Dataset]{%
            \label{fig:dataset_2015}
            \includegraphics[width=0.35\textwidth]{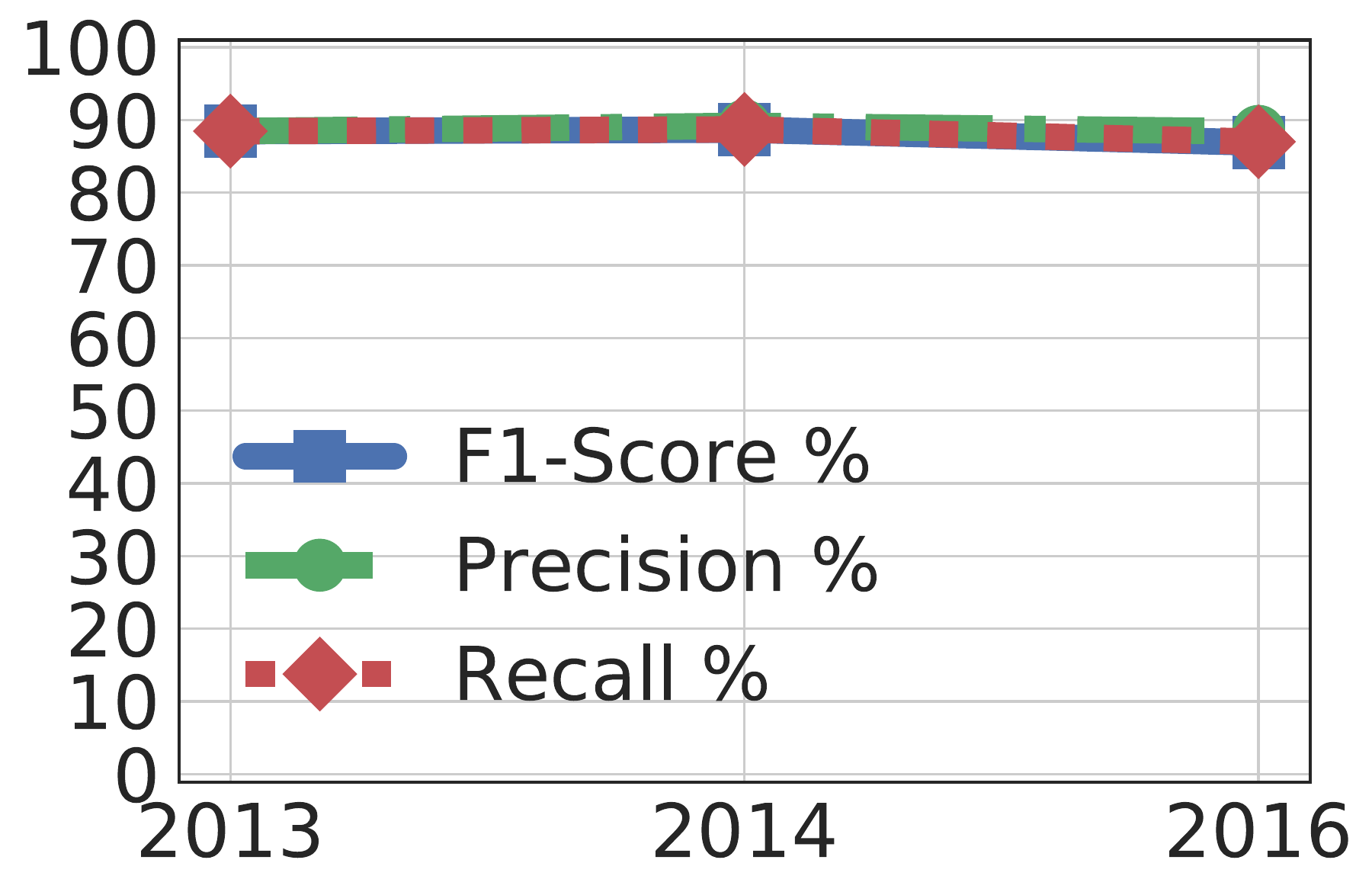}
        }
        \subfigure[\scriptsize 2016 Dataset] {%
           \label{fig:dataset_2016}
           \includegraphics[width=0.35\textwidth]{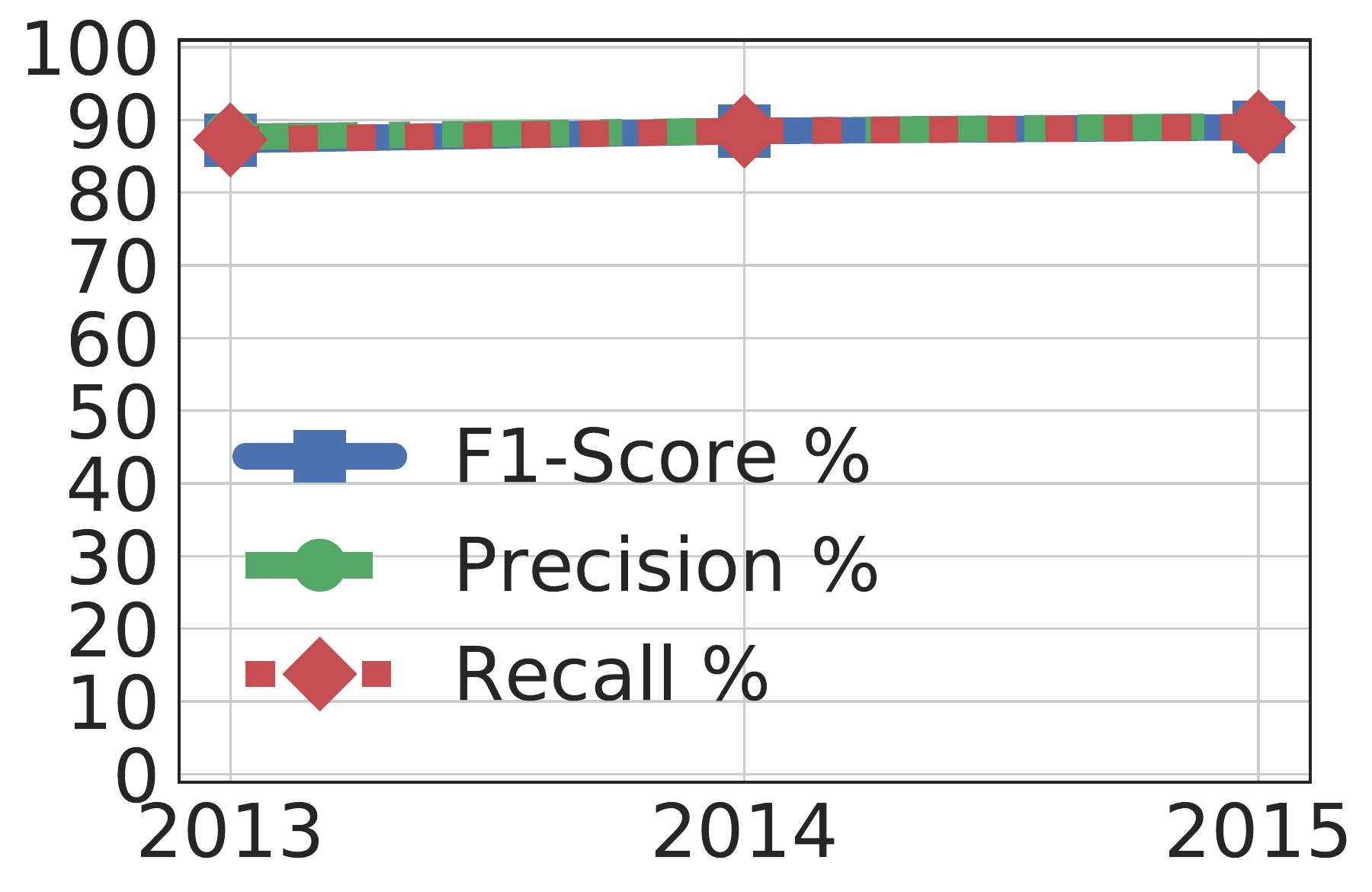}
        }
    \end{center}
    \caption{Detection vs Time}
   \label{fig:detection_overtime}
\end{figure*}

As we have seen in the previous section, \textsf{MalDozer} could detect new malware samples from unknown families using   samples from \textit{Drebin} dataset collected in the period of 2011/2012. In this section, we aim to answer the following quastion: \textit{Can \textsf{MalDozer} detect malicious and bengin apps collected in different years?}To answer this question, we evalute \textsf{MalDozer} on four datasets collected from \cite{DBLP:conf/msr/AllixBKT16} of four consecutive years: 2013, 2014, 2015, and 2016, as shown in Table \ref{tab:detection_time_dataset}, where, we train \textsf{MalDozer} in one year dataset and test it on the rest of the datasets. The results show that \textsf{MalDozer} detection is more resilient to API evolution over time compare to \cite{mariconti2017mamadroid}, as presented in Figure \ref{fig:detection_overtime}. Starting with 2013 dataset (Figure \ref{fig:dataset_2013}), we train \textsf{MalDozer} on 2013 samples and evaluate it on 2014, 2015, and 2016 ones. We notice a high detection rate in 2014 dataset since it is collected in the consecutive year of the training datset. However, the detection rate decreases in 2015 and 2016 datasets but it is above an acceptable detection rate (F1-Score=70\%). Similarly, we obtained the results of 2014 dataset, as depicted in Figure \ref{fig:dataset_2014}. Also, training \textsf{MalDozer} on 2015 or 2016 datasets exhibits very good results under all the datasets collected in other years, where we reach  F1-Score=90-92.5\%.

\begin{figure}[ht!]
\centering
\includegraphics[width=0.35\textwidth]{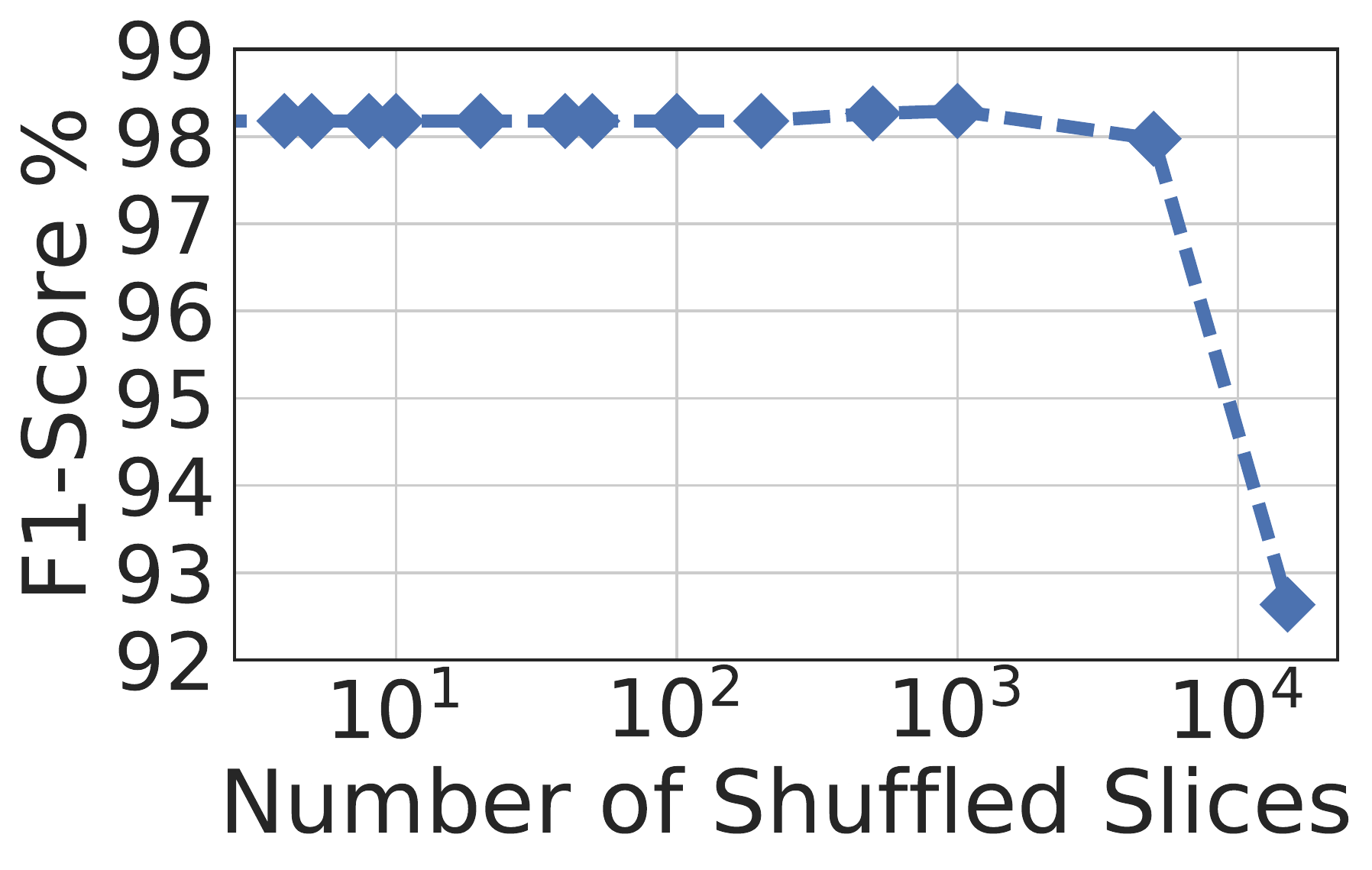}
\caption{Shuffle Rate vs F1-Score}
\label{fig:obfuscation_perturbation}
\end{figure}

\begin{table}[ht]
\centering
\begin{threeparttable}
\begin{tabular}{|l||c|c|c|c|}
\hline \hline
Dataset & \textbf{2013} & \textbf{2014} & \textbf{2015} & \textbf{2016} \\\hline\hline
\textbf{\#Malware} &  10k &  10k &  10k &  10k \\\hline
\textbf{\#Bengin}  &  10k &  10k &  10k &  10k \\\hline
\end{tabular}
\end{threeparttable}
\caption{Datasets for API Evolution Over Time Resiliency} 
\label{tab:detection_time_dataset}
\end{table}

\subsubsection{Resiliency against changing the order of API methods} \label{sec:model_obfuscation}
In this section, we evaluate the robustness of \textsf{MalDozer} against changing in the order of API method calls.  The latter could change for various reasons, for example: (i) We could use different dissassembly tools in the production, (ii) A malware developer could repackage the same malicious app multiple times. The previous scenarios could lead to losing the temporal relations among the API calls.  In case of the malware developer, he/she will be limited by keeping the same malicious semantics in the app. To validate the robustness of \textsf{MalDozer} against such methods order, we conduct the following experiment. First, we train our model on the training dataset. Afterward, we randomly shuffle the sequence of API method calls in the test dataset.  We divide the testing app sequence into $N$ blocks, then shuffle them and evaluate the F1-Score. We repeat until \textit{N} is equal to the number of sequences, i.e., one API call in each block. The result of this experiment is shown in Figure \ref{fig:obfuscation_perturbation}. The latter depicts the F1-Score versus the number of blocks, starting with four blocks and ending with 15K blocks, where each block contains one API call. Figure \ref{fig:obfuscation_perturbation} demonstrates the resiliency of \textsf{MalDozer} against changing the order of API method calls. We observe that even with completely random individual API method calls, \textsf{MalDozer} achieves 93\%. 

\subsection{Family Attribution Performance}
Family attribution is an important task for Android security, where \textsf{MalDozer} distinguishes itself from the existing malware detection solutions, since only few solutions provide this functionality. Starting with Malgenome dataset, \textsf{MalDozer} achieves a very good result, i.e., F1-Score of $99.18\%$. Similarly, \textsf{MalDozer} reaches an F1-Score of $98\%$ on Drebin dataset. The results per malware family attribution performance for Malgnome and Drebin are presented in Tables \ref{tab:malgenome_dataset} and \ref{tab:drebin_dataset}. \textsf{MalDozer} achieves good results in the case of \textsf{MalDozer} dataset, F1-Score of $85\%$. Our interpretation of this result comes from Tables \ref{tab:maldozer_dataset}, \ref{tab:malgenome_dataset} and \ref{tab:drebin_dataset}, which depict the detailed results per malware family. For example, the family \texttt{agent} unveils poor results because of the misslabeling,  since \texttt{agent} is a common name for many Android malware families. We believe that there is a lot of noise in the family labeling of the \textsf{MalDozer} dataset since we leverage only one security vendor for labeling. Despite this fact, \textsf{MalDozer} demonstrates acceptable results and robustness (Appendix Figure \ref{fig:attribution_confusion_matrices}).

\begin{table}[!h]
\centering
\begin{threeparttable}
\begin{tabular}{|c||c|c|c|}
\hline \hline 
& \textbf{F1\%} & \textbf{P\%} & \textbf{R\%}\\
%& \textbf{F1-Score\%} & \textbf{Precision\%} & \textbf{Recall\%}\\
\hline \hline
\textbf{2-Fold}	& 98.9834 & 99.0009 & 98.9847 	\\\hline
\textbf{3-Fold}	& 98.9910 & 99.0026 & 98.9847 \\\hline
\textbf{5-Fold}	& 99.0907 & 99.1032 & 99.0862  \\\hline
\textbf{10-Fold} & \textbf{99.1873} & \textbf{99.1873} & \textbf{99.1878} 	\\
\hline \hline
\end{tabular}
\end{threeparttable}
\caption{Attribution on Malgenome} 
\label{tab:attribution_scores_malgenome}
\end{table}

\begin{table}[!h]
\centering
\begin{threeparttable}
\begin{tabular}{|c||c|c|c|}
\hline \hline
& \textbf{F1\%} & \textbf{P\%} & \textbf{R\%}\\
%& \textbf{F1-Score\%} & \textbf{Precision\%} & \textbf{Recall\%}\\
\hline \hline
\textbf{2-Fold}	& 98.1192 & 98.1401 & 98.1334 	\\\hline
\textbf{3-Fold} & 98.6882 & 98.6998 & 98.6912 \\\hline
\textbf{5-Fold}	& 98.5824 & 98.5961 & 98.5839  \\\hline
\textbf{10-Fold} &\textbf{98.5198} & \textbf{98.5295} & \textbf{98.5196} 	\\
\hline \hline
\end{tabular}
\end{threeparttable}
\caption{Attribution on Drebin} 
\label{tab:attribution_scores_drebin}
\end{table}

\begin{table}[!h]
\centering
\begin{threeparttable}
\begin{tabular}{|c||c|c|c|}
\hline \hline
& \textbf{F1\%} & \textbf{P\%} & \textbf{R\%}\\
%& \textbf{F1-Score\%} & \textbf{Precision\%} & \textbf{Recall\%}\\
\hline \hline
\textbf{2-Fold}	& 89.3331 & 89.5044 & 89.3424 	\\\hline
\textbf{3-Fold} & 81.8742 & 82.7565 & 81.8109 \\\hline
\textbf{5-Fold}	&83.8518 & 84.1360 & 84.0061  \\\hline
\textbf{10-Fold} & \textbf{85.5233} & \textbf{85.6184} & \textbf{85.8479} 	\\
\hline \hline
\end{tabular}
\end{threeparttable}
\caption{Attribution on MalDozer} 
\label{tab:attribution_scores_maldozer}
\end{table}

\begin{table}[!h]
\centering
\begin{threeparttable}
\begin{tabular}{|c||c|c||c|}
\hline \hline
 &Malware Family &  \#Sample & F1-Score \\
\hline \hline
01 & FakeInst     &   4822 & 96.15\% \\
02 & Dowgin       &   2248 & 84.24\% \\
03 & SmsPay       &   1544 & 81.61\% \\
04 & Adwo         &   1495 & 87.79\% \\
05 & SMSSend      &   1088 & 81.48\% \\
06 & Wapsx        &    833 & 78.85\% \\
07 & Plankton     &    817 & 94.18\% \\
08 & Agent        &    778 & 51.45\% \\
09 & SMSReg       &    687 & 80.61\% \\
10 & GingerMaster &    533 & 76.39\% \\
11 & Kuguo        &    448 & 78.28\% \\
12 & HiddenAds    &    426 & 84.20\% \\
13 & Utchi        &    397 & 93.99\% \\
14 & Youmi        &    355 & 72.39\% \\
15 & Iop          &    344 & 93.09\% \\
16 & BaseBridge   &    341 & 90.50\% \\
17 & DroidKungFu  &    314 & 85.85\% \\
18 & SmsSpy       &    279 & 85.05\% \\
19 & FakeApp      &    278 & 93.99\% \\
20 & InfoStealer  &    253 & 82.82\% \\
21 & Kmin         &    222 & 91.03\% \\
22 & HiddenApp    &    214 & 76.71\% \\
23 & AppQuanta    &    202 & 99.26\% \\
24 & Dropper      &    195 & 77.11\% \\
25 & MobilePay    &    144 & 78.74\% \\
26 & FakeDoc      &    140 & 96.38\% \\
27 & Mseg         &    138 & 55.38\% \\
28 & SMSKey       &    130 & 81.03\% \\
29 & RATC         &    111 & 84.81\% \\
30 & Geinimi      &    106 & 95.58\% \\
31 & DDLight      &    104 & 90.55\% \\
32 & GingerBreak  &    103 & 84.87\% \\

\hline \hline
\end{tabular}
\end{threeparttable}
\caption{MalDozer  Android Malware Dataset} 
\label{tab:maldozer_dataset}
\end{table}

\begin{table}[!h]
\centering
\begin{threeparttable}
\begin{tabular}{|c||c|c||c|}
\hline \hline
 &Malware Family &  \#Sample & F1-Score\\
\hline \hline
01 & DroidKungFu3    &    309 & 99.83\% \\
02 & AnserverBot     &    187 & 99.19\% \\
03 & BaseBridge      &    121 & 98.37\% \\
04 & DroidKungFu4    &     96 & 99.88\% \\
05 & Geinimi         &     69 & 97.81\% \\
06 & Pjapps          &     58 & 95.65\% \\
07 & KMin            &     52 & 99.99\% \\
08 & GoldDream       &     47 & 99.96\% \\
09 & DroidDreamLight &     46 & 99.99\% \\
\hline \hline
\end{tabular}
\end{threeparttable}
\caption{Malgenome Attribution Dataset} 
\label{tab:malgenome_dataset}
\end{table}

\begin{table}[!h]
\centering
\begin{threeparttable}
\begin{tabular}{|c||c|c||c|}
\hline \hline
 &Malware Family &  \#Sample & F1-Score  \\
\hline \hline
01 & FakeInstaller      &    925 &  99.51\% \\
02 & DroidKungFu        &    666 &  98.79\% \\
03 & Plankton           &    625 &  99.11\% \\
04 & Opfake             &    613 &  99.34\% \\
05 & GinMaster          &    339 &  97.92\% \\
06 & BaseBridge         &    329 &  97.56\% \\
07 & Iconosys           &    152 &  99.02\% \\
08 & Kmin               &    147 &  99.31\% \\
09 & FakeDoc            &    132 &  99.24\% \\
10 & Geinimi            &     92 &  97.26\% \\
11 & Adrd               &     91 &  96.13\% \\
12 & DroidDream         &     81 &  98.13\% \\
13 & Glodream           &     69 &  90.14\% \\
14 & MobileTx           &     69 &  91.97\% \\
15 & ExploitLinuxLotoor &     69 &  99.97\% \\
16 & FakeRun            &     61 &  95.16\% \\
17 & SendPay            &     59 &  99.14\% \\
18 & Gappusin           &     58 &  97.43\% \\
19 & Imlog              &     43 &  98.85\% \\
20 & SMSreg             &     41 &  92.30\% \\
\hline \hline
\end{tabular}
\end{threeparttable}
\caption{Drebin Attribution Dataset} 
\label{tab:drebin_dataset}
\end{table}

\subsection{Run-Time Performance}
\begin{figure}[!htb]
\centering
     \begin{center}        
        \subfigure[\scriptsize Mixed APK]{%
            \label{fig:ben_fscore}
            \includegraphics[width=0.22\textwidth]{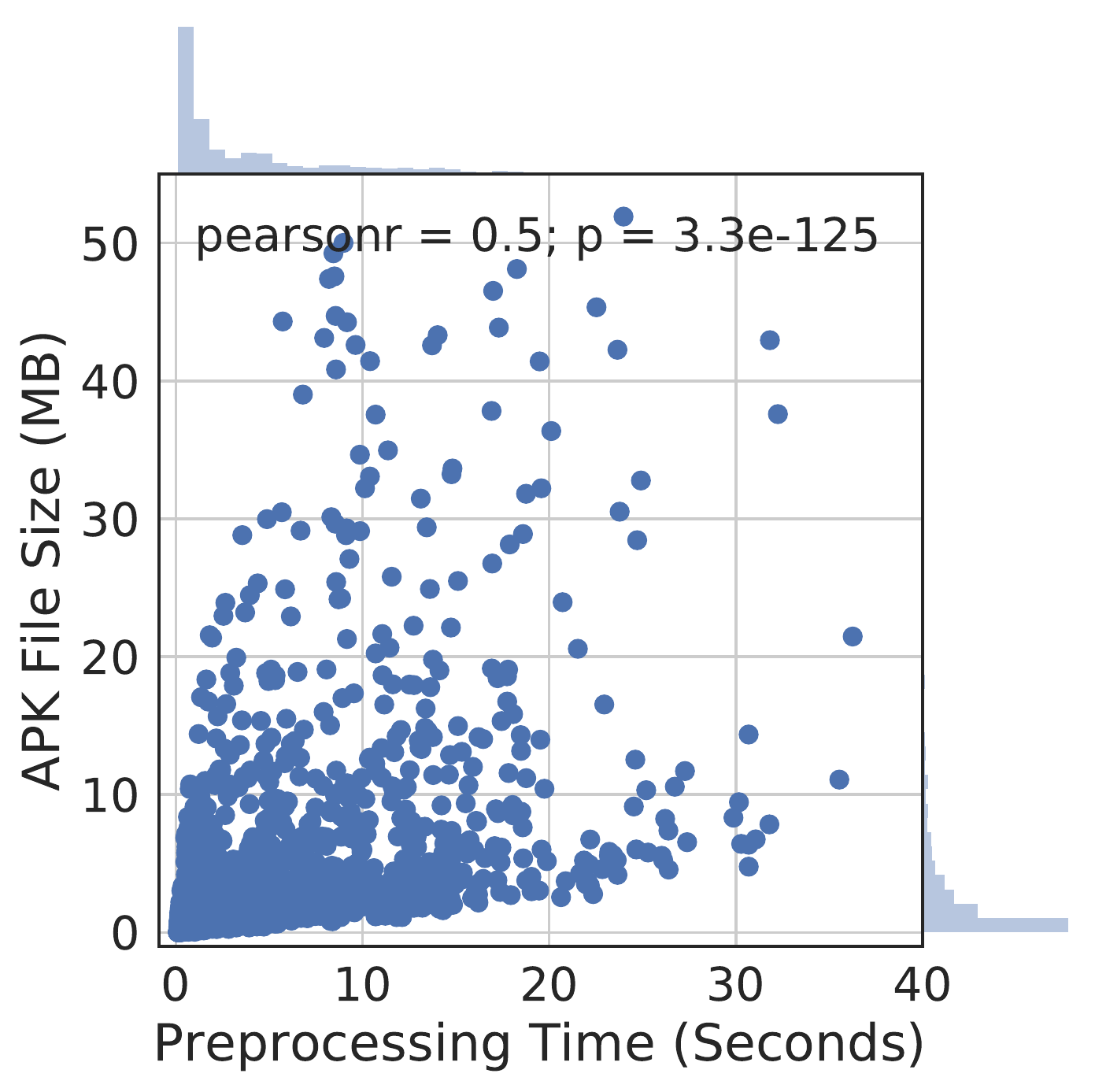}
        }
        \subfigure[\scriptsize Bengin APK]{%
            \label{fig:ben_fscore}
            \includegraphics[width=0.22\textwidth]{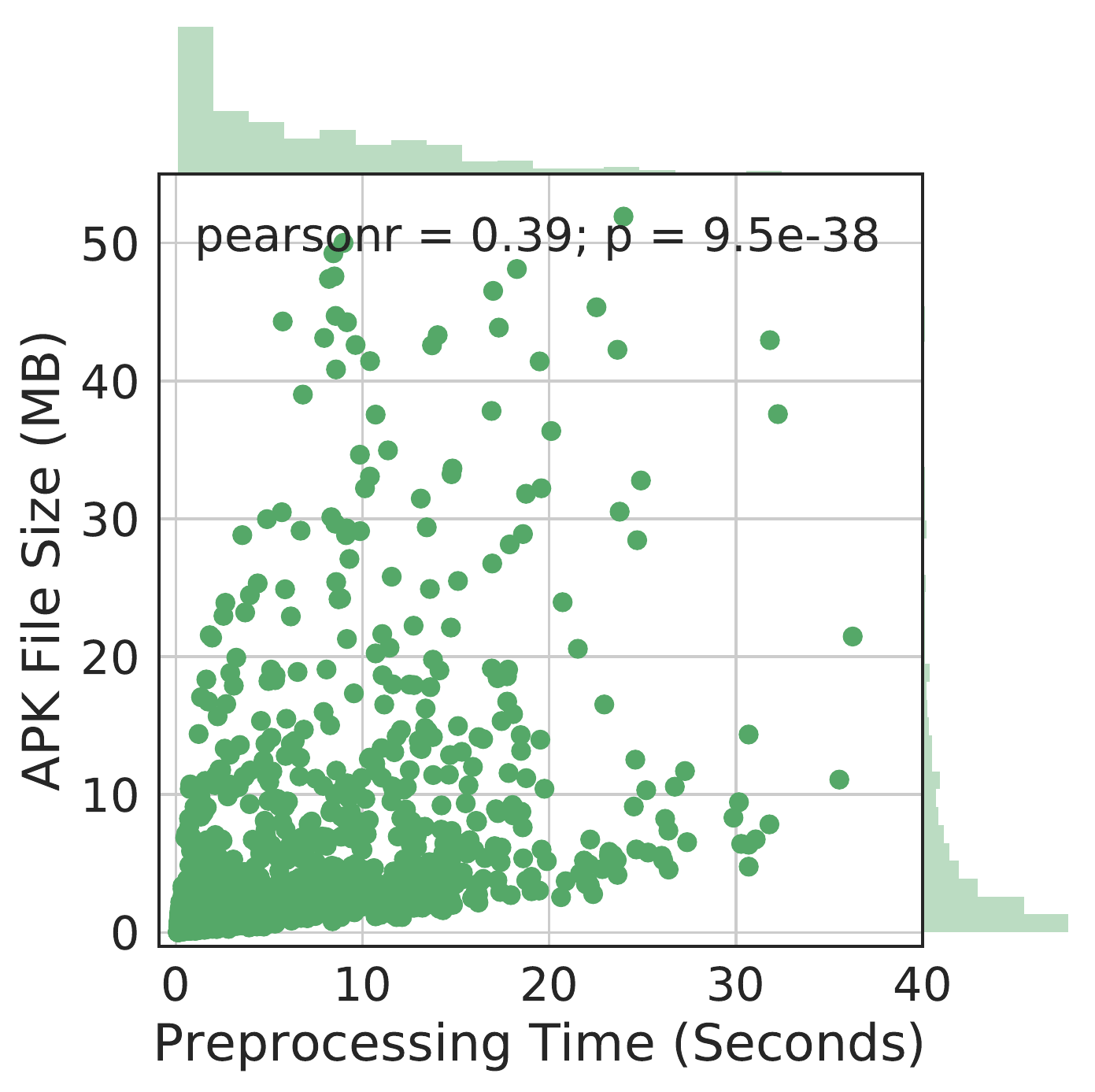}
        }\\
        \subfigure[\scriptsize Malware APK]{%
            \label{fig:ben_fscore}
            \includegraphics[width=0.22\textwidth]{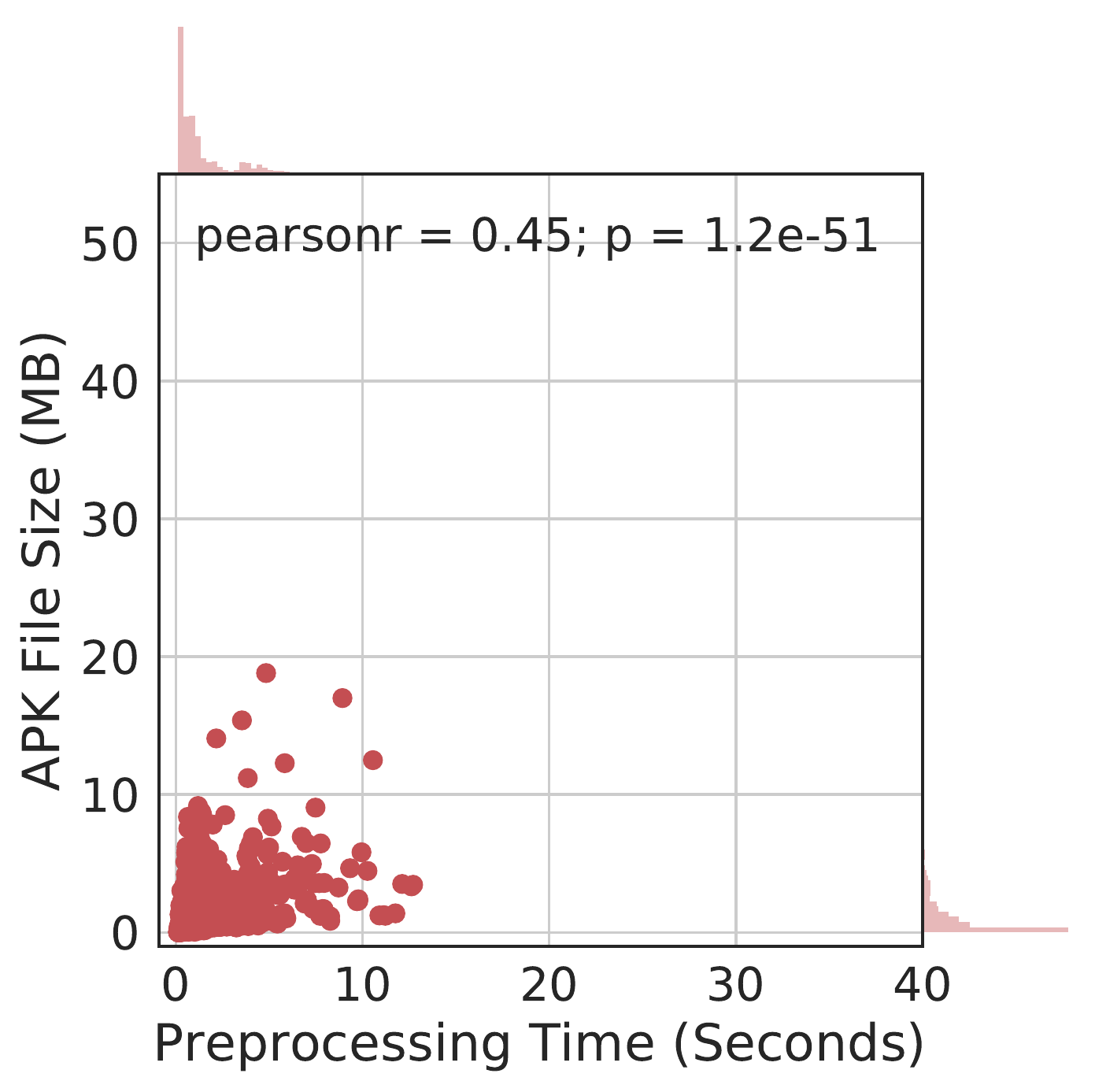}
        }
        \subfigure[\scriptsize Mixed DEX] {%
           \label{fig:ben_precision}
           \includegraphics[width=0.22\textwidth]{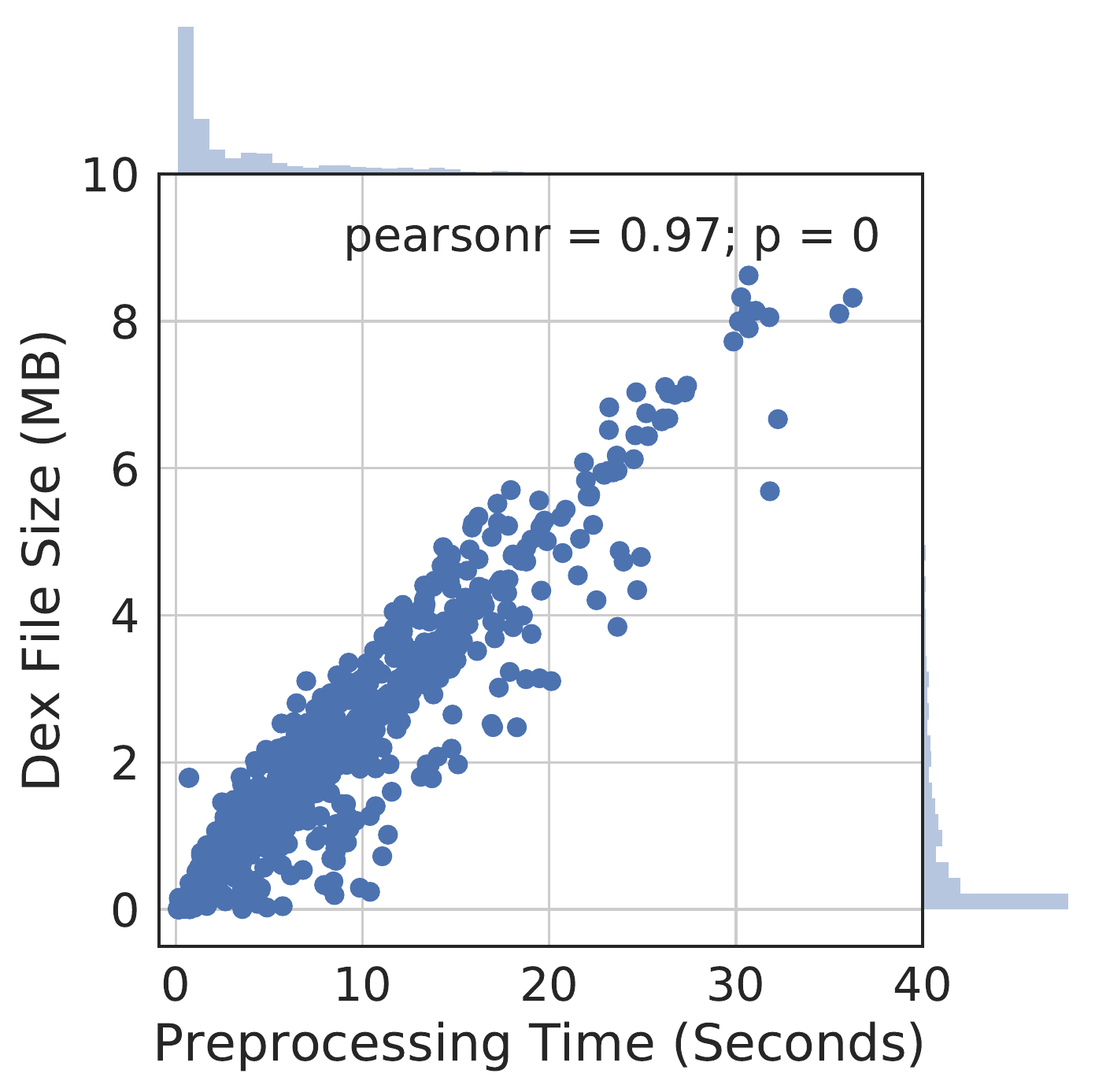}
        }\\
        \subfigure[\scriptsize Benign DEX] {%
           \label{fig:ben_precision}
           \includegraphics[width=0.22\textwidth]{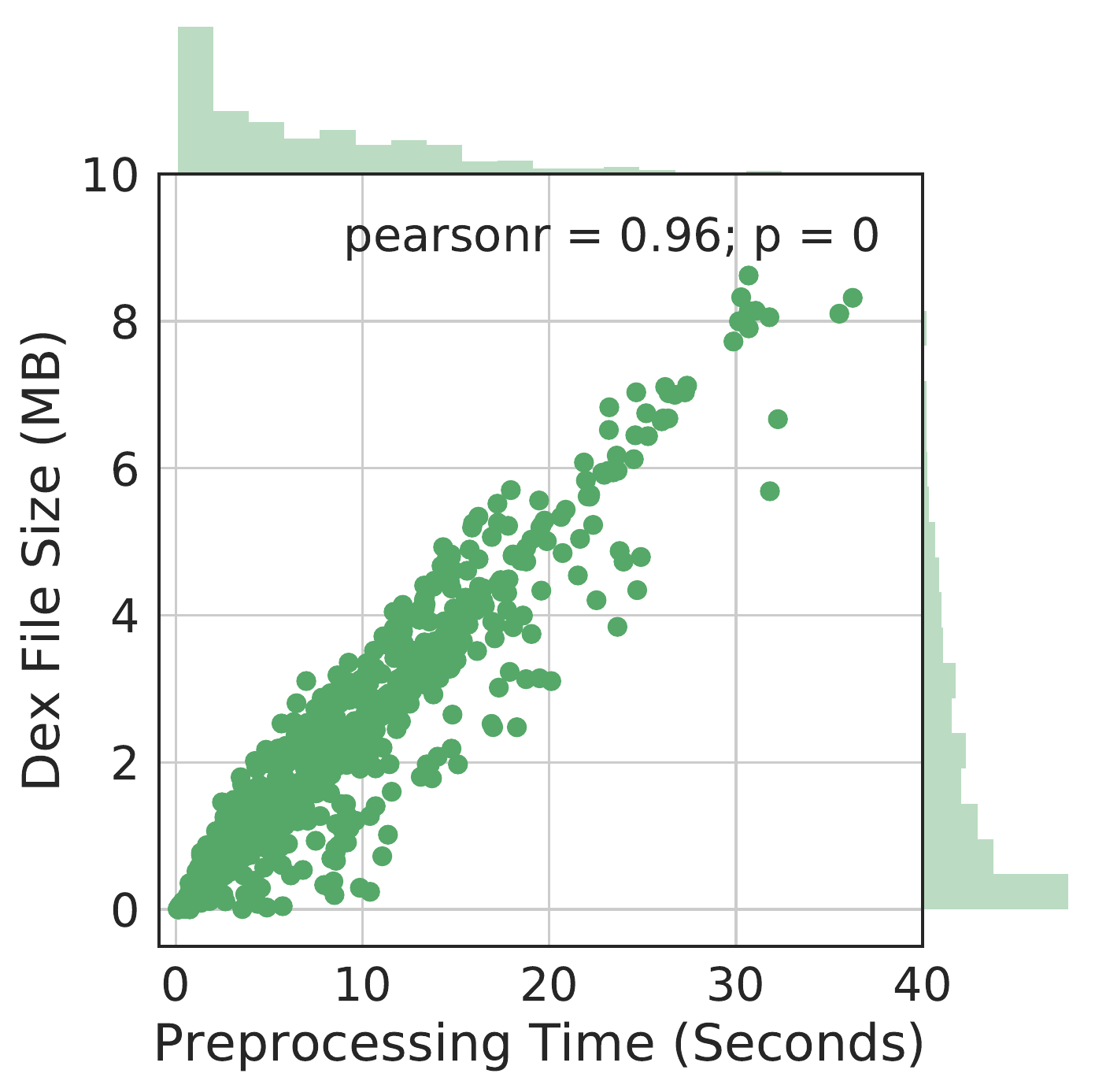}
        }
        \subfigure[\scriptsize Malware DEX] {%
           \label{fig:ben_precision}
           \includegraphics[width=0.22\textwidth]{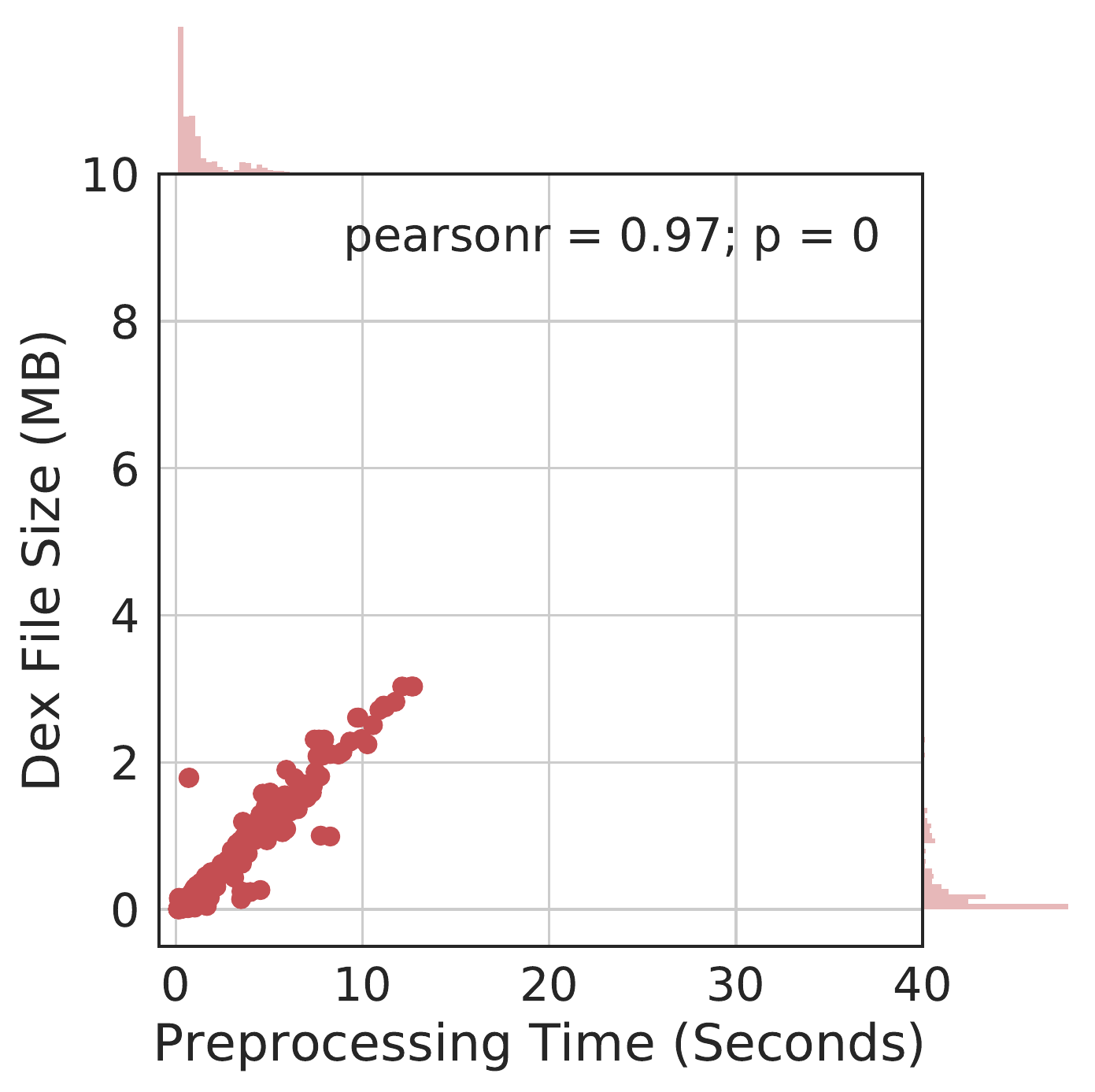}
        }
    \end{center}
    \caption{Preprocessing Time vs APK and DEX Sizes (IoT device)}
   \label{fig:runtime_preprocess_iot}
\end{figure}

\begin{figure}[!htb]
\centering
     \begin{center}
        \subfigure[Preprocess]{%
            \label{fig:process_time}
            \includegraphics[width=0.35\textwidth]{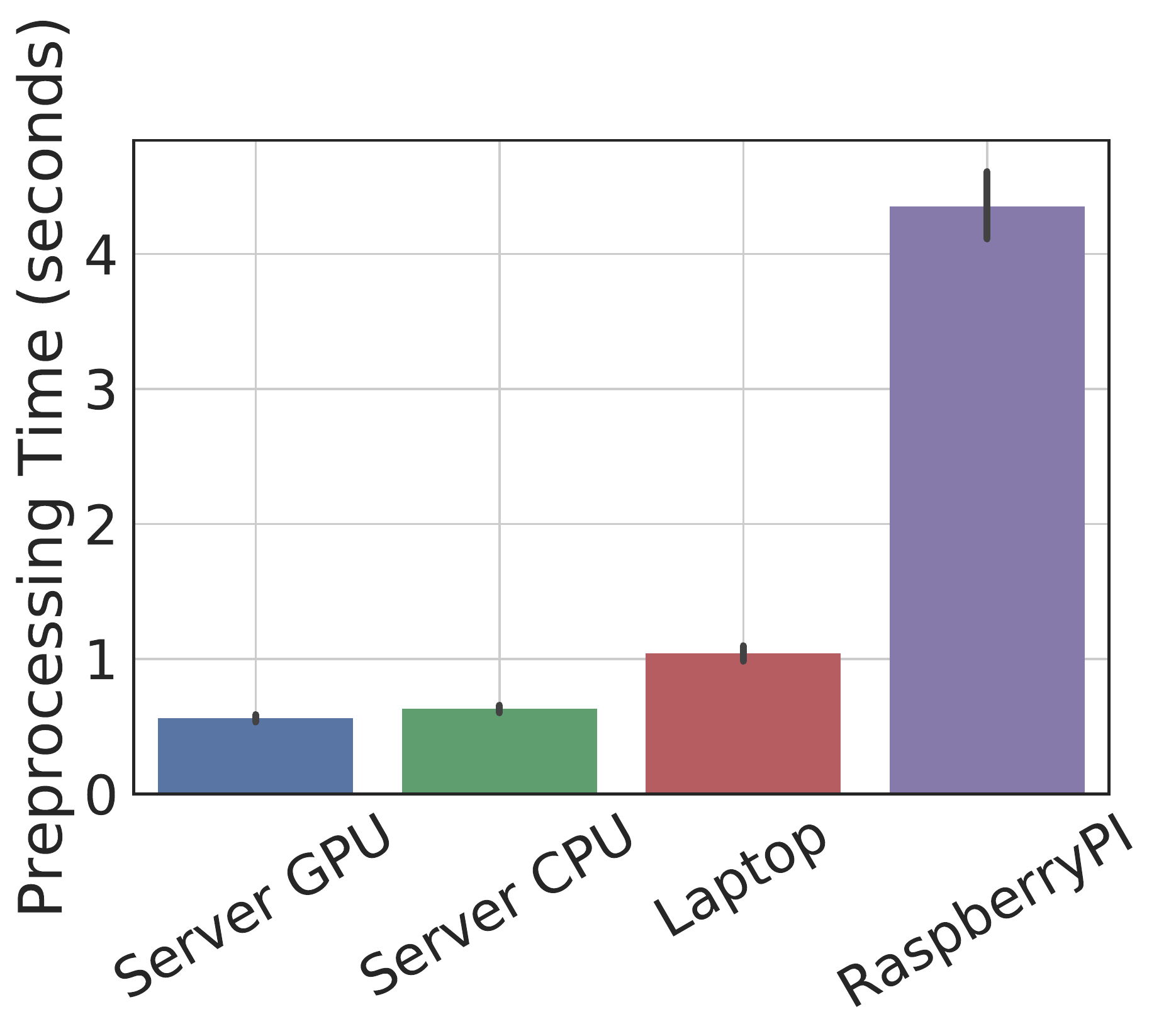}
        }
        \subfigure[Prediction] {%
           \label{fig:detection_time}
           \includegraphics[width=0.35\textwidth]{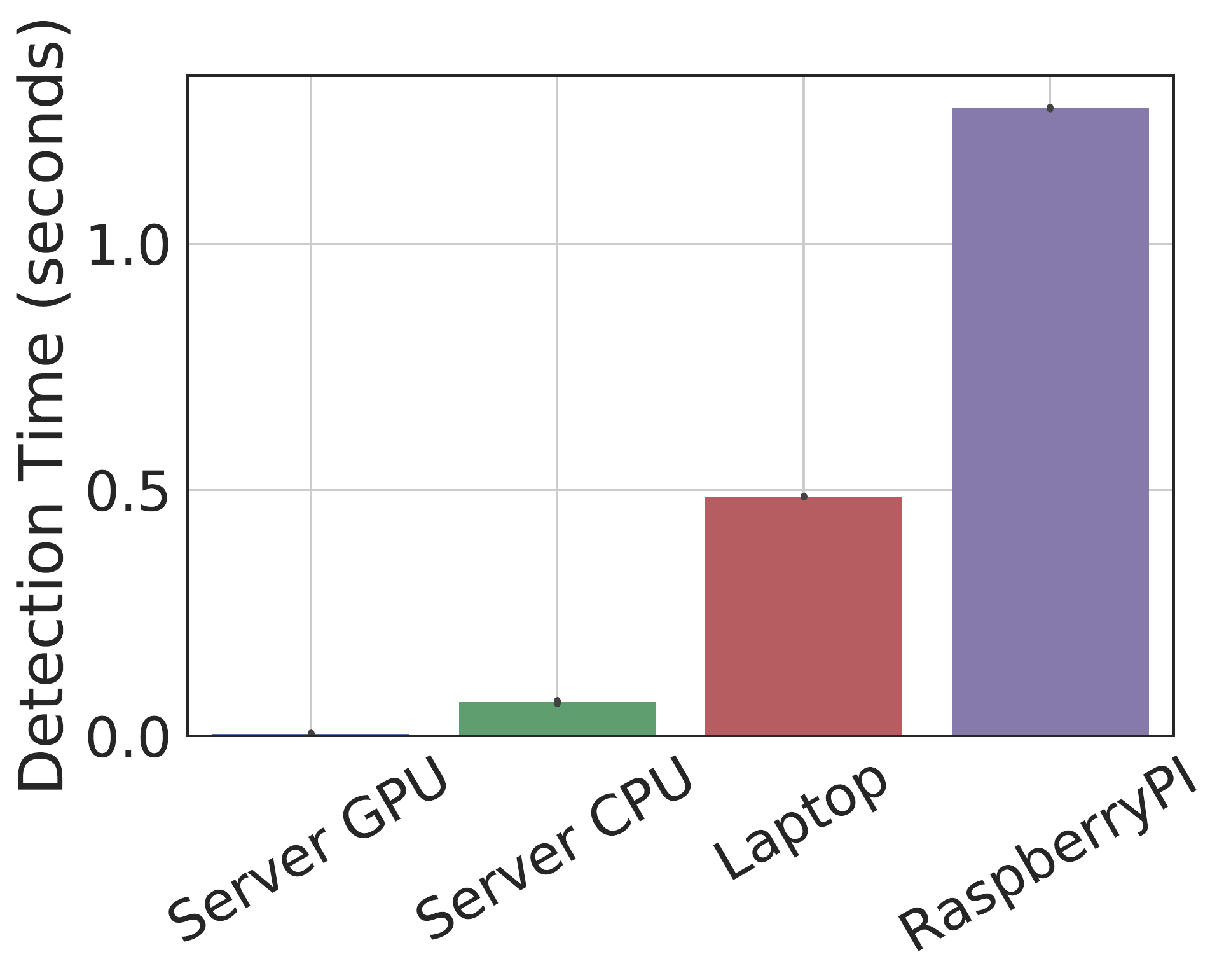}
        }
    \end{center}
    \caption{Run-Time vs. Hardware}
   \label{fig:snknown_need}
\end{figure}

In this section, we evaluate the efficiency of \textsf{MalDozer}, i.e., the runtime during the deployment phase. We divide the runtime into two parts: i) \textit{Preprocessing time}: the required time to extract and preprocess the sequences of Android API method calls. ii) \textit{Detection time:} time needed to make the prediction about a given sequence of API method calls. We analyze the detection time on the model complexity of different hardware. Figure \ref{fig:process_time} depicts the average preprocessing time along with its standard deviation, related to each hardware. The server machines and the laptop spend, on average, 1 second in the preprocessing time, which is very acceptable for production. Also, as mentioned previously, we do not optimize the current preprocessing workflow. In the IoT device \cite{rasp_2_iot}, the preprocessing takes, on average, about $4$ seconds, which is more than acceptable for such a small device. Figure \ref{fig:detection_time} presents the detection time on average that is related to each hardware. First, it is noticeable that the standard deviation is very negligible, i.e., the detection time is constant for all apps. Also, the detection time is very low for all the devices. As for the IoT device, the detection time is only $1.3$ seconds. Therefore, the average time that \textsf{MalDozer} needs to decide for a given app is $5.3$ seconds on average in case of IoT device, as we know that the preprocessing takes most of the time ($4/5.3$). Here, we ask the following two questions: (i) \textit{Which part in the preprocessing needs optimization?} (ii) \textit{Does the preprocessing time depend on the size of APK or DEX file?} To answer these questions, we randomly select 1K benign apps and 1K malware apps. We measure the preprocessing time and correlate it with the size of APK and DEX files. Figure \ref{fig:runtime_preprocess_iot} shows the experimentation results in the case of the IoT device \cite{rasp_2_iot}. The scattered charts depict the preprocessing time along with the size of the APK or DEX file for the mixed, only-benign, and only-malware datasets. From Figure \ref{fig:runtime_preprocess_iot}, it is clear that the preprocessing time is linearly related to the size of the DEX file. We perform the same experiment on Server and Laptop, and we get very similar results, as shown in Figures \ref{fig:runtime_preprocess_pc} and \ref{fig:runtime_preprocess_server}. Finally, we notice that the size of benign apps tend to be bigger than the malicious apps. Thus, the preprocessing time of the benign apps is longer.

\begin{scriptsize}
\begin{figure}[ht!]
     \begin{center}        
        \subfigure[\scriptsize Mixed APK]{%
            \label{fig:ben_fscore}
            \includegraphics[width=0.22\textwidth]{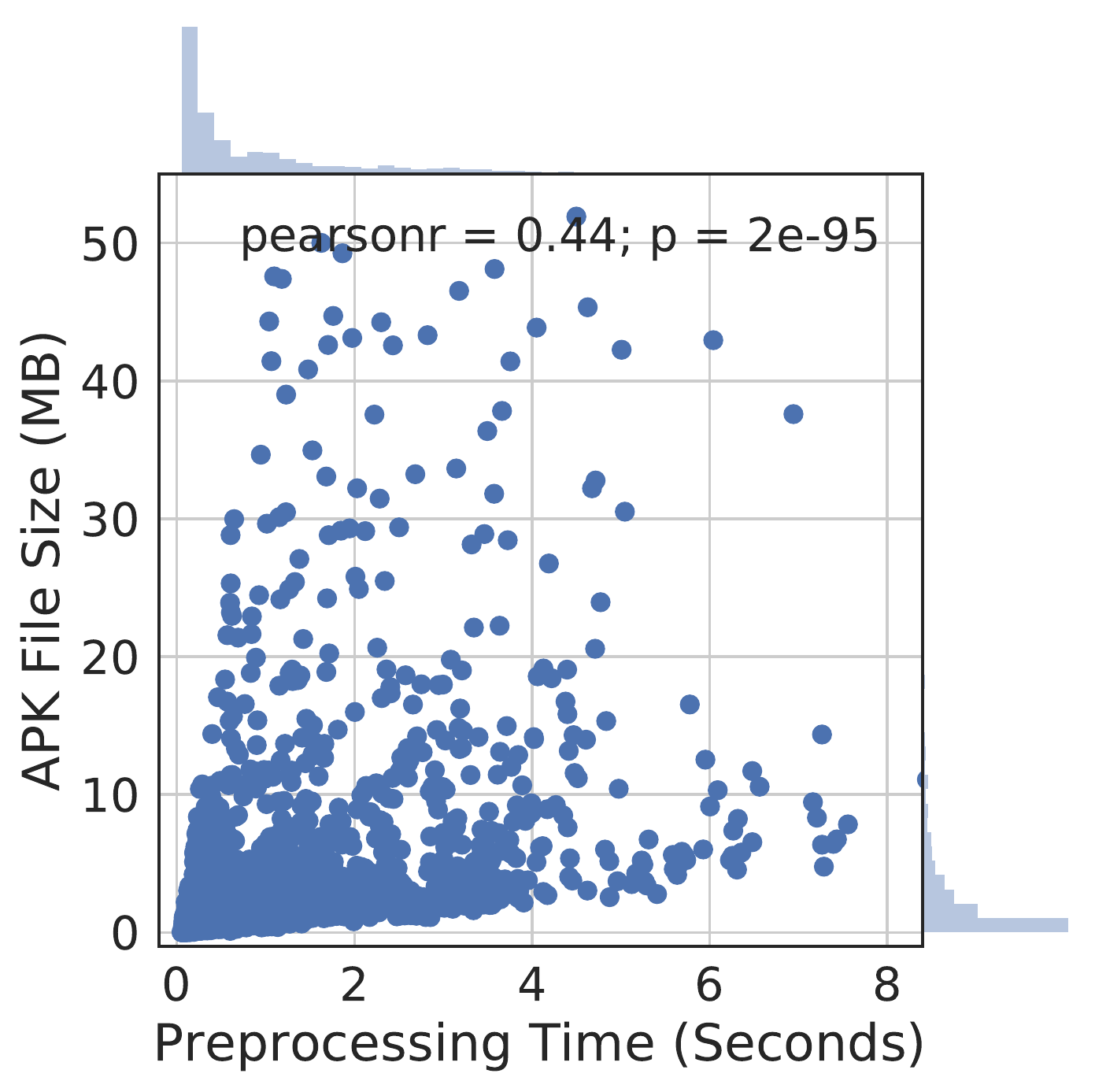}
        }
        \subfigure[\scriptsize Bengin APK]{%
            \label{fig:ben_fscore}
            \includegraphics[width=0.22\textwidth]{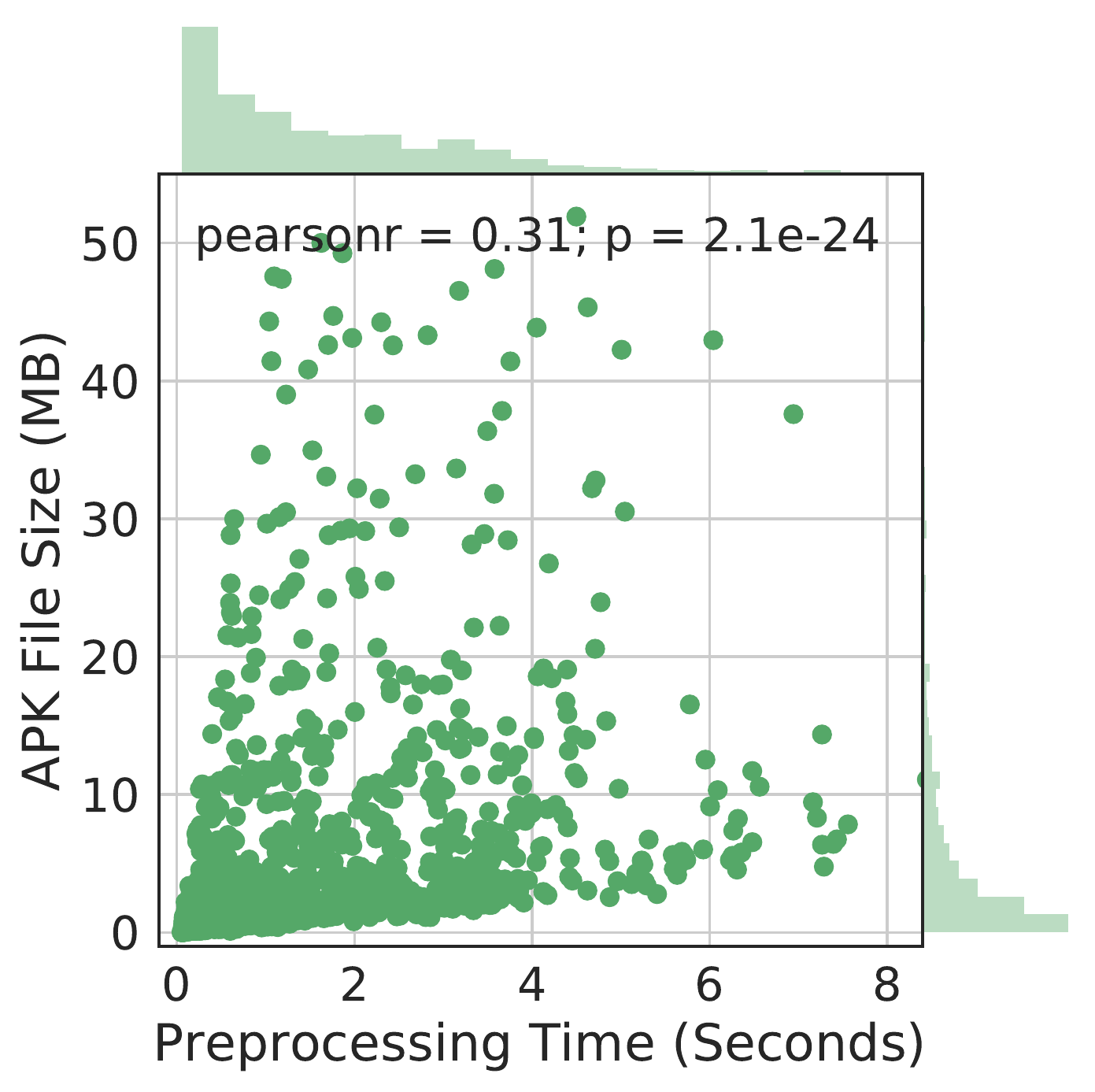}
        }\\
        \subfigure[\scriptsize Malware APK]{%
            \label{fig:ben_fscore}
            \includegraphics[width=0.22\textwidth]{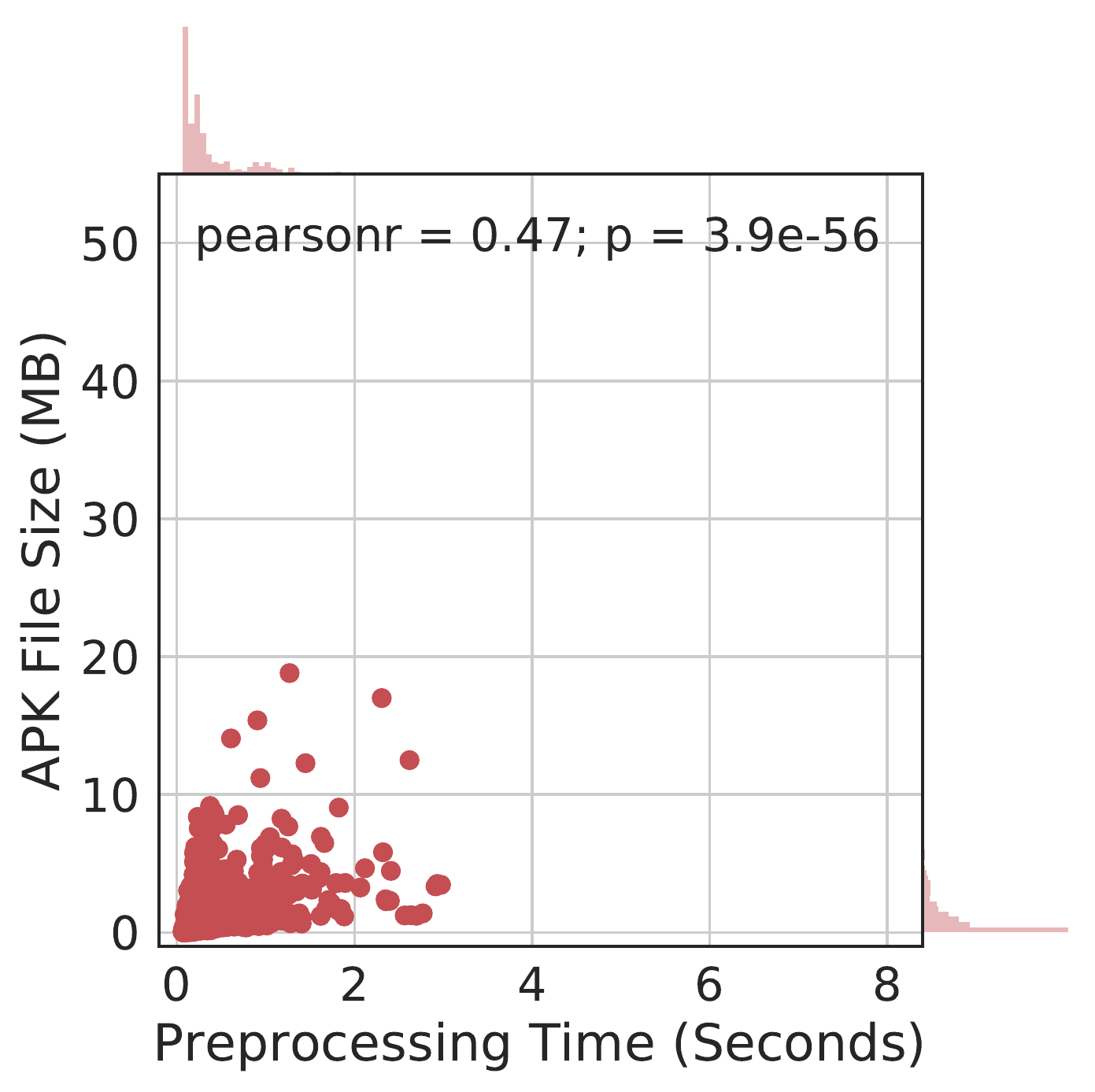}
        }
        \subfigure[\scriptsize Mixed Dex] {%
           \label{fig:ben_precision}
           \includegraphics[width=0.22\textwidth]{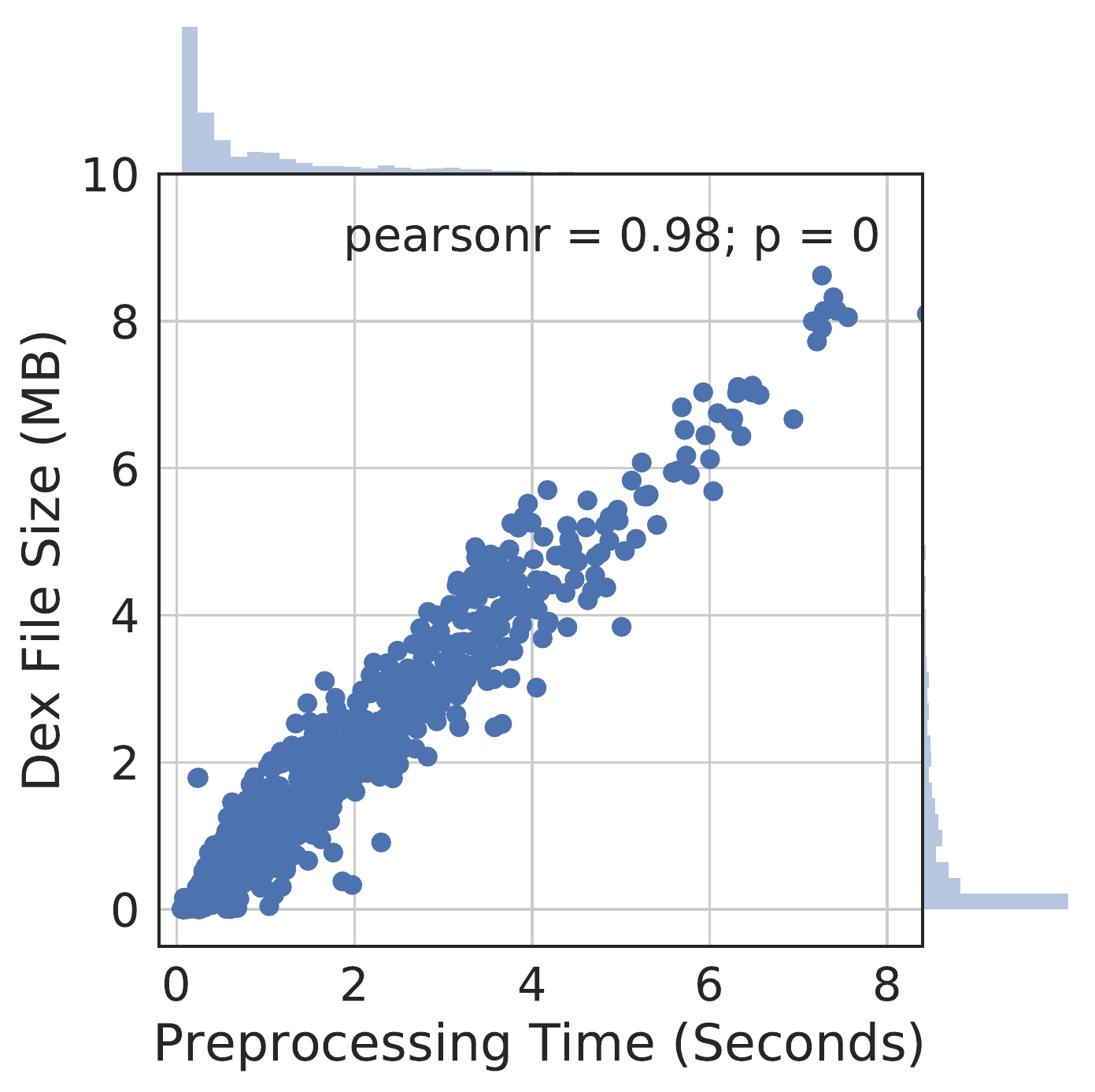}
        }\\
        \subfigure[\scriptsize Bengin Dex] {%
           \label{fig:ben_precision}
           \includegraphics[width=0.22\textwidth]{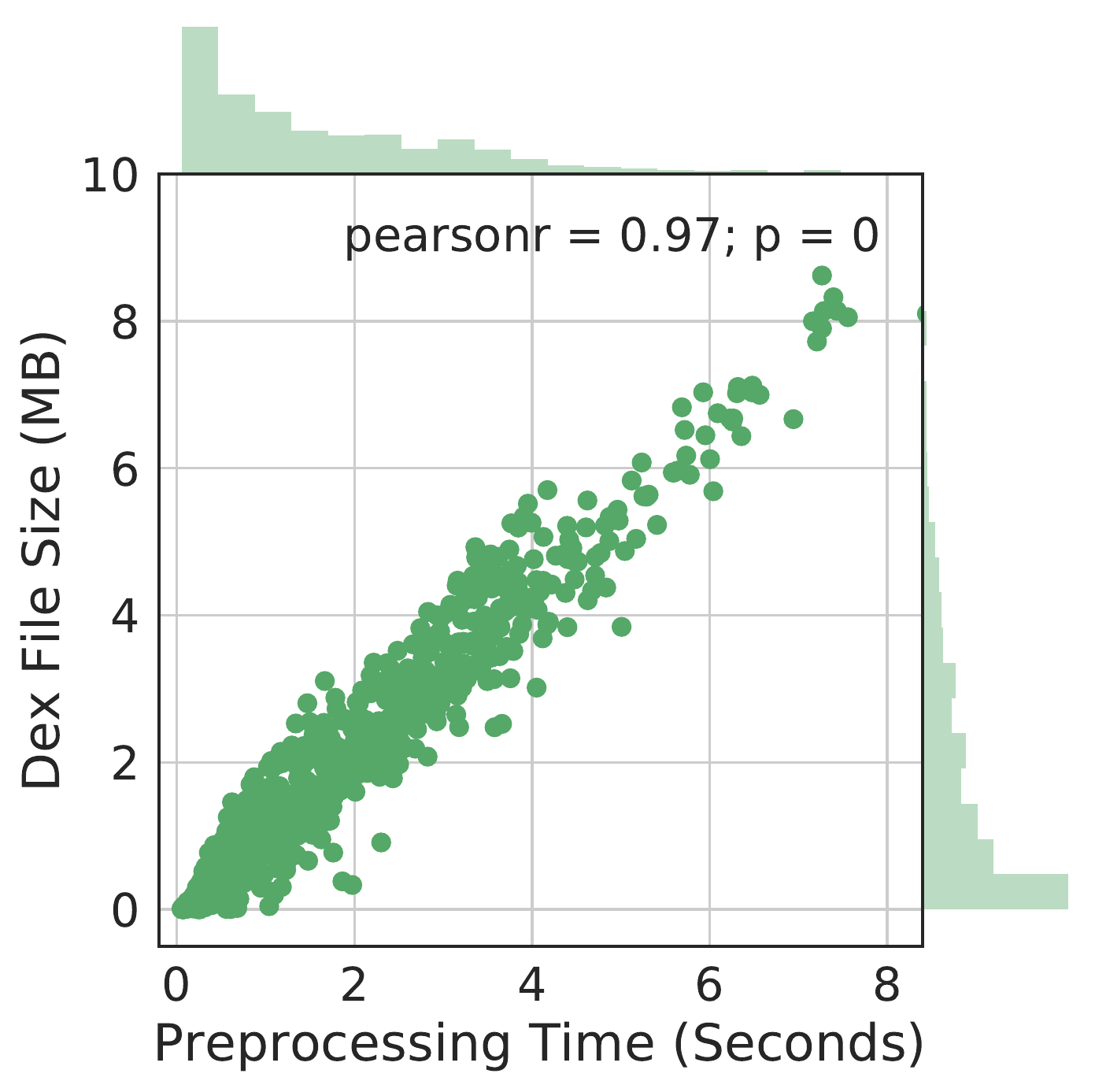}
        }
        \subfigure[\scriptsize Malware Dex] {%
           \label{fig:ben_precision}
           \includegraphics[width=0.22\textwidth]{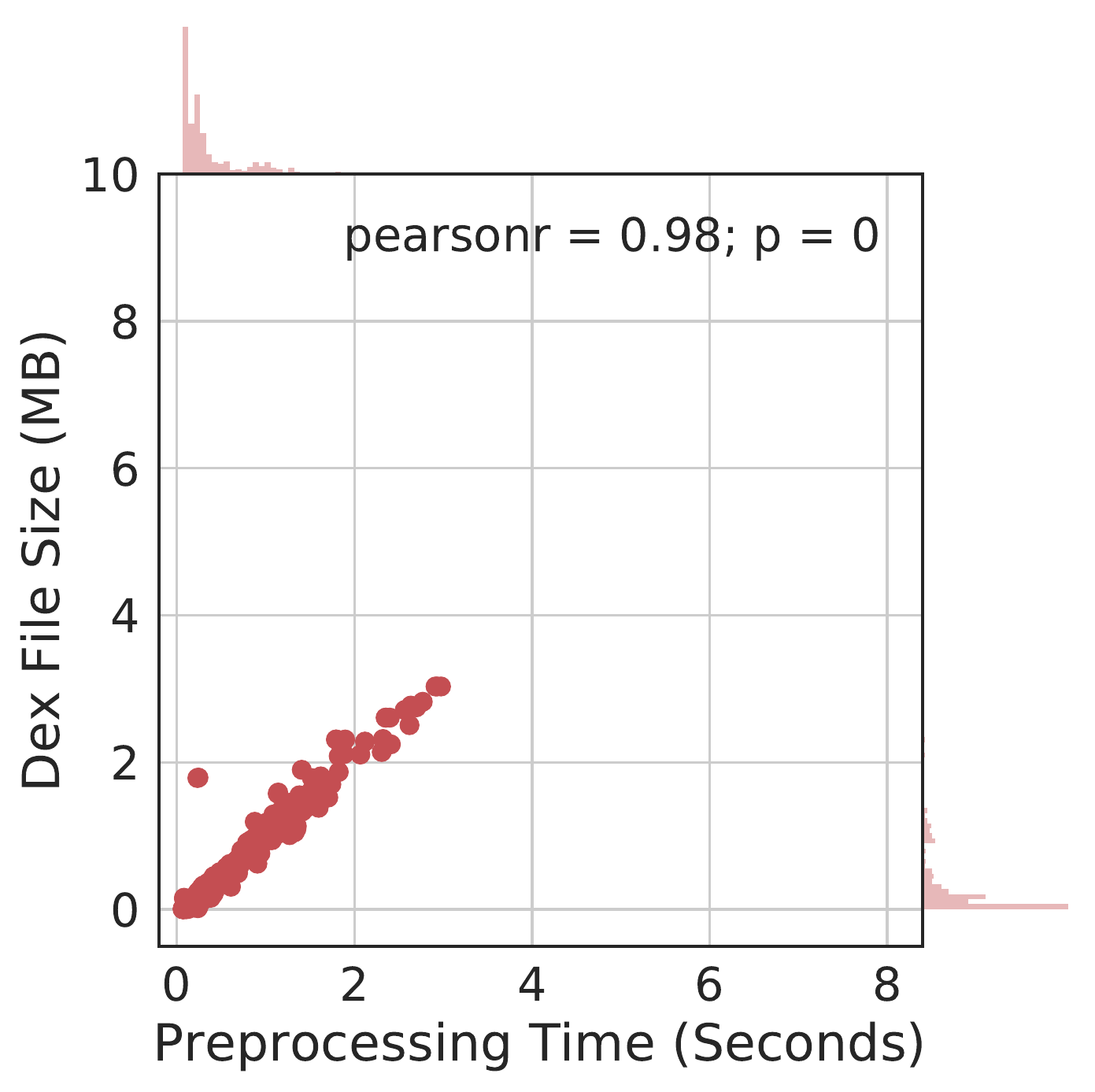}
        }
    \end{center}
    \caption{Preprocessing Time vs APK and Dex Sizes (Laptop)}
   \label{fig:runtime_preprocess_pc}
\end{figure}
\end{scriptsize}

\begin{scriptsize}
\begin{figure}[ht!]
     \begin{center}        
        \subfigure[\scriptsize Mixed APK]{%
            \label{fig:ben_fscore}
            \includegraphics[width=0.22\textwidth]{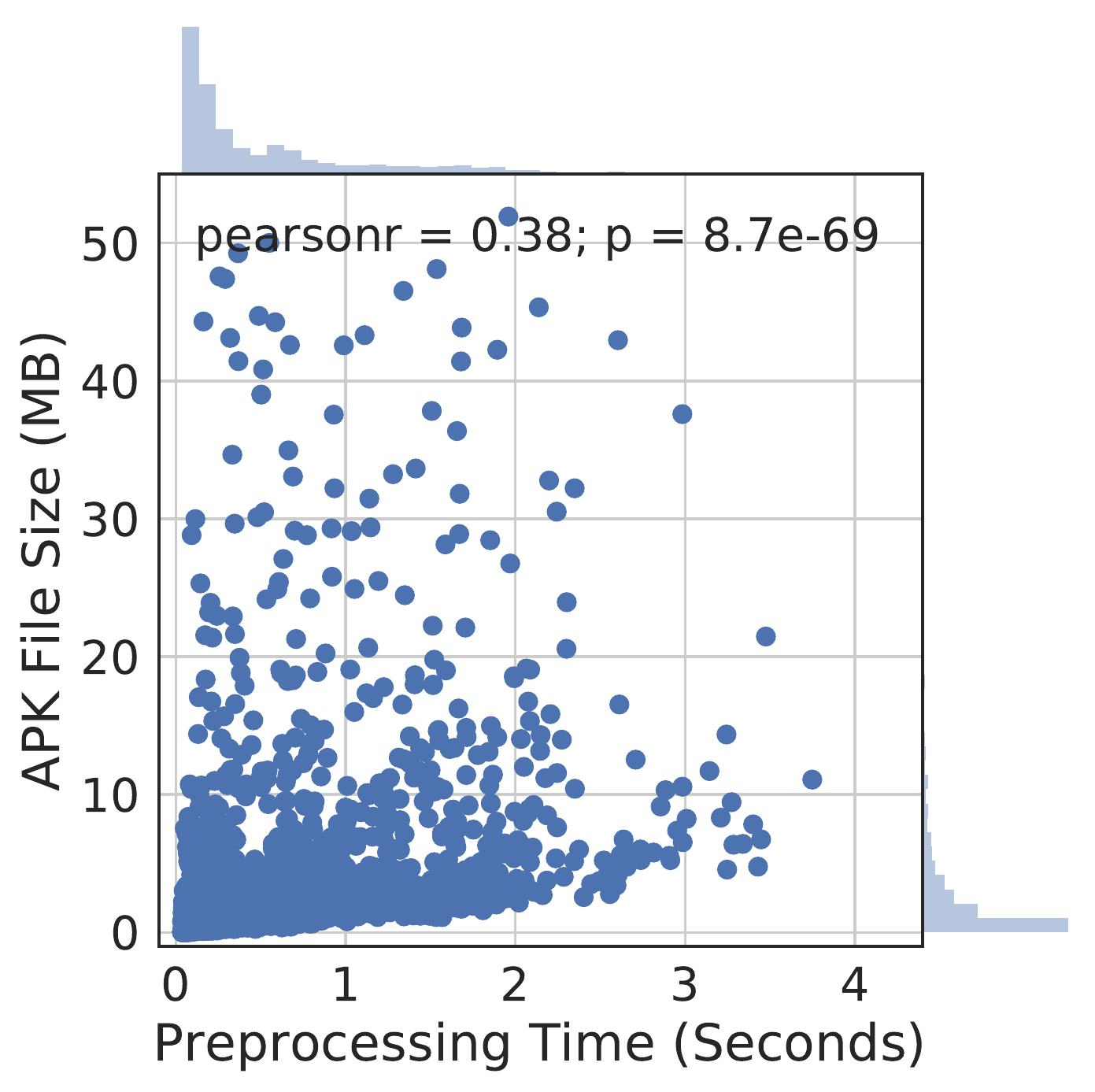}
        }
        \subfigure[\scriptsize Bengin APK]{%
            \label{fig:ben_fscore}
            \includegraphics[width=0.22\textwidth]{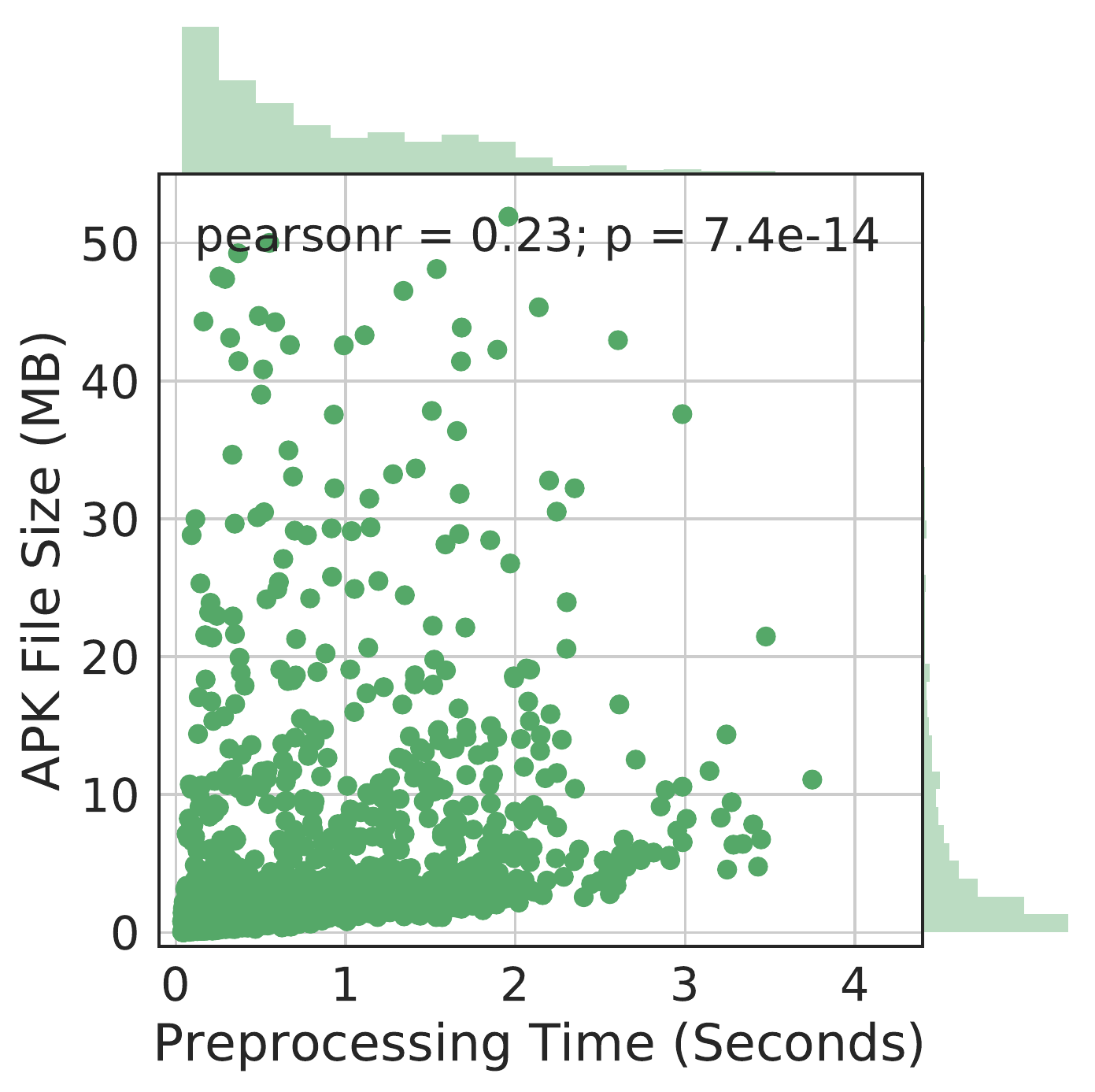}
        }\\
        \subfigure[\scriptsize Malware APK]{%
            \label{fig:ben_fscore}
            \includegraphics[width=0.22\textwidth]{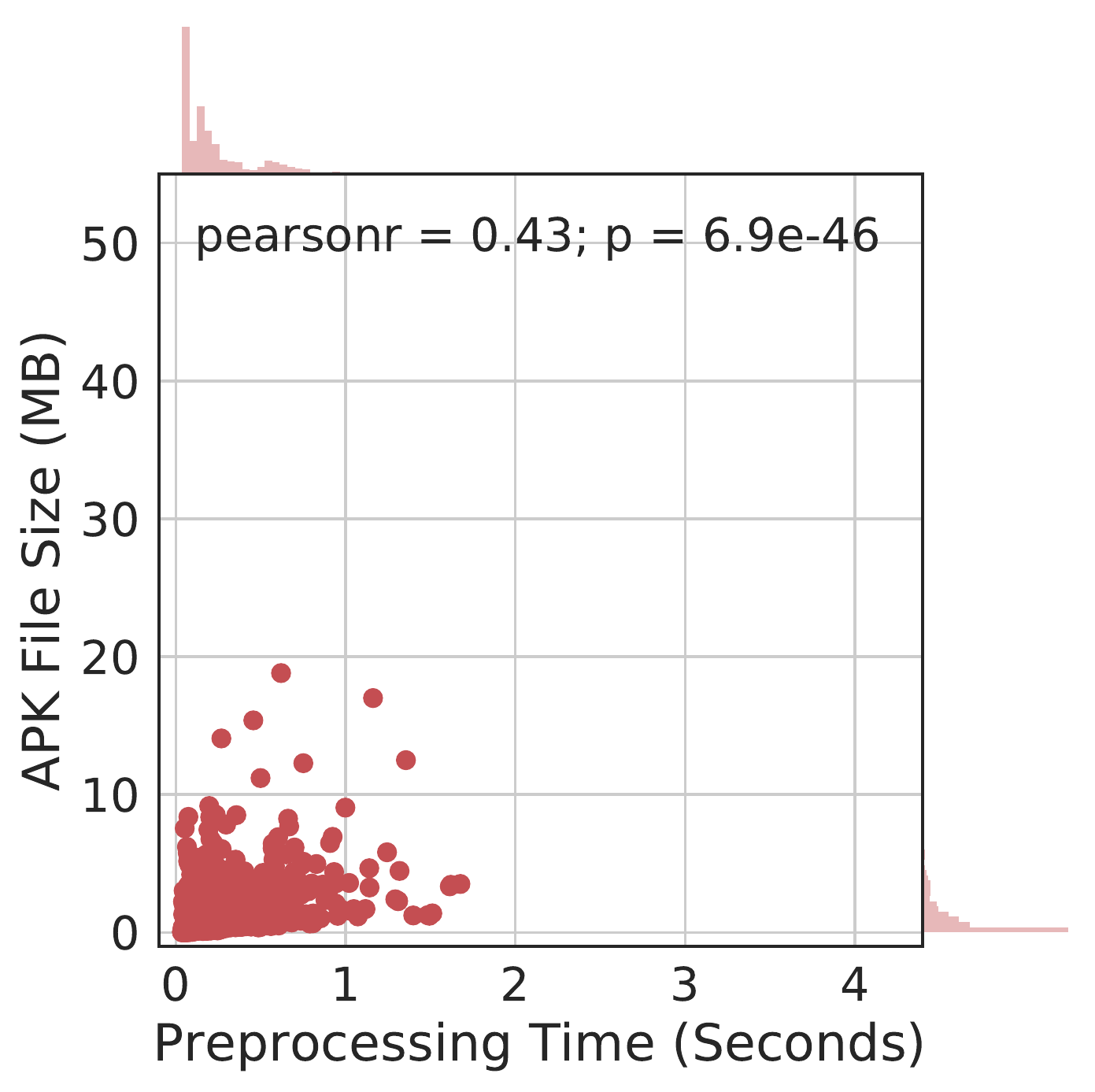}
        }
        \subfigure[\scriptsize Mixed Dex] {%
           \label{fig:ben_precision}
           \includegraphics[width=0.22\textwidth]{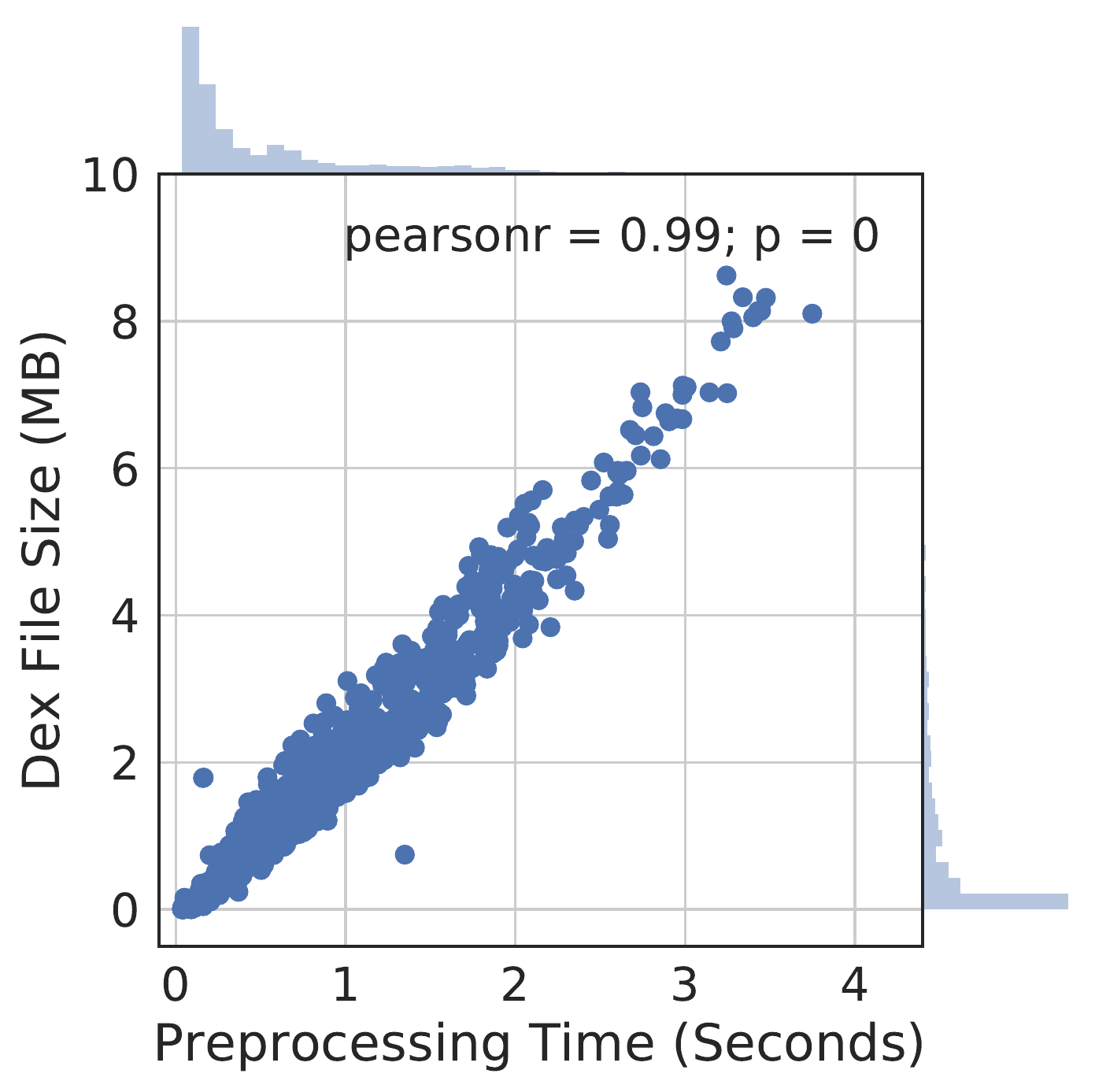}
        }\\
        \subfigure[\scriptsize Bengin Dex] {%
           \label{fig:ben_precision}
           \includegraphics[width=0.22\textwidth]{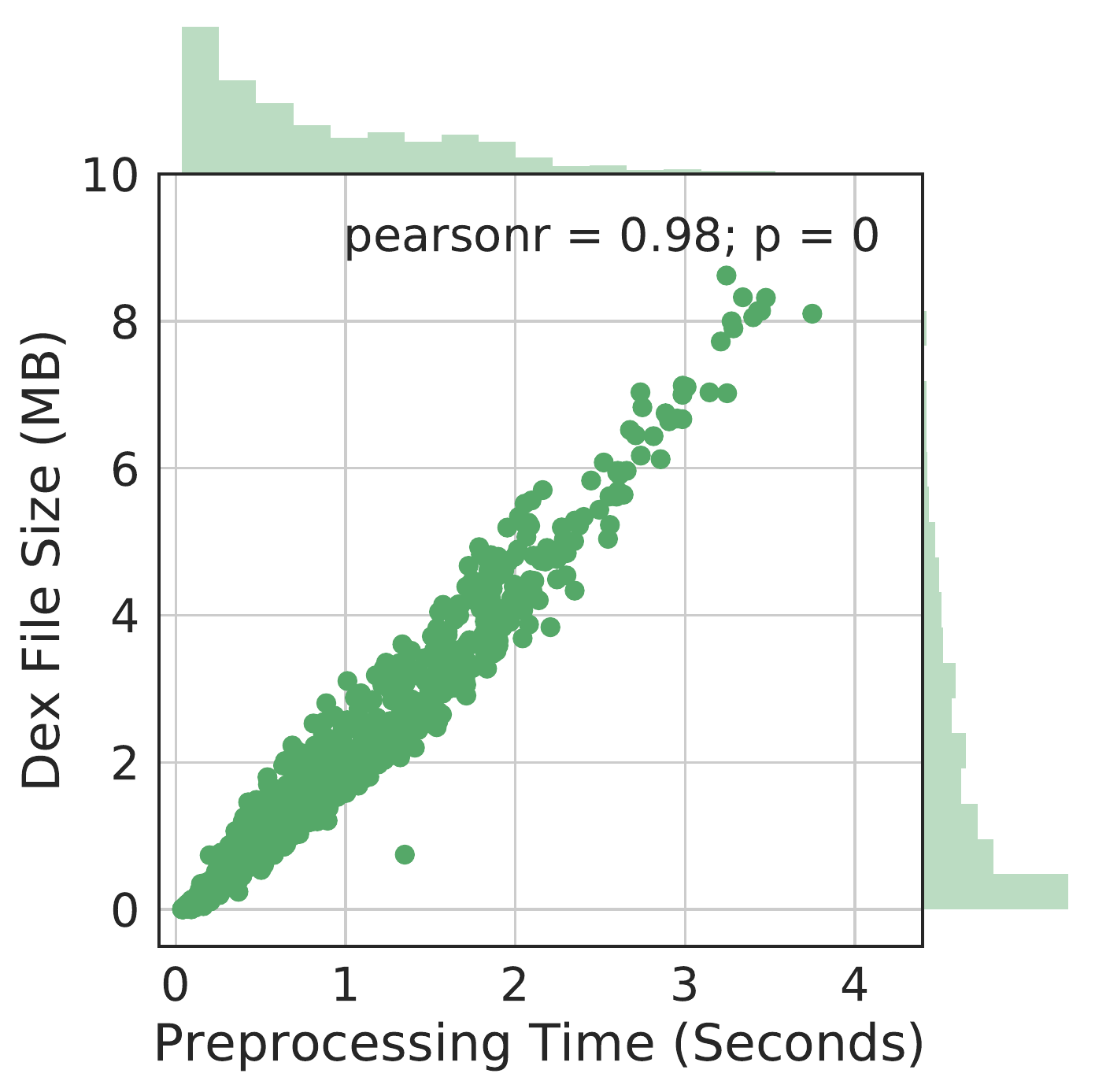}
        }
        \subfigure[\scriptsize Malware Dex] {%
           \label{fig:ben_precision}
           \includegraphics[width=0.22\textwidth]{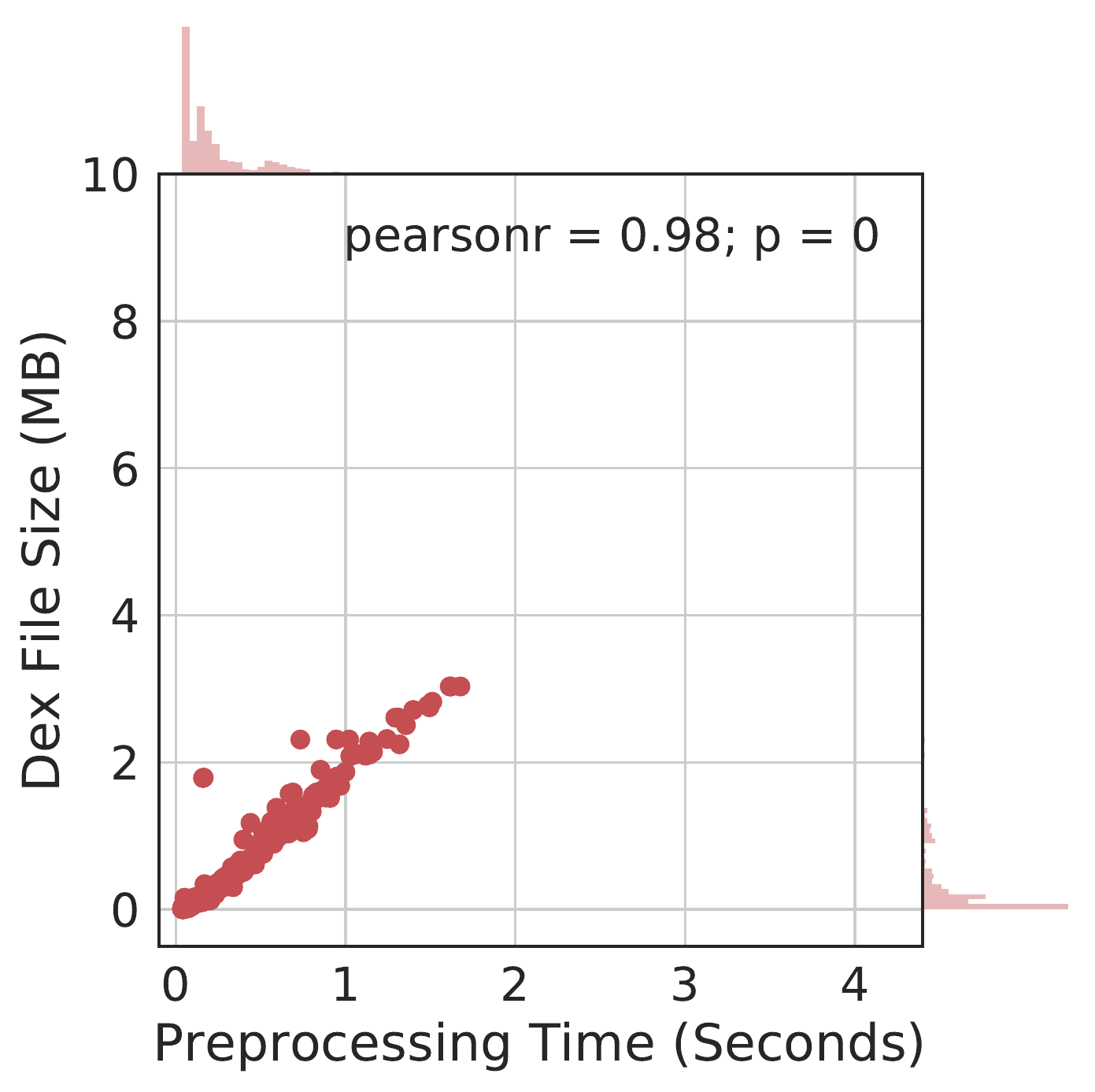}
        }
    \end{center}
    \caption{Preprocessing Time vs APK and Dex Sizes (Server)}
   \label{fig:runtime_preprocess_server}
\end{figure}
\end{scriptsize}

\begin{table}[ht]
\centering
\begin{threeparttable}
\begin{tabular}{|c||c|c|c|c|}
\hline \hline
 & \textbf{\#Params}& \textbf{F1\%} & \textbf{Word2Vec Size} \\
\hline \hline
\textbf{Model 01} &	6.6 Million & 98.95 & 100k \\\hline
\textbf{Model 02} &	4.6 Million & 95.84 & 70k \\\hline
\textbf{Model 03} &	3.4 Million & 93.81 & 50k \\\hline
\textbf{Model 04} & 1.5 Million  & 90.08  & 20k \\
\hline \hline
\end{tabular}
%\begin{tablenotes}
%\item[1] Word2Vec Size \item[2] Millions
%\end{tablenotes}
\end{threeparttable}
\caption{Model Complexity vs Detection Performance} 
\label{tab:model_complexity}
\end{table}

\begin{scriptsize}
\begin{figure}[ht!]
     \begin{center}        
        \subfigure[Server GPU]{%
            \label{fig:ben_fscore}
            \includegraphics[width=0.35\textwidth]{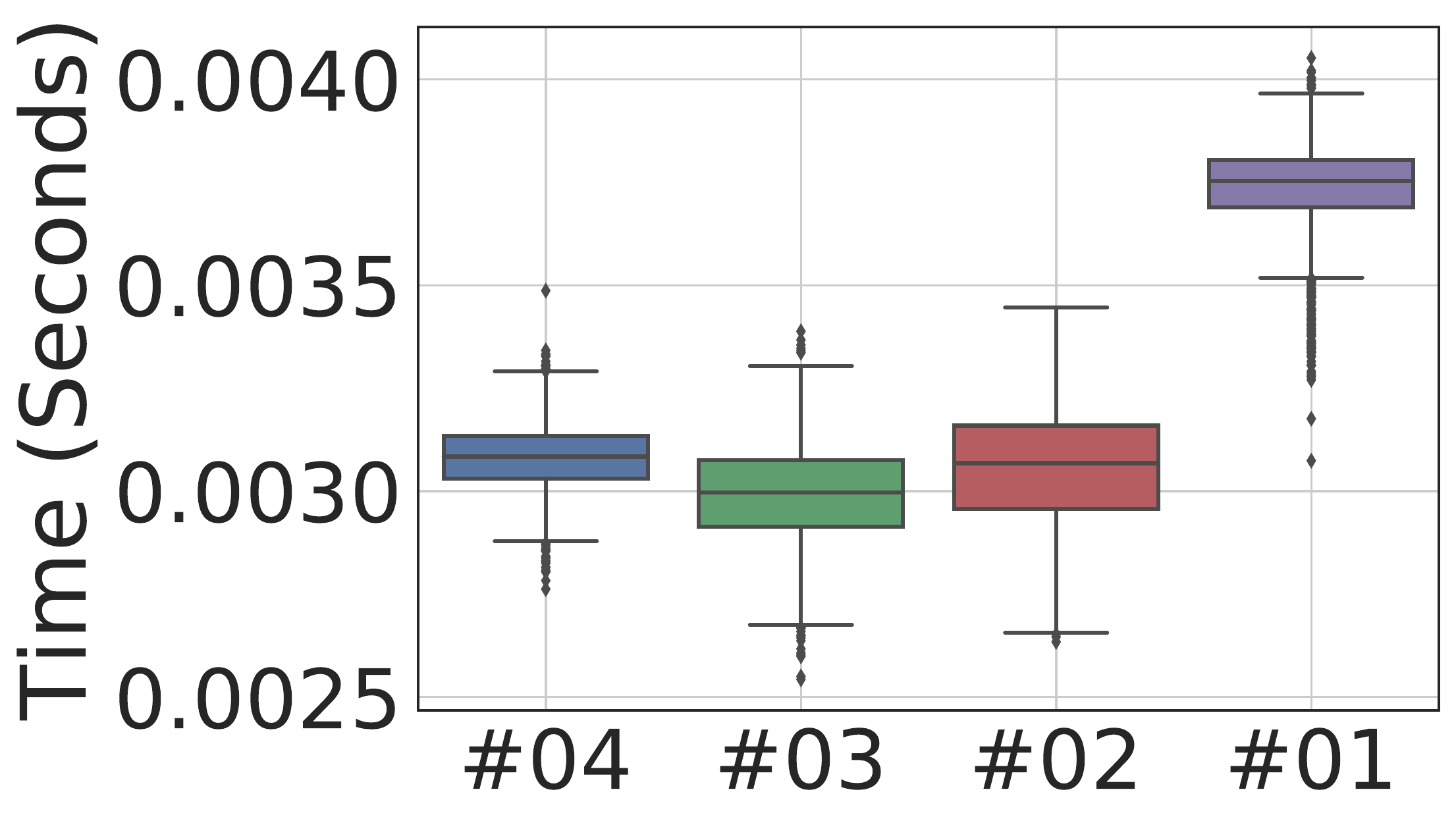}
        }\\
        \subfigure[Server CPU] {%
           \label{fig:ben_precision}
           \includegraphics[width=0.35\textwidth]{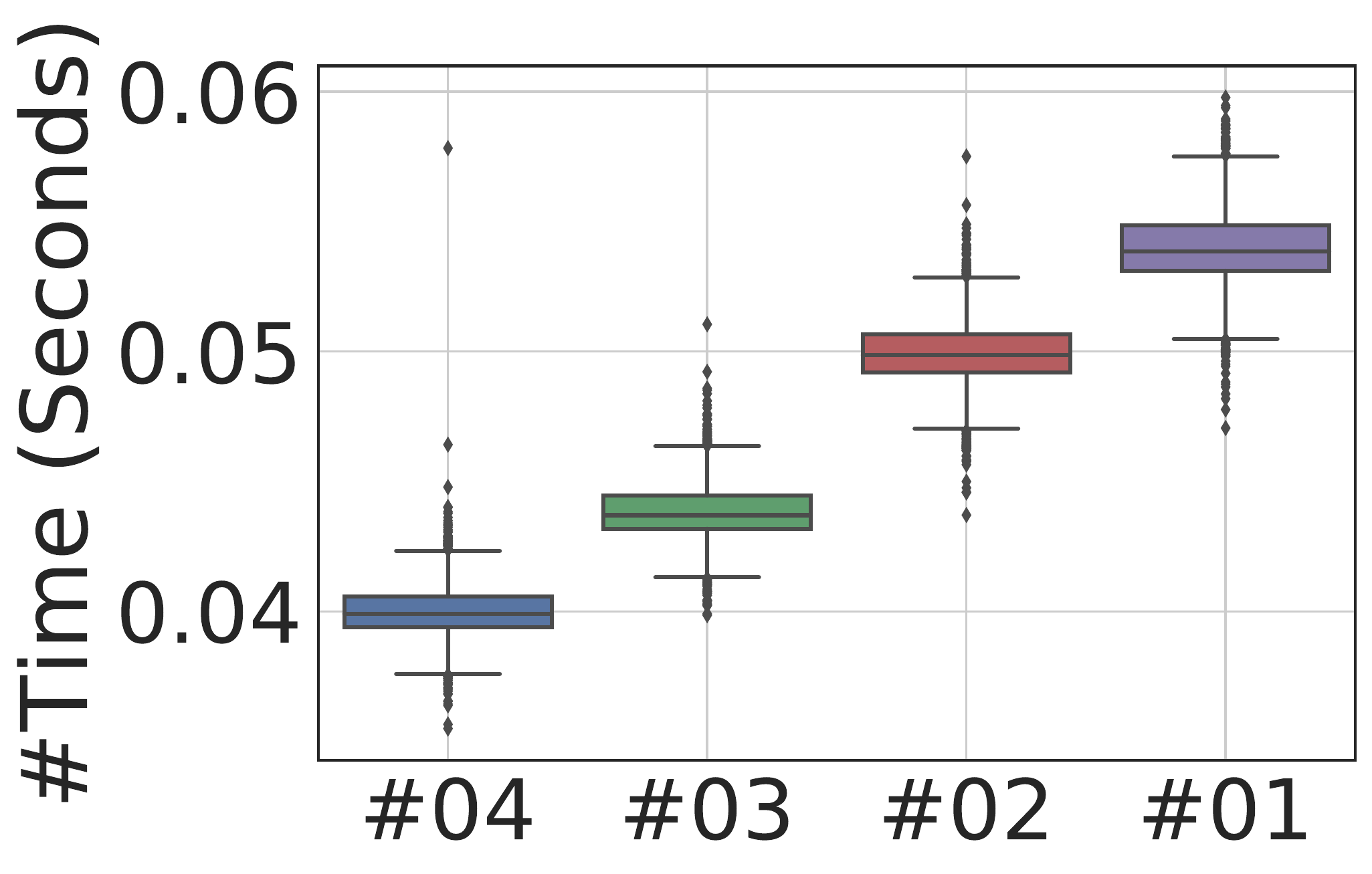}
        }\\
        \subfigure[Laptop]{%
            \label{fig:ben_fscore}
            \includegraphics[width=0.35\textwidth]{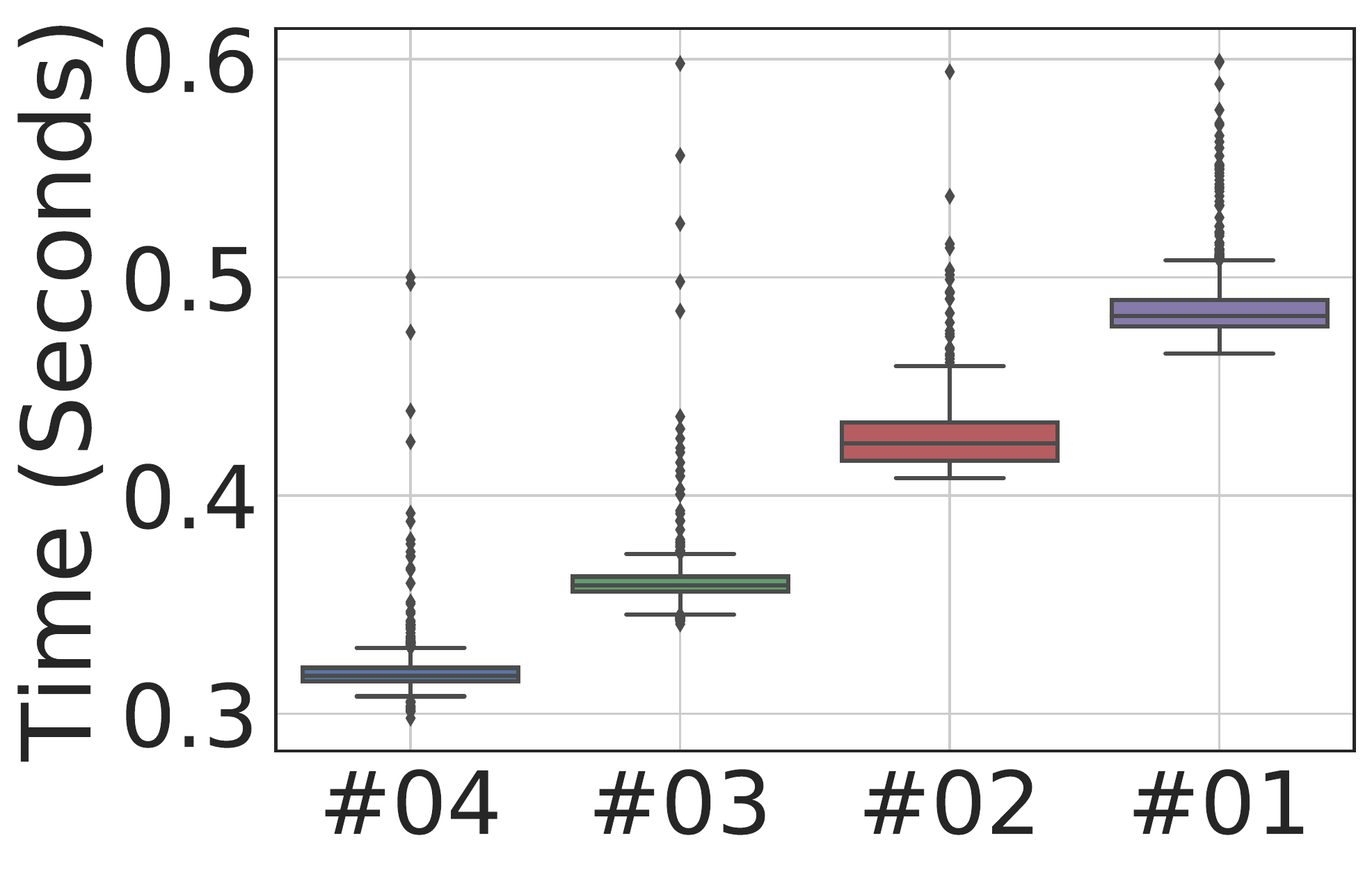}
        }\\
        \subfigure[IoT] {%
           \label{fig:ben_precision}
           \includegraphics[width=0.35\textwidth]{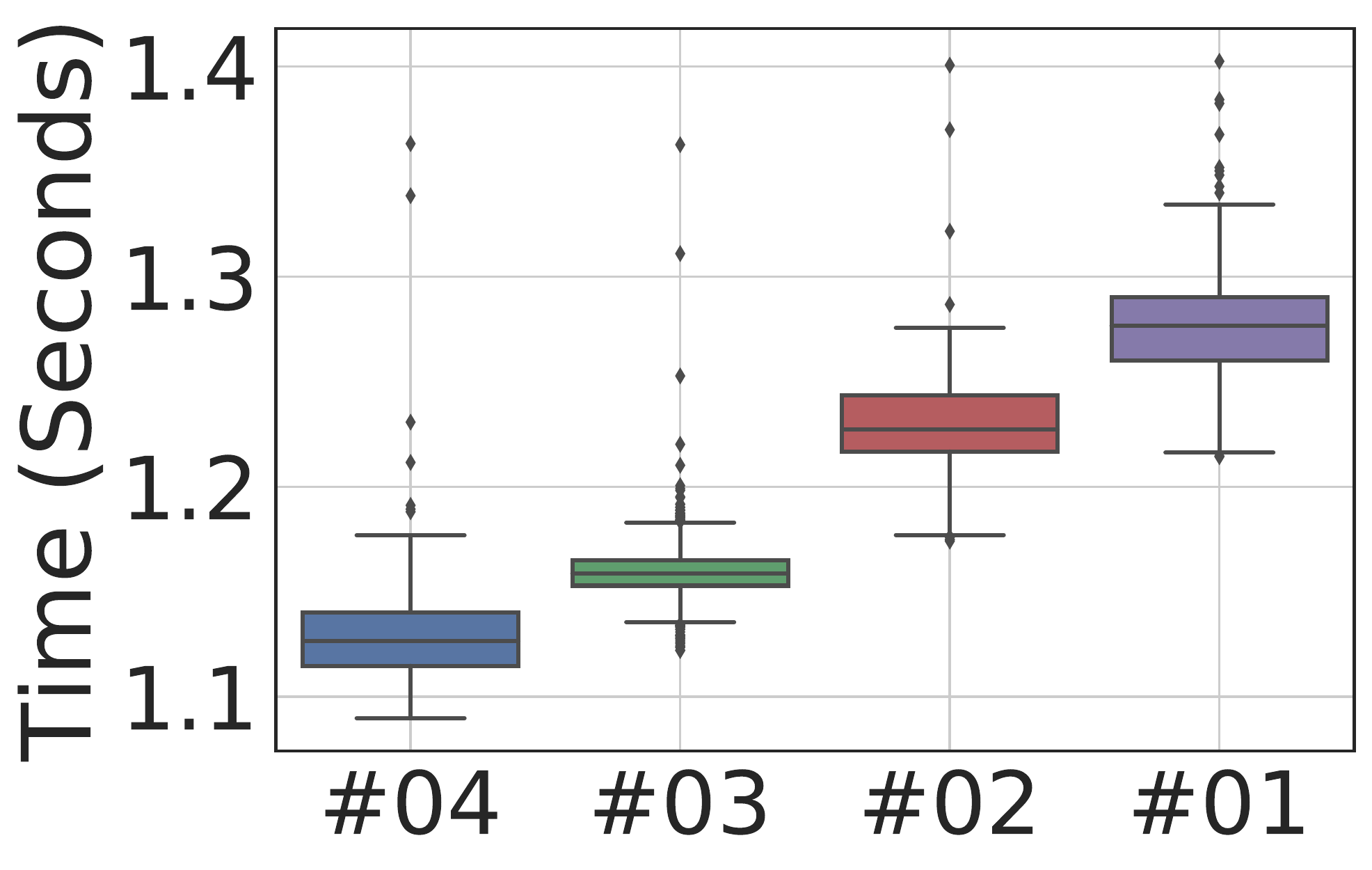}
        }\\
    \end{center}
    \caption{Detection Time vs Model Complexity}
   \label{fig:runtime_preprocess_all}
\end{figure}
\end{scriptsize}

%\begin{figure}[!htb]
%\begin{scriptsize}
%     \begin{center}        
%           \includegraphics[width=0.5\textwidth]{iot_rp2_detection_time_vs_model}
%    \end{center}
%    \caption{Detection Time vs Model Complexity (IoT)}
%   \label{fig:runtime_preprocess_iot}
%\end{scriptsize}
%\end{figure}

\subsubsection{Model Complexity Evaluation} \label{sec:model_complex}
In this section, we examine the effect of model complexity on the detection time. By model complexity, we mean the number of parameters in the model, as depicted in Table \ref{tab:model_complexity}. Many hyper-parameters could influence the complex nature of the model, but we primarily consider the word2vec embedding size. The latter is very important for the detection of the model, especially if we have a big dataset. Table \ref{tab:model_complexity} demonstrates the complexity of the  model versus the F1-Score. It is noticeable that the larger the number of parameters is, the more its performance increases. Based on our observation, bigger models are more accurate and more robust to changes, as will be discussed in Section \ref{sec:limitations}. Finally, Figure \ref{fig:runtime_preprocess_all} displays the execution time of the models in Table \ref{tab:model_complexity}  on the IoT device. The detailed execution related to all the hardware is presented in Figure \ref{fig:runtime_preprocess_all}.

\section{Discussion and Limitations} \label{sec:limitations}
In this paper, we have explored a new approach to capture Android apps behaviors using neural networks on API method calls. This approach achieves highly accurate malware detection and family attribution. Our detection technique is sample-based, i.e., the system could automatically recognize patterns in the training phase of new malware as well as benign apps from raw sequences of API method calls. Therefore, this allows our system to catch up with the rapid evolution of Android OS and malicious techniques by training it on the raw sequence of API methods of new apps, which contain a lot of information about the app's behaviors. Yet, this sequence is less affected by the obfuscation techniques.  Furthermore, our work pushes toward portable detection solutions, i.e., the solution should be used in app stores, mobile or IoT devices. A portable solution is a step towards ubiquitous security that enhances small devices security. In this context, MalDozer could resist to certain obfuscation techniques because we only consider the  API method calls. However, like all the detection schemes that are based on static analysis, \textsf{MalDozer} is not resilient against dynamic code loading and reflection obfuscation, where the app downloads a malicious code and executes it at runtime. Moreover, \textsf{MalDozer} does not  consider native codes. Finally, \textsf{MalDozer} uses many hyper-parameters for its deep learning model. The impact of using a lower number on accuracy detection and running time performance can be the subject of further investigations in future work. This work could also be extended to study the effect of a larger number of API method calls on detection accuracy, training time, and detection time. Furthermore, we plan to extend \textsf{MalDozer} to support the processing of native codes. In this context, MalDozer could resist certain obfuscation techniques because we only consider the API call methods. However, MalDozer is not immune to obfuscation (inherited from static analysis drawbacks) such as the use of reflection/dynamic code execution, as well as, other obfuscation techniques that rely on native code and non-framework service calls.

\section{Related work}
The Android malware analysis techniques can be classified as: \textit{static analysis}, \textit{dynamic analysis}, or \textit{hybrid analysis}. The static analysis methods \cite{arp2014drebin}, \cite{feng2014apposcopy}, \cite{yang2014apklancet}, 
\cite{mariconti2017mamadroid}, \cite{zhongyang2013droidalarm}, \cite{karbab2016cypider},  \cite{sanz2014anomaly}, \cite{karbab2016dna}, 
use static features that are extracted from the app, such as: requested permissions and APIs to detect malicious app. Some of these methods are generally not resistant to obfuscation. The dynamic analysis methods \cite{canfora2016acquiring}, \cite{karbab2016dysign}, \cite{spreitzenbarth2013mobile}, \cite{ali2016aspectdroid},\cite{zhang2013vetting}, \cite{amos2013applying}, \cite{wei2012android}, \cite{huang2014asdroid} 
aim to identify behavioral signature or behavioral anomaly of the running app. These methods are more resistant to obfuscation. On the other hand, the dynamic methods offer limited scalability as they incur additional cost in terms of processing and memory. The hybrid  analysis methods \cite{yuan2014droid}, \cite{grace2012riskranker}, \cite{bhandari2015draco}, \cite{vidas2014a5}, \cite{lindorfer2014andrubis} 
combine between both analyses to improve detection accuracy, which costs additional computational cost. Assuming that malicious apps of the same family share similar features, some methods  \cite{kim2015structural}, \cite{ali2015opseq}, \cite{deshotels2014droidlegacy}, \cite{zhou2012hey}, \cite{suarez2014dendroid}, \cite{kang2015detecting}, \cite{lin2013identifying}, \cite{faruki2015androsimilar}. 
measure the similarity between the features of two samples (similar malicious code). Some methods \cite{zhang2014semantics}, \cite{fan2016frequent}, \cite{meng2016semantic} 
employ semantics-aware features such as control-flow graphs \cite{christodorescu2005semantics}, data dependency graphs \cite{fredrikson2010synthesizing} and class dependence graphs \cite{deshotels2014droidlegacy}. The deep learning techniques are more suitable than conventional machine learning techniques for Android malware detection \cite{yuan2014droid}. Research work on deep learning for Android malare detection are recently getting more attention \cite{yuan2014droid}, \cite{yuan2016droiddetector}, \cite{hou2016deep4maldroid}, \cite{hou2016droiddelver}. 
Differently from the existing deep learning solutions, \textsf{MalDozer} offers many  advantages: (i) \textsf{MalDozer} provide automatic feature engineering for new types of malware in the training phase. (ii) \textsf{MalDozer} uses a minimal proprocessing, which fits small devices deployment. (iii) In addition to its high detection performance, \textsf{MalDozer} is able to attribute malware to its actual family with similar performance.

\subsection{MalDozer vs. MAMADroid}
In this section, we highlight the differences between \textsf{MalDozer} and MAMADroid \cite{mariconti2017mamadroid}, a recent and competitive malware detection solution that combines semantics-aware static features and machine learning. MAMADroid uses the sequence of abstracted API calls to build a behavioral model in the form of a Markov chain. The latter is used to extract features and perform classification. We compare our framework with MAMADroid with respect to the following points: (1) \textbf{Deployment:} Due to its large memory requirements, MAMADroid can only be deployed on high-power servers. In contrast, the detection component of \textsf{MalDozer} can efficiently run under multiple deployment architectures, including high-power and low-power machines as well as mobile and IoT devices. (2) \textbf{Design complexity:} The design of MAMADroid is complex since its preprocessing phase is composed of many steps. MAMADroid first constructs a call graph using flow-analysis tools such as: Soot \cite{vallee1999soot} and FlowDroid \cite{arzt2014flowdroid}. Soot converts the DEX file into Jimple and FlowDroid is used for Taint analysis. Next, the constructed graph is used to generate the sequence of abstracted API calls, which are used to build a Markov chain.  
%Also, Principal Component Analysis (PCA) is applied on the generated features. 
\textsf{MalDozer}, on the other hand, extracts the raw sequence of API method calls directly from the DEX file without any additional preprocessisng overhead. 
%(3)\textbf{Learning model:} \textsf{MalDozer} adopts a deep learning approach for dataset training and testing. Compared to the machine learning classifiers used in MAMADroid, the deep learning techniques are more suitable for Android malware detection \cite{yuan2014droid}. 
(3) \textbf{Adaptation to new APIs:} In MAMADroid, the release of new Android OS version with a new set of APIs implies that a manual intervention is needed to craft a new Markov model, and manually redefine the set of APIs to be fed to the classifier. \textsf{MalDozer} can automatically learn the new APIs directly from the DEX file, as they have a specific known format. 
%This makes \textsf{MalDozer} adaptive to changes in Android framework APIs.
%\textbf{Feature granularity:} MAMADroid uses two levels of abstraction, as it extracts from each API call its package and family. In addition, \textsf{MalDozer} uses fine-grained features as it goes deep and extracts the corresponding API method from each API call. Consequently, \textsf{MalDozer} can capture, more accurately, the behavior of the app.
(4)\textbf{Approach performance:} MAMADroid consumes a very large amount of memory (16GB of RAM) to extract the call graph from the app. Despite that, the graph extraction phase takes, in average, 25.4 seconds, and a maximum of 18 minutes. \textsf{MalDozer} is very lightweight. The average time to extract the features and classify the app in \textsf{MalDozer} is $5.3$ seconds on the most resource-constrained IoT device.   
(5)\textbf{Ability to analyze apks:} The Soot tool, used by MAMADroid, fails to process $4.6\%$ of the apps in the dataset of $43,940$ apps, which counts for more than $2,000$ apps. Such a failure does not happen in the case of \textsf{MalDozer} due to its simple preprocess. 
The above-discussed points clearly show the outperformance of \textsf{MalDozer} over MAMADroid.

\section{Conclusion}
We have presented \textsf{MalDozer}, an automatic, efficient and effective Android malware detection and attribution system. \textsf{MalDozer} relies on deep learning techniques and raw sequences of API method calls in order to identify Android malware. We have evaluated \textsf{MalDozer} on several small and large datasets, including \textit{Malgenome}, \textit{Drebin}, and our \textsf{MalDozer} dataset, in addition to a dataset of benign apps downloaded from Google Play. The evaluation results show that \textsf{MalDozer} is highly accurate in terms of malware detection as well as their attribution to corresponding families. Moreover, \textsf{MalDozer} can efficiently run under multiple deployment architectures, ranging from servers to small IoT devices. This work represents a step towards practical, automatic and effective Android malware detection and family attribution.

%\section*{Acknowledgements}
%The authors would like to thank the anonymous reviewers  and Timothy Vidas for  their  insightful  comments  that allowed us to significantly improve this paper.

%\section*{References}
\bibliography{references}

\begin{thebibliography}{10}
\expandafter\ifx\csname url\endcsname\relax
  \def\url#1{\texttt{#1}}\fi
\expandafter\ifx\csname urlprefix\endcsname\relax\def\urlprefix{URL }\fi
\expandafter\ifx\csname href\endcsname\relax
  \def\href#1#2{#2} \def\path#1{#1}\fi

\bibitem{delmastro2016people}
F.~Delmastro, V.~Arnaboldi, M.~Conti, People-centric computing and
  communications in smart cities, IEEE Communications Magazine.

\bibitem{Gilchrist:2016:III:2994178}
A.~Gilchrist, Industry 4.0: The Industrial Internet of Things, 2016.

\bibitem{Yan:2008:ITR:1796470}
L.~Yan, Y.~Zhang, al, The Internet of Things: From RFID to the Next-Generation
  Pervasive Networked Systems, Auerbach Publications, 2008.

\bibitem{Ericsson2016}
{Ericsson Mobility Report - https://tinyurl.com/gmnezg6} (2016).

\bibitem{Smartphone2016}
{Smartphone os market share, 2017 q1 - https://tinyurl.com/y8tqgjfu} (2017).

\bibitem{Android2016}
{Android Things on the Intel Edison board - https://tinyurl.com/gl9gglk}
  (2016).

\bibitem{IoT2016}
{Android Things OS - https://tinyurl.com/z3lf3ha} (2016).

\bibitem{rasp_3_iot}
{RASPBERRY PI 3- https://tinyurl.com/ho7ngty} (2017).

\bibitem{brillokey}
{Android Things - http://tinyurl.com/q5ko3zu} (2016).

\bibitem{android_auto}
{Android Auto - http://tinyurl.com/hdsunht} (2016).

\bibitem{android_wear}
{Android Wear - http://tinyurl.com/qfa55o4} (2016).

\bibitem{rasp_2_iot}
{RASPBERRY PI 2 - https://tinyurl.com/q65phuy} (2017).

\bibitem{malgenome_dataset}
{MalGenome Dataset - http://tinyurl.com/combopx} (2015).

\bibitem{Drebin_Dataset}
{Drebin Dataset - http://tinyurl.com/pdsrtez} (2015).

\bibitem{google_play}
{Google Play - https://play.google.com/} (2016).

\bibitem{Mikolov2013Distributed}
T.~Mikolov, I.~Sutskever, al, {Distributed Representations of Words and Phrases
  and their Compositionality}, NIPS Neural Inf. Process. Syst.

\bibitem{android_ndk}
{Android NDK - http://tinyurl.com/ppn559l} (2016).

\bibitem{oha_handset}
{Open Handset Alliance - https://tinyurl.com/2u76za} (2016).

\bibitem{android_arch}
{Android Platform Architecture - https://tinyurl.com/hc7s4or} (2017).

\bibitem{Kim2014Convolutional}
Y.~Kim, Convolutional neural networks for sentence classification, CoRR.

\bibitem{Pennington201GloVe}
J.~Pennington, R.~Socher, al., {GloVe: Global Vectors for Word Representation},
  in: Conf. Empir. Methods Nat. Lang. Process., 2014.

\bibitem{Goodfellow-et-al-2016}
I.~Goodfellow, Y.~Bengio, Al, Deep Learning, MIT Press, 2016.

\bibitem{tensorflow}
{Tensorflow - https://www.tensorflow.org} (2017).

\bibitem{arp2014drebin}
D.~Arp, M.~Spreitzenbarth, H.~Malte, H.~Gascon, al., {DREBIN: Effective and
  Explainable Detection of Android Malware in Your Pocket.}, in: Symp. Netw.
  Distrib. Syst. Secur., 2014.

\bibitem{contagiominidump}
{Contagiominidump - https://contagiominidump.blogspot.ca} (2017).

\bibitem{playdrone}
{playdrone dataset - https://archive.org/details/playdrone-apks} (2017).

\bibitem{DBLP:conf/msr/AllixBKT16}
K.~Allix, al, Androzoo: collecting millions of android apps for the research
  community, in: Proceedings of the 13th International Conference on Mining
  Software Repositories, {MSR}, 2016.

\bibitem{mariconti2017mamadroid}
E.~Mariconti, L.~Onwuzurike, P.~Andriotis, E.~De~Cristofaro, G.~Ross,
  G.~Stringhini, Mamadroid: Detecting android malware by building markov chains
  of behavioral models, in: NDSS, 2017.

\bibitem{feng2014apposcopy}
Y.~Feng, S.~Anand, al., Apposcopy: Semantics-based detection of android malware
  through static analysis, in: Proceedings of the 22nd ACM International
  Symposium on Foundations of Software Engineering, 2014.

\bibitem{yang2014apklancet}
W.~Yang, J.~Li, Y.~Zhang, al., Apklancet: tumor payload diagnosis and
  purification for android applications, in: Proceedings of the 9th ACM
  symposium on Information, computer and communications security, 2014.

\bibitem{zhongyang2013droidalarm}
Y.~Zhongyang, Z.~Xin, al., Droidalarm: an all-sided static analysis tool for
  android privilege-escalation malware, in: Proceedings of the 8th ACM SIGSAC
  symposium on Information, computer and communications security, 2013.

\bibitem{karbab2016cypider}
E.~B. Karbab, M.~Debbabi, A.~Derhab, D.~Mouheb, {Cypider: Building
  Community-Based Cyber-Defense Infrastructure for Android Malware Detection},
  in: ACM Computer Security Applications Conference (ACSAC), 2016.

\bibitem{sanz2014anomaly}
B.~Sanz, I.~Santos, al., Anomaly detection using string analysis for android
  malware detection, in: International Joint Conference
  SOCO’13-CISIS’13-ICEUTE’13, Springer, 2014.

\bibitem{karbab2016dna}
E.~B. Karbab, M.~Debbabi, D.~Mouheb,
  \href{http://linkinghub.elsevier.com/retrieve/pii/S1742287616300469}{{Fingerprinting
  Android packaging: Generating DNAs for malware detection}}, Digital
  Investigation\href {http://dx.doi.org/10.1016/j.diin.2016.04.013}
  {\path{doi:10.1016/j.diin.2016.04.013}}.
\newline\urlprefix\url{http://linkinghub.elsevier.com/retrieve/pii/S1742287616300469}

\bibitem{canfora2016acquiring}
G.~Canfora, E.~Medvet, al., {Acquiring and Analyzing App Metrics for Effective
  Mobile Malware Detection}, in: Proc. 2016 ACM Int. Work. Secur. Priv. Anal.,
  2016.

\bibitem{karbab2016dysign}
E.~M.~B. Karbab, M.~Debbabi, S.~Alrabaee, D.~Mouheb, {DySign: Dynamic
  fingerprinting for the automatic detection of android malware}, 2016 11th
  International Conference on Malicious and Unwanted Software, MALWARE 2016
  (2016) 139--146\href {http://arxiv.org/abs/1702.05699}
  {\path{arXiv:1702.05699}}, \href
  {http://dx.doi.org/10.1109/MALWARE.2016.7888739}
  {\path{doi:10.1109/MALWARE.2016.7888739}}.

\bibitem{spreitzenbarth2013mobile}
M.~Spreitzenbarth, F.~Freiling, F.~Echtler, al., Mobile-sandbox: having a
  deeper look into android applications, in: Proceedings of the 28th Annual ACM
  Symposium on Applied Computing, 2013.

\bibitem{ali2016aspectdroid}
A.~Ali-Gombe, al, Aspectdroid: Android app analysis system, in: ACM Conference
  on Data and Application Security and Privacy, 2016.

\bibitem{zhang2013vetting}
Y.~Zhang, M.~Yang, B.~Xu, Z.~Yang, al., {Vetting undesirable behaviors in
  android apps with permission use analysis}, in: ACM conference on Computer
  and communications security (CCS), 2013.

\bibitem{amos2013applying}
B.~Amos, al, {Applying machine learning classifiers to dynamic android malware
  detection at scale}, in: Wirel. Commun. Mob. Comput. Conf., 2013.

\bibitem{wei2012android}
T.-E.~E. Wei, C.-H.~H. Mao, A.~B. Jeng, H.-M.~M. Lee, H.-T.~T. Wang, D.-J.~J.
  Wu, in: Trust. Secur. Priv. Comput. Commun., 2012.

\bibitem{huang2014asdroid}
J.~Huang, X.~Zhang, al., Asdroid: Detecting stealthy behaviors in android
  applications by user interface and program behavior contradiction, in:
  Proceedings of the 36th International Conference on Software Engineering,
  2014.

\bibitem{yuan2014droid}
Z.~Yuan, Y.~Lu, Z.~Wang, Y.~Xue, {Droid-Sec: deep learning in android malware
  detection}, in: ACM SIGCOMM Comput. Commun. Rev., 2014.

\bibitem{grace2012riskranker}
M.~Grace, Y.~Zhou, al., Riskranker: scalable and accurate zero-day android
  malware detection, in: Proceedings of the 10th international conference on
  Mobile systems, applications, and services, 2012.

\bibitem{bhandari2015draco}
S.~Bhandari, R.~Gupta, al., {DRACO: DRoid analyst combo an android malware
  analysis framework}, in: Proc. 8th Int. Conf. Secur. Inf. Networks, 2015.

\bibitem{vidas2014a5}
T.~Vidas, J.~Tan, J.~Nahata, al., A5: Automated analysis of adversarial android
  applications, in: Proceedings of the 4th ACM Workshop on Security and Privacy
  in Smartphones $\&$ Mobile Devices, 2014.

\bibitem{lindorfer2014andrubis}
M.~Lindorfer, M.~Neugschwandtner, al., Andrubis--1,000,000 apps later: A view
  on current android malware behaviors, in: Building Analysis Datasets and
  Gathering Experience Returns for Security (BADGERS), IEEE, 2014.

\bibitem{kim2015structural}
J.~Kim, al, {Structural information based malicious app similarity calculation
  and clustering}, in: Proc. 2015 Conf. Res. Adapt. Converg. Syst., 2015.

\bibitem{ali2015opseq}
A.~Ali-Gombe, al, Opseq: android malware fingerprinting, in: Proceedings of the
  5th Program Protection and Reverse Engineering Workshop, 2015.

\bibitem{deshotels2014droidlegacy}
L.~Deshotels, V.~Notani, A.~Lakhotia, Droidlegacy: Automated familial
  classification of android malware, in: Proceedings of ACM SIGPLAN on Program
  Protection and Reverse Engineering Workshop 2014, 2014.

\bibitem{zhou2012hey}
Y.~Zhou, Z.~Wang, al, Hey, you, get off of my market: detecting malicious apps
  in official and alternative android markets., in: NDSS, 2012.

\bibitem{suarez2014dendroid}
G.~Suarez-Tangil, J.~E. Tapiador, al., Dendroid: A text mining approach to
  analyzing and classifying code structures in android malware families, Expert
  Systems with Applications.

\bibitem{kang2015detecting}
H.~Kang, J.-w. Jang, A.~Mohaisen, H.~K. Kim, Int. J. Distrib. Sens. Networks.

\bibitem{lin2013identifying}
Y.-D. Lin, Y.-C. Lai, C.-H. Chen, H.-C. Tsai, Identifying android malicious
  repackaged applications by thread-grained system call sequences, computers
  $\&$ security.

\bibitem{faruki2015androsimilar}
P.~Faruki, V.~Laxmi, A.~Bharmal, M.~S. Gaur, V.~Ganmoor, {AndroSimilar: Robust
  signature for detecting variants of Android malware}, J. Inf. Secur. Appl.

\bibitem{zhang2014semantics}
M.~Zhang, al., Semantics-aware android malware classification using weighted
  contextual api dependency graphs, in: Proceedings of the 2014 ACM SIGSAC
  Conference on Computer and Communications Security, 2014.

\bibitem{fan2016frequent}
M.~Fan, J.~Liu, X.~Luo, K.~Chen, al., Frequent subgraph based familial
  classification of android malware, in: Software Reliability Engineering
  (ISSRE), 2016 IEEE 27th International Symposium on, 2016.

\bibitem{meng2016semantic}
G.~Meng, Y.~Xue, Z.~Xu, al, Semantic modelling of android malware for effective
  malware comprehension, detection, and classification, in: Proceedings of the
  25th International Symposium on Software Testing and Analysis, 2016.

\bibitem{christodorescu2005semantics}
M.~Christodorescu, S.~Jha, Seshia, Semantics-aware malware detection, in: IEEE
  Symposium on Security and Privacy (SP), 2005.

\bibitem{fredrikson2010synthesizing}
M.~Fredrikson, S.~Jha, M.~Christodorescu, R.~Sailer, X.~Yan, Synthesizing
  near-optimal malware specifications from suspicious behaviors, in: IEEE
  Symposium on Security and Privacy (SP), 2010.

\bibitem{yuan2016droiddetector}
Z.~Yuan, Y.~Lu, Y.~Xue, Droiddetector: android malware characterization and
  detection using deep learning, Tsinghua Science and Technology.

\bibitem{hou2016deep4maldroid}
S.~Hou, A.~Saas, al., Deep4maldroid: A deep learning framework for android
  malware detection based on linux kernel system call graphs, in: International
  Conference on Web Intelligence Workshops, 2016.

\bibitem{hou2016droiddelver}
S.~Hou, A.~Saas, al.s, Droiddelver: An android malware detection system using
  deep belief network based on api call blocks, in: International Conference on
  Web-Age Information Management, 2016.

\bibitem{vallee1999soot}
R.~Vall{\'e}e-Rai, P.~Co, E.~Gagnon, L.~Hendren, P.~Lam, V.~Sundaresan, Soot-a
  java bytecode optimization framework, in: Proceedings of the 1999 conference
  of the Centre for Advanced Studies on Collaborative research, 1999.

\bibitem{arzt2014flowdroid}
S.~Arzt, S.~Rasthofer, C.~Fritz, al., Flowdroid: Precise context, flow, field,
  object-sensitive and lifecycle-aware taint analysis for android apps, Acm
  Sigplan Notices.

\end{thebibliography}

\clearpage
\newpage
\appendix
\section{Confusion Matrices}

%\subsection{Detection Results}
%\subsection{Attribution Results}

\begin{table}[!h]
\centering
\begin{threeparttable}
\begin{tabular}{|c||c|c|c||c|c|}
\multicolumn{3}{c}{\textbf{2 Fold}} & \multicolumn{3}{c}{\textbf{3 Fold}} \\\hline
& PB\tnote{1}  & PM  &  &  PB & PM \\\hline
AB\tnote{2}		&  37,604  	& 23   	 & AB	&  36,882  	& 745   \\\hline
AM 	&  107  	& 1,151  & AM 	&  40  		& 1,218   \\
\hline %\hline
\multicolumn{3}{c}{\textbf{5 Fold}}& \multicolumn{3}{c}{\textbf{10 Fold}} \\\hline
  & PB & PM   & &  PB & PM \\\hline
AB	&  37,591  	& 36   	 & AB	&  37,612  	& 15   \\\hline
AM 	&  40  		& 1,218  & AM 	&  44  		& 1,214   \\
\hline %\hline
\end{tabular}
\begin{tablenotes}
\item[1] PB: Predicted Benign, PM: Predicted Malware
\item[2] AB: Actual Benign, PM: Actual Malware
\end{tablenotes}
\end{threeparttable}
\caption{Detailed Detection Result on Malgenome} 
\label{tab:detecton_confmatrix_malgenome}
\end{table}

\begin{table}[!h]
\centering
\begin{threeparttable}
\begin{tabular}{|c||c|c|c||c|c|}
\multicolumn{3}{c}{\textbf{2 Fold}}& \multicolumn{3}{c}{\textbf{3 Fold}} \\\hline
  & PB\tnote{1}  & PM  & &  PB & PM \\\hline
AB\tnote{2}		&  37,578  	& 49   	 & AB	&  37,435  	& 192   \\\hline
AM 				&  426  	& 5,129  & AM 	&  233  		& 5,322   \\
\hline %\hline
\multicolumn{3}{c}{\textbf{5 Fold}}& \multicolumn{3}{c}{\textbf{10 Fold}} \\\hline
  & PB & PM   & &  PB & PM \\\hline
AB	&  37,509  	& 118   	 & AB	&  37,457  	& 170   \\\hline
AM 	&  261  		& 5,294  & AM 	&  168  	& 5,387   \\
\hline %\hline
\end{tabular}
\begin{tablenotes}
\item[1] PB: Predicted Benign, PM: Predicted Malware
\item[2] AB: Actual Benign, PM: Actual Malware
\end{tablenotes}
\end{threeparttable}
\caption{Detailed Detection Result on Drebin} 
\label{tab:detecton_confmatrix_drebin}
\end{table}

\begin{table}[!h]
\centering
\begin{threeparttable}
\begin{tabular}{|c||c|c|c||c|c|}
\multicolumn{3}{c}{\textbf{2 Fold}}& \multicolumn{3}{c}{\textbf{3 Fold}} \\\hline 
  & PB\tnote{1}  & PM  & &  PB & PM \\\hline
AB\tnote{2}		&  37,246  	& 381   	 & AB	&  36,872  	& 755   \\\hline
AM 				&  1,421  	& 18,668  & AM 	&  618  		& 19,471  \\
\hline %\hline
\multicolumn{3}{c}{\textbf{5 Fold}}& \multicolumn{3}{c}{\textbf{10 Fold}} \\\hline
  & PB & PM   & &  PB & PM \\\hline
AB	&  36,779  		& 848   	 & AB	&  37,193  	& 434   \\\hline
AM 	&  436  		& 19,653  & AM 	&  611  	& 19,478   \\
\hline %\hline
\end{tabular}
\begin{tablenotes}
\item[1] PB: Predicted Benign, PM: Predicted Malware
\item[2] AB: Actual Benign, PM: Actual Malware
\end{tablenotes}\end{threeparttable}
\caption{Detailed Detection Result on MalDozer} 
\label{tab:detecton_confmatrix_maldozer}
\end{table}

\begin{table}[!h]
\centering
\begin{threeparttable}
\begin{tabular}{|c||c|c|c||c|c|}
\multicolumn{3}{c}{\textbf{2 Fold}}& \multicolumn{3}{c}{\textbf{3 Fold}} \\\hline 
  & PB\tnote{1}  & PM  & &  PB & PM \\\hline
AB\tnote{2}		&  36,673  	& 954   	 & AB	&  36,118  	& 1,509   \\\hline
AM 				&  1,821  	& 31,245  & AM 	&  2,006  		& 31,060  \\
\hline %\hline
\multicolumn{3}{c}{\textbf{5 Fold}} & \multicolumn{3}{c}{\textbf{10 Fold}} \\\hline
  & PB & PM   & &  PB & PM \\\hline
AB	&  36,621  		& 1,006   	 & AB	&  36,426  	& 1,201   \\\hline
AM 	&  1,585  		& 31,481  & AM 	&  1,417  	& 31,649   \\
\hline %\hline
\end{tabular}
\begin{tablenotes}
\item[1] PB: Predicted Benign, PM: Predicted Malware
\item[2] AB: Actual Benign, PM: Actual Malware
\end{tablenotes}\end{threeparttable}
\caption{Detailed Detection Result on All Dataset} 
\label{tab:detecton_confmatrix_all}
\end{table}

\begin{figure}[ht!]
     \begin{center}        
        \subfigure[Malgenome]{%
            \label{fig:ben_fscore}
            \includegraphics[width=0.47\textwidth]{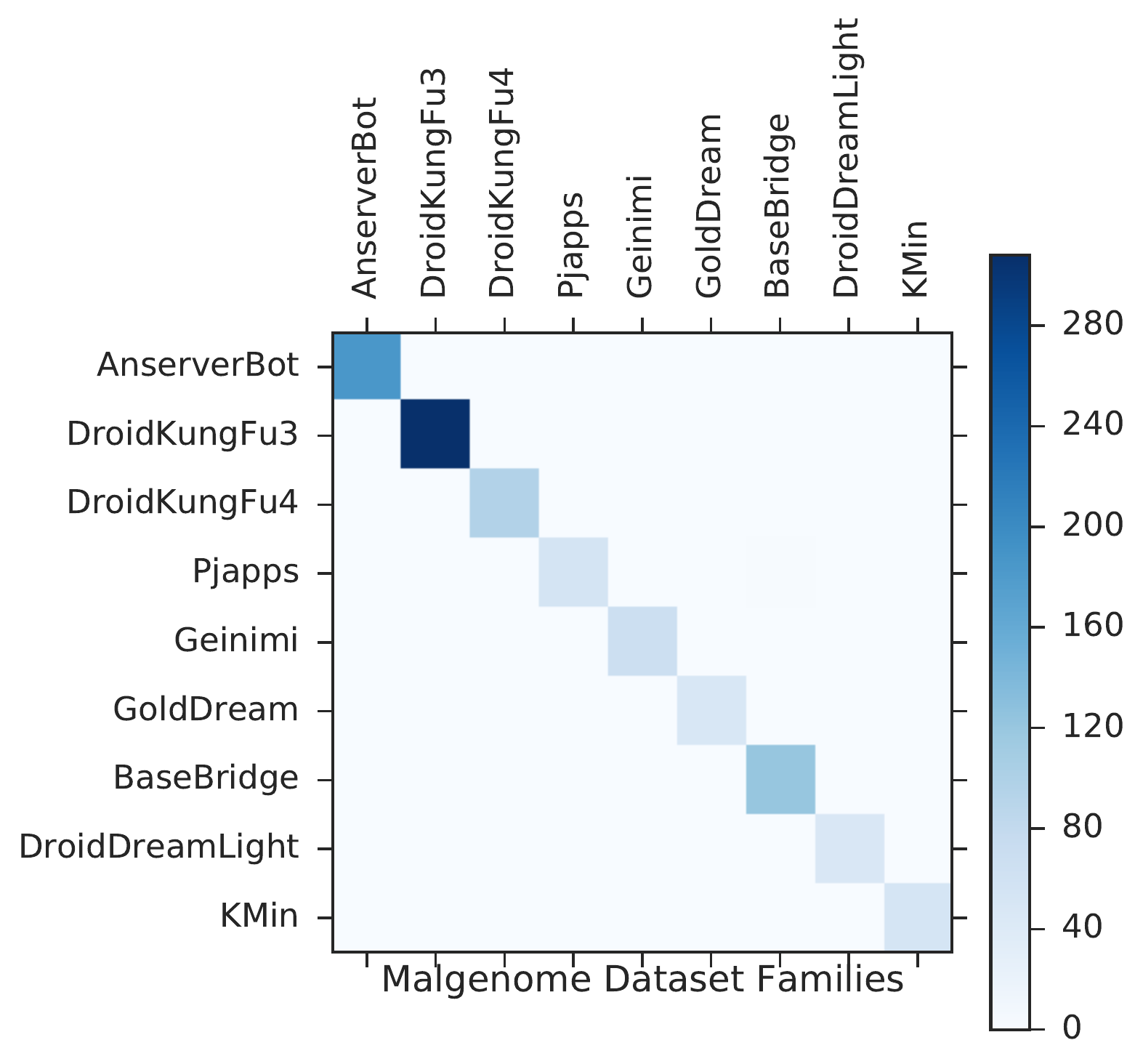}
        }\\
        \subfigure[Drebin] {%
           \label{fig:ben_precision}
           \includegraphics[width=0.47\textwidth]{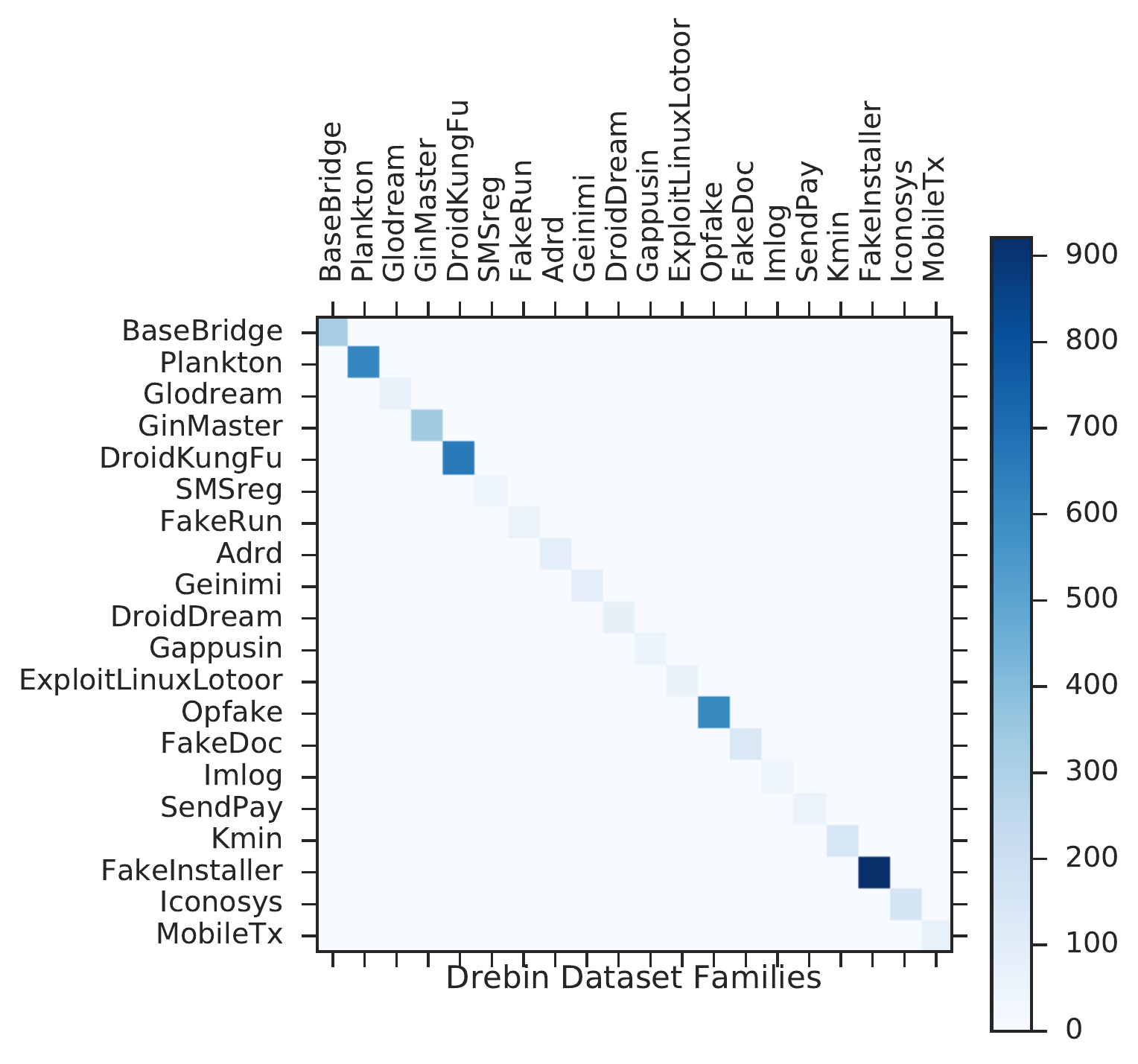}
        }\\
        \subfigure[MalDozer]{%
            \label{fig:ben_recall}
            \includegraphics[width=0.47\textwidth]{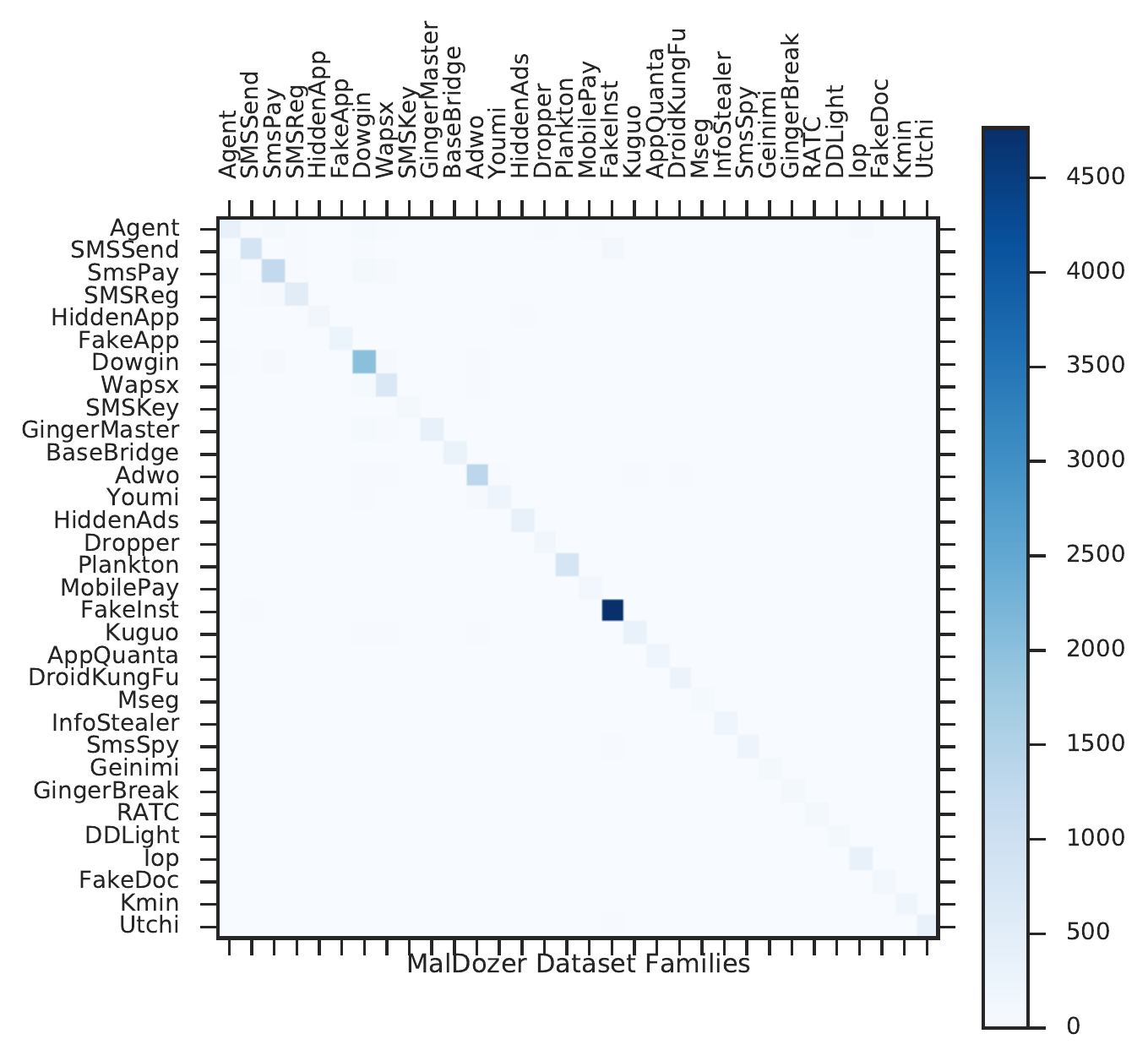}
        }
    \end{center}
    \caption{attribution confusion matrices}
   \label{fig:attribution_confusion_matrices}
\end{figure}

\end{document}